\documentclass[prd,aps,showpacs,onecolumn,12pt]{revtex4}
\usepackage{amssymb}
\usepackage{amsmath}
\usepackage{graphicx}
\usepackage{bm}

\begin{document}

\title{Bloch electrons interacting with an external electromagnetic field
and Bloch electrons in interaction. }
\author{Pierre Gosselin$^{1}$ and Herv\'{e} Mohrbach$^{2}$}

\address{$^1$Institut Fourier, UMR 5582 CNRS-UJF UFR de Math\'ematiques, Universit\'e Grenoble I,
BP74, 38402 Saint Martin
d'H\`eres, Cedex, France \\
$^2$ Laboratoire de Physique Mol\'eculaire et des Collisions,
ICPMB-FR CNRS 2843, Universit\'e Paul Verlaine-Metz, 57078 Metz
Cedex 3, France}

\begin{abstract}
We apply a general method developed recently for the derivation of the
diagonal representation of an arbitrary matrix valued quantum Hamiltonian to
the particular case of Bloch electrons in an external electromagnetic field.
We find the diagonal representation as a series expansion to the second
order in $\hbar .$ This result is the basis for the determination of the
effective in-band Hamiltonian of interacting Bloch electrons living in
different energy bands. Indeed, the description of effects such as magnetic
moment-moment interactions mediated by the magnetic part of the full
electromagnetic interaction requires a computation to second order in $\hbar
$. It is found that the electronic current is made of two contributions: the
first one comes from the velocity and the second one is a magnetic moment
current similar to the spin current for Dirac particles. This last
contribution is responsible for the interaction between magnetic moments
similarly to the spin-spin interaction in the Breit Hamiltonian for Dirac
electrons in interaction.
\end{abstract}

\maketitle

\section{Introduction:}

The properties of electrons in solids are usually described in the framework
of Bloch theory of electrons in a periodic potential. In particular, the
study of the dynamics and transport properties of Bloch electrons
perturbated by external fields led to important results for the
understanding of metals, semiconductors and insulators properties \cite%
{MERMIN}. For weak fields such that interband transitions called Zener
tunnelling are negligible, the dynamics of a Bloch electron in a given $n$th
band is usually based on the following semiclassical equations of motion
\begin{eqnarray}
\dot{\mathbf{r}} &=&\partial \mathcal{E}(\mathbf{k})/\hbar \partial \mathbf{k%
}  \notag \\
\hbar \dot{\mathbf{k}} &=&-e\mathbf{E}-e\dot{\mathbf{r}}\times \mathbf{B(r)}
\label{blochequations}
\end{eqnarray}%
where $\mathbf{E}$ and $\mathbf{B}$ are the electric and magnetic fields
respectively and $\mathcal{E}(\mathbf{k})=\mathcal{E}_{0}(\mathbf{k})-%
\mathbf{m(k).B}$ is the energy of the band including a correction due to the
orbital magnetic moment $\mathbf{m(k)}$. In the band energy $\mathcal{E}_{0}$
of the unperturbated crystal, the electron momentum $\mathbf{K}$ has been
substituted by the gauge covariant momentum $\mathbf{k}=\mathbf{K}+e\mathbf{A%
}(\mathbf{R})/\hbar $. This substitution has been first justified by Peierls
in the context of the tight binding model and for this reason is called the
Peierls substitution \cite{PEIERLS}. A full justification was later given by
Kohn \cite{KOHN}. A simpler version of the proof was later provided by
Blount, Roth and Wannier and Fredkin who could derived an approximate band
energy operator as an asymptotic series expansion in the fields strength as
well as general expressions for the first few terms in this series \cite%
{Blount}\cite{ROTH}\cite{WANNIER}. Their methods, although different, are
all based on some approximate unitary transformation of the initial
Hamiltonian of Bloch electrons in an electromagnetic field which eliminates
the interband matrix elements and leads to an effective diagonal-in-band
energy operator. The principal advantage of the Blount's method is that it
is an application of a general diagonalization scheme applicable to any kind
of matrix valued Hamiltonian whereas the two other ones are specific to the
solid states. In particular, Blount has also considered the case of a Dirac
particle in an electromagnetic field whose Hamiltonian was diagonalized to
the second order in the fields strength and first order in their first
derivatives. Later on, Weigert and Littlejohn developer a systematic method
to diagonalize general quantum Hamiltonian in a series expansion in $\hbar $%
\ \cite{LITTLEJOHN} instead in fields strength. an obvious advantage of this
kind of expansion is obviously that it can be valid for strong external
field but also that the semiclassical limit is readily obtained.
Unfortunately, the method \cite{LITTLEJOHN} involves formal series expansion
in terms of symbols of operators which makes the method very complicated for
practical applications. It is worth mentioning that recently a variant of
the Foldy Wouthuysen transformation valid for strong fields and based also
on an expansion in $\hbar $\ of the Dirac Hamiltonian was presented \cite%
{SILENKO}.

This is not the end story with regards to the Bloch equations $\left( \ref%
{blochequations}\right) $. Indeed, Karplus, Luttinger and Kohn \cite{KARPLUS}
predicted very early a spontaneous Hall effect in ferromagnetic materials
due to a corrective term to the velocity in Eq. $\left( \ref{blochequations}%
\right) $, known as the anomalous velocity. Later, Adams and Blount \cite%
{ADAMS}, by interpreting this term as resulting from the noncommutativity of
the intraband coordinate operators, derived new semiclassical equations of
motion for Bloch electrons with an anomalous velocity. However these
equations turn out to be correct only for external electric fields. It is
only recently that the correct equations of motion of Bloch electrons in the
presence of both electric and magnetic fields and including anomalous
velocity were derived by Chang and Niu \cite{NIU}. Indeed, using a
time-dependent variational principle in a Lagrangian formulation and a
description of the electron in terms of wave packets, they found the
following new equations of motion in the presence of electromagnetic fields
\begin{eqnarray}
\dot{\mathbf{r}} &=&\partial \mathcal{E}(\mathbf{k})/\hbar \partial \mathbf{k%
}-\dot{\mathbf{k}}\times \Theta (\mathbf{k})  \notag \\
\hbar \dot{\mathbf{k}} &=&-e\mathbf{E}-e\dot{\mathbf{r}}\times \mathbf{B(r)}
\label{niuequations}
\end{eqnarray}%
The correction term to the velocity $-\dot{\mathbf{k}}\times \Theta $ is the
anomalous velocity which is due to the presence of a Berry curvature $\Theta
(\mathbf{k})$ of electronic Bloch state in the given $n$th band, associated
to the electron motion in the $n$th energy band. For crystals with
simultaneous time-reversal and spatial inversion symmetry, the Berry
curvature and the magnetic moment vanish identically throughout the
Brillouin zone. This is the case of most applications in solid state
physics, but there are situations where these symmetries are not
simultaneously present like in GaAs where inversion symmetry is broken or in
ferromagnets which break time reversal symmetries. In the same way, the
presence of a strong magnetic field, the magnetic bloch bands corresponding
to the unperturbated system breaks the time inversion symmetries. In all
these cases, the dynamical and transport properties must be described by the
full equations of motion given by Eq. $\left( \ref{niuequations}\right) $.
Notice that\ even for crystals with simultaneous time-reversal and spatial
inversion symmetry, energy bands degeneracies can lead to a non vanishing
curvature and magnetization; a typical example is provided by the graphene
\cite{NIUGRAPHENE} This case is due to the presence of a topological Berry
phase associated to bands degeneracies \cite{BERRY}. This particular
situation is not considered in this paper.

The method developed in \cite{NIU} is by construction limited to the
semiclassical level, but the description of phenomena such as the
electromagnetic interaction of Bloch electrons, as will be discussed in this
paper, requires a theory which goes beyond the semiclassical approximation.
Recently we came back to the initial considerations of Blount and others
with regards to the diagonalization procedure for an arbitrary matrix valued
Hamiltonian $H$ (applicable to any kind of quantum system which has an
energy band spectrum) in the presence of external fields and a first
original method based on a differential equation of the diagonal in-band
energy operator with respect to Planck constant $\hbar $ was proposed in
\cite{SERIESPIERRE}. In this approach, where $\hbar $ is promoted as a
(formal) running parameter noted $\alpha $, one has to diagonalize $H$ at
the scale $\alpha $ where it is assumed that the canonical dynamical
operators satisfy the algebra $\left[ R_{\alpha }^{i},P_{\alpha }^{j}\right]
=i\alpha \delta _{ij}$. Relating two diagonalization processes for close
values of $\alpha ,$ leads to a differential equation of the required
diagonal Hamiltonian $\varepsilon _{\alpha }$ with respect to $\alpha $.
This differential equation has to be supplemented by an additional equation
which is the consequence of the unitarity condition of the matrix $U_{\alpha
}$ diagonalizing $H$ at the scale $\alpha .$ The resolution of this
differential equation can then be performed by a systematic series expansion
in $\hbar $, and in this way, at least in principle, an exact
diagonalization of arbitrary Hamiltonians can be achieved. This approach
reveals that the diagonal energy operator is most naturally expressible in
terms of covariant (noncanonical ) coordinates $\mathbf{r}=\mathbf{R+}\emph{A%
}_{\mathbf{R}}$ and momentum operators $\mathbf{p}=\mathbf{P+}\emph{A}_{%
\mathbf{P}}$ which are both corrected by Berry connections terms $\emph{A}_{%
\mathbf{R/P}}$ and which satisfy a non-commutative algebra. Particle motion
in this noncanonical phase space is obviously drastically modified (as in
Eq. $\left( \ref{niuequations}\right) $). Particularly interesting is the
fact that, in the semiclassical limit, which is often enough to get physical
insight to the problem considered, the diagonal Hamiltonian is obtained by a
straightforward integration of this differential equation. In this limiting
case, the differential approach turns out to be so powerful that the general
diagonal representation for an arbitrary matrix valued Hamiltonian in terms
of covariant operators and commutators between Berry connections could be
given (actually, this general formula was first derived by a direct
diagonalization procedure in \cite{PIERRESEMIDIAG}). This result allowed us
to deduce effective semiclassical Hamiltonians and to predict new phenomena
in various physical situations. First, the study of Bloch electron in
magnetic Bloch bands \cite{PIERREEUROPHYS} showed that besides the position
operator which get a Berry-phase contribution (as already shown by \cite{NIU}%
), the momentum in the band energy $\mathcal{E}(\mathbf{k})$ also has to be
replaced by a new Berry-dependent momentum operator $\mathbf{k}=\mathbf{K}+e%
\mathbf{A}/\hbar -\emph{A}_{\mathbf{P}}$ instead of the Peierls
substitution. It turns out that this result is essential for the correct
derivation of the full equations of motion for Bloch electrons Eq. $\left( %
\ref{niuequations}\right) $. Likewise, for electrons in graphene in a
magnetic field, it was also observed that it is in terms of $\mathbf{k}$
that the semiclassical quantification of the orbit has to be achieved \cite%
{PIERREGRAPHENE}.

At the semiclassical level other systems were also investigated, like Dirac
electrons in electromagnetic \cite{SERIESPIERRE} and gravitational fields
\cite{PIERREELECGRAVIT} with the discovery of a spin-magnetotorsion
coupling. The study of the photon in a static gravitational field when
polarization effect are taken into account predicts the gravitational
birefringence phenomenon where an helicity dependent anomalous velocity
deviates the photon from the usual Einsteinien geodesics \cite%
{PIERREPHOTONGRAVIT}. This kind of polarization effects which are called
spin Hall effect of light have been recently observed \cite{SPINHALLSCIENCE}%
. Despite these results, the applicability of the differential approach
beyond the semiclassical turns out to be very complicated. Even the
deduction of the diagonal representation of a generic Hamiltonian at second
order in $\hbar $ is a prohibitively difficult problem, although a solution
was found for simple practical applications like a photon in an homogeneous
isotropic media and Bloch electron in uniform electric field \cite%
{SERIESPIERRE}. But clearly, one can not expect to use this method in the
case of several Bloch electrons in electromagnetic interaction, a problem we
would like to consider in this paper.\

However, very recently, a new general and powerful method for the
diagonalization of an arbitrary matrix valued Hamiltonian has been proposed
by Gosselin and Mohrbach (referred as GM) \cite{DIAGOEXACT}.It leads to a
particularly compact and elegant exact expression for the required diagonal
energy operator.\textbf{\ }This approach is therefore particularly well
adapted to problems in solid state physics and to Dirac particles in
external fields. This last case can be considered intuitively as a simple
two bands versions (particles and anti-particles) of Bloch electrons in a
crystal. The philosophy behind this approach consists in mapping the initial
quantum system to a classical one which can be diagonalized and then to
return to the full quantum system. This method is not based on a
differential equation for the diagonal energy operator with respect to $%
\hbar $, but it also requires the introduction of new mathematical objects
like non-commuting operators which evolve with $\hbar $\ promoted as a
running variable. This new mathematical construction leads us to define a
differential calculus on a non-commutative space showing some similarities
with the stochastic calculus as both stress the role of second order terms.
This approach allows us to write both the diagonal Hamiltonian $\varepsilon
\left( \mathbf{x}\right) $ and the transforming matrix $U\left( \mathbf{x}%
\right) $ (where $\mathbf{x=(r,p)}$ is the phase space of the covariant
dynamical operators) as a result of the application of integro-differential
operators on $\varepsilon _{0}\left( \mathbf{X}_{0}\right) $ and $%
U_{0}\left( \mathbf{X}_{0}\right) $ respectively, i.e. $\varepsilon \left(
\mathbf{x}\right) =\widehat{O}\left( \varepsilon _{0}\left( \mathbf{X}%
_{0}\right) \right) $ and $U\left( \mathbf{x}\right) =\widehat{N}\left(
U_{0}\left( \mathbf{X}_{0}\right) \right) $. Matrices with the subscribe $0$
correspond to the operators replaced by classical commuting variables $%
\mathbf{X}_{0}\mathbf{=}\left( \mathbf{R}_{0}\mathbf{,P}_{0}\right) .$ The
only requirement of the method is the knowledge of $U_{0}\left( \mathbf{X}%
_{0}\right) $ at $\alpha =0$ which gives the diagonal form $\varepsilon
_{0}\left( \mathbf{X}_{0}\right) .$ Generally, these equations do not allow
to find directly $\varepsilon \left( \mathbf{x}\right) $, $U\left( \mathbf{x}%
\right) $, however, they allow us to produce the solutions for $\varepsilon
\left( \mathbf{x}\right) $ and $U\left( \mathbf{x}\right) $ recursively in a
series expansion in $\hbar .$ But contrary to the procedure in \cite%
{SERIESPIERRE}, it turns out that the expansion in a series of $\hbar $\ is
much more easier to obtain than by the successive integration of the
differential equation and therefore more convenient for getting higher order
contributions. Remarkably, it was also found that the exact expression for $%
\varepsilon \left( \mathbf{x}\right) $ is actually an exact solution of the
differential equation of \cite{SERIESPIERRE}. This result obviously places
the present approach on a firm base\cite{DIAGOEXACT}. Another interesting
feature of this approach is the confirmation of the fundamental role played
by Berry curvatures in these systems since the method results in an
effective diagonal Hamiltonian with Berry phase corrections as well as
noncommutative coordinates and momentum covariant operators as in previous
approaches \cite{SERIESPIERRE}\cite{PIERRESEMIDIAG}\textbf{).}

Although similar in spirit to Blount's method and in particular to Weigert
and Littlejohn one \cite{LITTLEJOHN}, the approach proposed in \cite%
{DIAGOEXACT} is essentially different as it is based on a very new
mathematical formulation. (The general method of \cite{LITTLEJOHN} leads
also to a diagonal in-band energy representation as\ a formal series
expansion written in terms of symbols of operators which makes the method
very complicated for practical applications). In our opinion this new
approach is more tractable for applications and, as an illustration of this
statement, a general in-band energy for any arbitrary Hamiltonian to the
second order in $\hbar $ was achieved in \cite{DIAGOEXACT} (higher order
expressions becomes again very cumbersome but could in principle be
computed). This expression will be the starting point for a straightforward
study of a single Bloch electron in an external electromagnetic field.
Another purpose of the present work is the adaptation of GM's results to the
case of Bloch electrons in interaction. Note that both problems can be
transposed to the case of Dirac electrons which are actually treated by the
same method in the another paper \cite{PIERREDIRAC}. It is obviously the
Coulomb (electrostatic) interaction between Bloch (Dirac) electrons that
dominates over the magnetic one, so that a first order diagonalization seems
to be sufficient. But, in the presence of non vanishing electronic magnetic
moments, other effects like moment-moment interactions mediated by the
magnetic part of the full electromagnetic self-interaction are expected.
This comes out by analogy with the spin-spin interaction in the Breit
Hamiltonian of non-relativistic Dirac particles \cite{BREIT} which is
recover in the non-relativistic limit \cite{PIERREDIRAC}. Obviously these
kind of interactions are of second order in $\hbar $ and a diagonalization
procedure which goes to this order is necessary.

Before starting, two points are worth mentioning. As already noted in \cite%
{LITTLEJOHN} but also in \cite{PIERRESEMIDIAG}\cite{SERIESPIERRE}, there is
certain latitude in finding $U$ which reflects a kind of gauge invariance of
the method. Because of this freedom two equivalent diagonalizing operators
leads to two different forms for the diagonal Hamiltonian but of course to
the same eigenvalues. This is similar to the Schrodinger equation in a
magnetic field where the Hamiltonian and the wave function gauge dependence
combines itself to give gauge independent energy levels. In our case it
turns out that this freedom is only present at the second order in $\hbar $,
because the first order diagonalization is performed with the zero order
matrix $U_{0}\left( \mathbf{X}_{0}\right) $ which can be uniquely defined.
Actually this gauge dependence can be included in the gauge covariant
dynamical operator $\mathbf{x=(r,p)}$. Then the diagonal in-bands energy
operator is uniquely defined when it is written in terms of $\mathbf{x}$
instead of the canonical operators $\mathbf{X.}$ As a particular gauge
choice can be made on the ground of simplicity and convenience, here as in
previous works \cite{SERIESPIERRE}\cite{DIAGOEXACT} the reality condition of
diagonal elements (the anti-hermitian diagonal elements are setting to zero)
of $U$ is imposed.

The second point we would like to mention is that, in order to simplify the
expressions, only time independent electromagnetic fields are considered,
but results can be easily extended to include time dependent interactions.
In addition, as the ultimate goal is to consider Bloch electrons in
interaction through an internal electromagnetic field, the time dependence
of the vector potential can be safely neglected as all retarded effects in
the electromagnetic interaction are negligible owing to the fact that Bloch
electrons are non-relativistic.

The paper is organized as follows. In the next section we give a reminder of
the diagonalization procedure of \cite{DIAGOEXACT} for an arbitrary matrix
valued Hamiltonian. We provide some detailed formulas for the diagonalized
energy operator at the second order in $\hbar $ as well as for the Berry
phases at this order. Section 3 applies this formalism to the case of a
Bloch electron in an external electromagnetic field. The diagonalized
Hamiltonian at the second order in $\hbar $ is written in terms of the
transformed dynamical variables and magnetization operators. In section 4,
we consider the case of $P$ Bloch electrons interacting through an internal
electromagnetic field. Diagonalizing the matter part of the Hamiltonian to
the second order in $\hbar $ and solving for the electromagnetic field
yields the $P$ particles effective Hamiltonian. Last section is for the
conclusion.

\section{Diagonalization of an arbitrary matrix valued Hamiltonian}

To start with, an outline of the approach developed by Gosselin and Mohrbach
to diagonalize formally an arbitrary matrix valued Hamiltonian is given,
with the notations of GM \cite{DIAGOEXACT}. Let consider an arbitrary
quantum mechanical system whose state space is a tensor product $L^{2}\left(
\mathcal{R}^{3}\right) \otimes \mathcal{V}$\ with $\mathcal{V}$\ some
internal space. In other words, the Hamiltonian of this system can be
written as a matrix $H\left( \mathbf{R,P}\right) $\ of size $\dim \mathcal{V}
$\ whose elements are operators depending on a couple of canonical variables
$\mathbf{X}=\left( \mathbf{R,P}\right) .$ The archetype example is usually
the Dirac Hamiltonian with $\mathcal{V}=C^{4}$, but as shown in \cite%
{PIERRESEMIDIAG} the following set up fits with all system presenting an
energy band spectrum, as for Bloch electron (where $\mathcal{V}$ correspond
to the energy band indices), a system which is the main concern of the
present paper.

In \cite{DIAGOEXACT}, a method to find an unitary matrix $U\left( \mathbf{X}%
\right) $ to diagonalize any arbitrary matrix-valued quantum Hamiltonian $%
H\left( \mathbf{X}\right) $ such that $\varepsilon \left( \mathbf{X}\right)
=UH\left( \mathbf{X}\right) U^{+}$ is the diagonal in-band energy operator
was achieved.

As explained in the introduction, the principle of this method is to link
continuously an Hamiltonian in which the variables are considered as
classical (i.e. $\hbar =0$) to the true Hamiltonian we aim at diagonalizing
(that is at scale $\hbar $). The idea is to diagonalize the Hamiltonian for $%
\hbar =0$, which appears in general to be much easier, and then to come back
to scale $\hbar $\ to obtain the required Band Hamiltonian. To perform this
program, we have to proceed in an indirect way. Actually, we first need to
introduce a family of canonical variables $\left( \mathbf{R}_{\alpha }%
\mathbf{,P}_{\alpha }\right) $\ indexed by a continuous parameter $\alpha
\in \left[ 0,\hbar \right] $, such that the commutators are given by $\left[
\mathbf{R}_{\alpha }^{i}\mathbf{,P}_{\alpha }^{i}\right] =i\alpha $\ and
then, to introduce for any arbitrary function $F\left( \mathbf{R}_{\alpha }%
\mathbf{,P}_{\alpha }\right) $\ both a notion of differentiation and
integration describing the variation of $F\left( \mathbf{R}_{\alpha }\mathbf{%
,P}_{\alpha }\right) $\ as $\alpha $ varies. These notions have to take into
account the fact that the commutation relations are depending on $\alpha $.
It leads us naturally to introduce the notion of infinitesimal non
commutative canonical variables $\left( d\mathbf{R}_{\alpha }\mathbf{,dP}%
_{\alpha }\right) $\ as well as a form of differential calculus presenting
some formal analogy with the non commutative stochastic calculus. The
introduction of these differentials notion, will allow to connect ultimately
our Hamiltonian at scales $0$\ and $\hbar $.

More precisely now, we introduce a space of non commuting infinitesimal
operators $dX_{\alpha }^{i}\equiv \left\{ dR_{\alpha }^{i},dP_{\alpha
}^{i}\right\} $ $\ i=1,2,3$ indexed by a continuous parameter $\alpha $,
that satisfy the following infinitesimal Heisenberg algebra with a reversed
sign $\left[ dR_{\alpha }^{i},dP_{\alpha \prime }^{j}\right] =-id\alpha
\delta _{\alpha ,\alpha \prime }\delta _{ij}\ $and\ $\left[ dR_{\alpha
}^{i},dR_{\alpha \prime }^{j}\right] =\left[ dP_{\alpha }^{i},dP_{\alpha
\prime }^{j}\right] =0.$ From it, we define a set of running coordinate and
momentum operators by writing the following formal sums $R_{\alpha
}^{i}=R^{i}-\int_{\alpha }^{\hbar }dR_{\lambda }^{i}$, and $P_{\alpha
}^{i}=P^{i}-\int_{\alpha }^{\hbar }dP_{\lambda }^{i}$ with the choice of
convention $dR_{\alpha }^{i}=R_{\alpha }^{i}-R_{\alpha -d\alpha }^{i}$ and $%
dP_{\alpha }^{i}=P_{\alpha }^{i}-P_{\alpha -d\alpha }^{i}$, so that the
running operators satisfy $\left[ R_{\alpha }^{i},P_{\alpha }^{j}\right]
=i\alpha \delta _{ij}$ and $\left[ R_{\alpha }^{i},R_{\alpha }^{j}\right] =%
\left[ P_{\alpha }^{i},P_{\alpha }^{j}\right] =0.$ For $\alpha =\hbar $ we
recover the usual canonical operators $R^{i}\equiv R_{\hbar }^{i}$and $%
P^{i}\equiv P_{\hbar }^{i}$ which evidently satisfy the canonical Heisenberg
algebra. The differential of an arbitrary function $F\left( \mathbf{X}%
_{\alpha },\alpha \right) $ where $X_{\alpha }^{i}\equiv \left\{ R_{\alpha
}^{i},P_{\alpha }^{i}\right\} $ on this space is given by

\begin{eqnarray}
dF\left( \mathbf{X}_{\alpha },\alpha \right) &=&\sum_{i=1}^{6}\nabla
_{X_{\alpha }^{i}}F\left( \mathbf{X}_{\alpha },\alpha \right) dX_{\alpha
}^{i}-\frac{1}{4}\sum_{i,j=1}^{6}\nabla _{X_{\alpha }^{i}}\nabla _{X_{\alpha
}^{j}}F\left( \mathbf{X}_{\alpha },\alpha \right) \left( dX_{\alpha
}^{i}dX_{\alpha }^{j}+dX_{\alpha }^{i}dX_{\alpha }^{j}\right)  \notag \\
&&+\left( \frac{\partial F\left( \mathbf{X}_{\alpha },\alpha \right) }{%
\partial \alpha }+\left\langle F\left( \mathbf{X}_{\alpha },\alpha \right)
\right\rangle \right) d\alpha  \label{dalphaY}
\end{eqnarray}
with $i$,$j=1..6$. We also assume that $X_{\alpha }^{i}\equiv R_{\alpha
}^{i} $ for $i=1,2,3$ and $X_{\alpha }^{i}\equiv P_{\alpha }^{i}$ for $%
i=4,5,6$ .The notation $\left\langle F\left( \mathbf{X}_{\alpha },\alpha
\right) \right\rangle $ (which in \cite{SERIESPIERRE} was corresponding to
the operation $-\frac{i}{2}Asym\nabla _{R_{i}}\nabla _{P^{i}}F\left( \mathbf{%
X}_{\alpha },\alpha \right) $) is defined as a specific procedure on a
series expansion of $F$ in the variables $R_{\alpha }^{i}$, $P_{\alpha }^{i}$
in the following way : let $F$ be a sum of monomials of the kind $%
M_{1}\left( \mathbf{R}_{\alpha }\right) M_{2}\left( \mathbf{P}_{\alpha
}\right) M_{3}\left( \mathbf{R}_{\alpha }\right) ....$ the $M_{i}$ being
arbitrary monomials in $R_{\alpha }$ or $P_{\alpha }$ alternatively. Let the
operator $\nabla _{R_{i}}\nabla _{P^{i}}$ acts on such an expression by
deriving all combinations of one monomial in $\mathbf{R}_{\alpha }$ and one
monomial in $\mathbf{P}_{\alpha }$. For each of these combinations, insert a
$dR_{\alpha }^{i}$ at the place where the derivative $\nabla _{R_{i}}$ is
acting and in a same manner a $dP_{\alpha }^{j}$ at the place where the
derivative $\nabla _{P^{i}}$ is acting. This leads to an expression with two
kind of terms, one kind being proportional to the $dR_{\alpha
}^{i}dP_{\alpha }^{j},$ and the second proportional to $dP_{\alpha
}^{j}dR_{\alpha }^{i}.$ Then rewrite this expression in terms of $dR_{\alpha
}^{i}dP_{\alpha }^{j}+dP_{\alpha }^{j}dR_{\alpha }^{i}$ and $dR_{\alpha
}^{i}dP_{\alpha }^{j}-dP_{\alpha }^{j}dR_{\alpha }^{i}=-i\delta ^{ij}d\alpha
$. Then $\left\langle F\left( \mathbf{X}_{\alpha },\alpha \right)
\right\rangle $ is defined as minus the contributions of terms proportional
to $-i\delta ^{ij}d\alpha $ in the computation in the procedure just
considered. This definition implies a procedure which is clearly dependent
of the symmetrization chosen for the expansion of $F$.

To make the definition of $\left\langle F\left( \mathbf{X}_{\alpha },\alpha
\right) \right\rangle $\ clearer, consider some important practical
examples. If the function $F$\ has the following form $F=\frac{1}{2}\left(
A\left( \mathbf{R}_{\alpha }\right) B\left( \mathbf{P}_{\alpha }\right)
+B\left( \mathbf{P}_{\alpha }\right) A\left( \mathbf{R}_{\alpha }\right)
\right) $\ which corresponds to a frequent choice of symmetrization in $%
R_{\alpha }$\ and $P_{\alpha }$, then $\left\langle F\left( \mathbf{X}%
_{\alpha },\alpha \right) \right\rangle =\frac{i}{4}\left[ A\left( \mathbf{R}%
_{\alpha }\right) ,B\left( \mathbf{P}_{\alpha }\right) \right] $. Another
choice of symmetrization leads in general to a different result. For
instance, if we rewrite the same function $F$ in a fully symmetrized form in
$R_{\alpha }$\ and $P_{\alpha }$\ (that is invariant by all permutations in $%
R_{\alpha }$\ and $P_{\alpha }$) which is also often used, we have now have
a different result since $\left\langle F\left( \mathbf{X}_{\alpha },\alpha
\right) \right\rangle =0$.

Nevertheless, this dependence of $\left\langle F\left( \mathbf{X}_{\alpha
},\alpha \right) \right\rangle $ in the symmetrization choice is not
astonishing at all. \ Actually changing the symmetrization of a function $%
F\left( \mathbf{X}_{\alpha },\alpha \right) $ introduces some explicit terms
in $\alpha $ which changes also the term $\partial _{\alpha }Fd\alpha $
present in the differential Eq. $\left( \ref{dalphaY}\right) $. As a
consequence, neither the partial derivative with respect to $\alpha $, nor
the bracket are invariant by a change of form. But, what is invariant is the
sum $\partial _{\alpha }F+\left\langle F\right\rangle $. This assertion is
shown in \cite{DIAGOEXACT}.

Now, what really matters for us in Eq. $\left( \ref{dalphaY}\right) $ is
this invariant term proportional to $d\alpha $ this is why we define an
expectation operation $\mathcal{E}\left( .\right) $ so that
\begin{equation}
\mathcal{E}\left( dF\left( \mathbf{X}_{\alpha },\alpha \right) \right) =%
\mathcal{E}\left( \left( \frac{\partial F\left( \mathbf{X}_{\alpha },\alpha
\right) }{\partial \alpha }+\left\langle F\left( \mathbf{X}_{\alpha },\alpha
\right) \right\rangle \right) d\alpha \right)  \label{epsidf}
\end{equation}

Combining the expectation operator as well as the differential allows to
relate a function evaluated at the physical scale $\hbar $, $F\left( \mathbf{%
X}_{\hbar },\hbar \right) $ to that same function evaluated at the scale $0$%
, $F\left( \mathbf{X}_{0},0\right) $ that is when the canonical variables
are considered as classical ones. Actually by integration of the previous
relation one has:
\begin{equation}
F\left( \mathbf{X}_{\hbar },\hbar \right) =\mathcal{E}\left\{ F\left(
\mathbf{X}_{0},0\right) +\int_{0}^{\hbar }\left( \left( \frac{\partial
F\left( \mathbf{X}_{\alpha },\alpha \right) }{\partial \alpha }+\left\langle
F\left( \mathbf{X}_{\alpha },\alpha \right) \right\rangle \right) d\alpha
\right) \right\}
\end{equation}
(we use a property that the expectation operator satisfies $\mathcal{E}%
\left( F\left( \mathbf{X}_{\hbar },\hbar \right) \right) =F\left( \mathbf{X}%
_{\hbar },\hbar \right) $).

Here recall that we denote by $\mathbf{X}_{\alpha }$ the dynamical variables
when\ $\mathbf{P}$ and $\mathbf{R}$ are considered as commuting classical
variables that is when $\hbar =\alpha $. This last formula is the starting
point of the method. Iterating this relation, by successive differentiations
and integrations, one can relate a full quantum function $F\left( \mathbf{X}%
_{\hbar },\hbar \right) $ to the same function evaluated with classical
variables $\mathbf{X}_{0}$. This will prove very useful in a problem of
diagonalization of a matricial Hamiltonian since in that case, the
diagonalization when the canonical variables $\mathbf{X}_{0}$ commute
reduces to the diagonalization of an usual matrix of finite size. Thus
diagonalizing our Hamiltonian at scale $0$, that is finding an $F\left(
\mathbf{X}_{0},0\right) $ is in general an easier task.

However, this not the end of the story. Having found a way to relate $%
F\left( \mathbf{X}_{\hbar },\hbar \right) $ to its classical counterpart is
not enough since we want ultimately to recover expressions of interest
evaluated at $\mathbf{X}_{\hbar }$ not at the $\hbar =0$ scale. This kind of
coming back process after a \textquotedblright classical\textquotedblright\
diagonalization is performed by an other operation defined in \cite%
{DIAGOEXACT}. It allows in the previous integral relation to replace, inside
the expectation, $\mathbf{X}_{0}$ and $\mathbf{X}_{\alpha }$ by $\mathbf{X}%
_{\hbar }$ at the price of a modification of the expression inside the
integral. Define the exponentiated Bracket plus Shift operator (EBS) between
$\alpha _{2}$ and $\alpha _{1}$ as acting on any function $F\left( \mathbf{X}%
_{\alpha _{2}},\alpha _{2}\right) $ to yield an other function depending on $%
\left( \mathbf{X}_{\alpha _{1}},\alpha _{2}\right) $, $\alpha _{1}>\alpha
_{2}$ :
\begin{eqnarray*}
\exp \left( -\left\langle .\right\rangle _{\alpha _{2}\rightarrow \alpha
_{1}}^{S}\right) &\equiv &T\exp \left( -\int_{\alpha _{2}}^{\alpha _{1}}S_{%
\mathbf{X}_{\alpha _{1}}}\left\langle .\right\rangle _{\alpha }S_{\mathbf{X}%
_{\alpha }}d\alpha \right) \\
&=&\sum \int_{\alpha _{2}<\beta _{n}<...\beta _{1}<\alpha _{1}}\left[ S_{%
\mathbf{X}_{\alpha _{1}}}\left\langle .\right\rangle _{\beta _{n}}S_{\mathbf{%
X}_{\beta _{n}}}\right] ...\left[ \left\langle .\right\rangle _{\beta
_{1}}S_{\mathbf{X}_{\beta _{1}}}\right] d\beta _{1}...d\beta _{n}
\end{eqnarray*}
where the Shift operation $S_{\mathbf{X}_{\alpha _{1}}}$ sets the dynamical
variables $\mathbf{X}_{\alpha }$ to $\mathbf{X}_{\alpha _{1}}$\ and
satisfies $S_{\mathbf{X}_{\alpha }}S_{\mathbf{X}_{\beta }}=S_{\mathbf{X}%
_{\alpha }}$ whatever the values of $\alpha $ and $\beta $. Apart from the
repeated application of the Bracket $\left\langle .\right\rangle _{\alpha }$
the EBS operation has the virtue to shift progressively the variables from $%
\mathbf{X}_{0}$, the \textquotedblright classical
variables\textquotedblright\ to $\mathbf{X=X}_{\hbar }$ the full quantum
variables. Actually its main property is the following :
\begin{equation*}
\mathcal{E}F\left( \mathbf{X}_{\alpha _{2}},\alpha _{2}\right) =\mathcal{E}%
\exp \left( -\left\langle .\right\rangle _{\alpha _{2}\rightarrow \alpha
_{1}}^{S}\right) F\left( \mathbf{X}_{\alpha _{2}},\alpha _{2}\right)
\end{equation*}
so that it can of course be specialized to :
\begin{equation*}
\mathcal{E}F\left( \mathbf{X}_{0},0\right) =\mathcal{E}\exp \left(
-\left\langle .\right\rangle _{0\rightarrow \hbar }^{S}\right) F\left(
\mathbf{X}_{0},0\right)
\end{equation*}
These two formulas can be understood intuitively as follows. The EBS
operation changes the function (by the action of the bracket defined above)
but also changes progressively the variables from $\mathbf{X}_{\alpha _{2}}$
to $\mathbf{X}_{\alpha _{1}}$ (through the shift operator). As a
consequence, and despite the appearances, the expression in the right hand
side $\mathcal{E}\exp \left( -\left\langle .\right\rangle _{\alpha
_{2}\rightarrow \alpha _{1}}^{S}\right) F\left( \mathbf{X}_{\alpha
_{2}},\alpha _{2}\right) $ is a function of $\mathbf{X}_{\alpha _{1}}$.
Moreover, both the EBS operation and the shift of variable compensate each
over to produce the equality with the left hand side.

With this mathematical construction in hand it was possible to write the
solution of our diagonalization procedure for a general matrix valued
Hamiltonian through an unitary transformation $U$ as the solution of the
following system of integro-differential equations. Introducing $U_{0}\left(
\mathbf{X}_{0}\right) $ the diagonalization matrix when $\hbar =0$, such
that the classical energy matrix obtained as $\varepsilon _{0}\left( \mathbf{%
X}_{0}\right) =U_{0}H_{0}\left( \mathbf{X}_{0}\right) U_{0}^{+}$ is a
diagonal matrix, we could write :
\begin{eqnarray}
\varepsilon \left( \mathbf{X}\right) &=&\mathcal{E}\left( \left[ \mathcal{T}%
\exp \left[ \int_{0<\alpha <\hbar }e^{-\left\langle .\right\rangle _{\alpha
\rightarrow \hbar }^{S}}O_{\alpha }e^{-\left\langle .\right\rangle
_{0\rightarrow \alpha }^{S}}d\alpha \right] \right] \varepsilon _{0}\left(
\mathbf{X}_{0}\right) \right)  \label{EE} \\
U\left( \mathbf{X}\right) &=&\mathcal{E}\left( \left[ \mathcal{T}\exp \left[
\int_{0<\alpha <\hbar }e^{-\left\langle .\right\rangle _{\alpha \rightarrow
\hbar }^{S}}N_{\alpha }e^{-\left\langle .\right\rangle _{0\rightarrow \alpha
}^{S}}d\alpha \right] \right] U_{0}\left( \mathbf{X}_{0}\right) \right)
\label{U}
\end{eqnarray}
$\mathcal{T}$ is the notation for the time ordered product, the operator $%
\exp \left( -\left\langle .\right\rangle _{\alpha _{2}\rightarrow \alpha
_{1}}^{S}\right) $ acts as explained above, and $O_{\alpha }$ and $N_{\alpha
}$ act in the \ following way :

The operation $O_{\alpha }\varepsilon _{0}\left( \mathbf{X}_{\alpha }\right)
=\left( \frac{\partial }{\partial \alpha }+\left\langle .\right\rangle
\right) \varepsilon _{\alpha }\left( \mathbf{X}_{\alpha }\right) $ can also
be written as $O_{\alpha }\varepsilon _{0}\left( \mathbf{X}_{\alpha }\right)
=\left( T_{\alpha }+M_{\alpha }\right) \varepsilon _{0}\left( \mathbf{X}%
_{\alpha }\right) $ with a translation\ operator $T$ and a
\textquotedblright magnetization\textquotedblright\ $M$ operator (this
terminology is explained in \cite{SERIESPIERRE}) where
\begin{eqnarray}
T_{\alpha }\varepsilon _{0}\left( \mathbf{X}_{\alpha }\right) &=&\mathcal{P}%
_{+}\left\{ \frac{1}{2}\left( \mathcal{A}_{\alpha }^{R_{l}}\nabla
_{R_{l}}\varepsilon _{0}\left( \mathbf{X}_{\alpha }\right) +\nabla
_{R_{l}}\varepsilon _{0}\left( \mathbf{X}_{\alpha }\right) \mathcal{A}%
_{\alpha }^{R_{l}}+\mathcal{A}_{\alpha }^{P_{l}}\nabla _{P_{l}}\varepsilon
_{0}\left( \mathbf{X}_{\alpha }\right) +\nabla _{P_{l}}\varepsilon
_{0}\left( \mathbf{X}_{\alpha }\right) \mathcal{A}_{0}^{P_{l}}\right)
\right\}  \notag \\
M_{\alpha }\varepsilon _{0}\left( \mathbf{X}_{\alpha }\right) &=&\frac{i}{4}%
\left\{ \mathcal{P}_{+}\left\{ \left[ \varepsilon _{0}\left( \mathbf{X}%
_{\alpha }\right) ,\mathcal{A}_{\alpha }^{R_{l}}\right] \mathcal{A}_{\alpha
}^{P_{l}}-\left[ \varepsilon _{0}\left( \mathbf{X}_{\alpha }\right) ,%
\mathcal{A}_{\alpha }^{P_{l}}\right] \mathcal{A}_{\alpha }^{R_{l}}\right\}
+H.C.\right\}  \notag \\
&&+\mathcal{P}_{+}\left[ U_{\alpha }\left( \left( \frac{\partial }{\partial
\alpha }+\left\langle .\right\rangle \right) H\left( \mathbf{X}_{\alpha
}\right) \right) U_{\alpha }^{+}\right]  \label{TM}
\end{eqnarray}
These equations require some explanations. First, the operators $\mathcal{A}%
_{\alpha }^{\mathbf{X}}$ are given by
\begin{eqnarray}
\mathcal{A}_{\alpha }^{\mathbf{R}}\left( \mathbf{X}_{\alpha }\right) &=&i%
\left[ U_{\alpha }\left( \mathbf{X}_{\alpha }\right) \mathbf{\nabla }_{%
\mathbf{P}}U_{\alpha }^{+}\left( \mathbf{X}_{\alpha }\right) \right]
\label{Ber} \\
\text{and }\mathcal{A}_{\alpha }^{\mathbf{P}}\left( \mathbf{X}_{\alpha
}\right) &=&-i\left[ U_{\alpha }\left( \mathbf{X}_{\alpha }\right) \mathbf{%
\nabla }_{\mathbf{R}}U_{\alpha }^{+}\left( \mathbf{X}_{\alpha }\right) %
\right]  \notag
\end{eqnarray}
and $\mathcal{P}_{+}$ and $\mathcal{P}_{-}$ are respectively the projectors
on the diagonal and off-diagonal elements of matrices. From Eq. $\left( \ref%
{TM}\right) $ one can deduce the following relation
\begin{equation*}
\mathcal{E}\left( \mathcal{T}\exp \left[ \int_{0<\alpha <\hbar }T_{\alpha
}d\alpha \right] \varepsilon _{0}\left( \mathbf{X}_{0}\right) \right) =%
\mathcal{E}\varepsilon _{0}\left( \mathbf{x}\right)
\end{equation*}
where $\mathbf{x=}\left( \mathbf{r},\mathbf{p}\right) $ and $\mathbf{r}$ and
$\mathbf{p}$ are new covariant coordinate and momentum operators defined in
the following way :
\begin{eqnarray}
\mathbf{r} &\equiv &\mathbf{R+}\emph{A}^{\mathbf{R}}  \notag \\
\mathbf{p} &\equiv &\mathbf{P+}\emph{A}^{\mathbf{P}}  \label{RP1}
\end{eqnarray}
The Berry connections terms being defined as :
\begin{eqnarray*}
\mathbf{\emph{A}^{\mathbf{R}}} &=&\int_{0<\alpha <\hbar }\mathcal{P}_{+}%
\left[ \mathcal{A}_{\alpha }^{\mathbf{R}}\right] d\alpha +\int_{0<\alpha
<\hbar }\frac{1}{2}\left[ \left[ \mathcal{P}_{+}\left[ \mathcal{A}_{\alpha
}^{\mathbf{X}}\right] .\mathbf{\nabla }_{\mathbf{X}}\int_{0<\alpha
_{1}<\alpha }\mathcal{P}_{+}\left[ \mathcal{A}_{\alpha _{1}}^{\mathbf{R}}%
\right] \right] +H.C.\right] d\alpha _{1}d\alpha +... \\
\mathbf{\emph{A}^{\mathbf{P}}} &=&\int_{0<\alpha <\hbar }\mathcal{P}_{+}%
\left[ \mathcal{A}_{\alpha }^{\mathbf{P}}\right] d\alpha +\int_{0<\alpha
<\hbar }\frac{1}{2}\left[ \left[ \mathcal{P}_{+}\left[ \mathcal{A}_{\alpha
}^{\mathbf{X}}\right] .\mathbf{\nabla }_{\mathbf{X}}\int_{0<\alpha
_{1}<\alpha }\mathcal{P}_{+}\left[ \mathcal{A}_{\alpha _{1}}^{\mathbf{P}}%
\right] \right] +H.C.\right] d\alpha _{1}d\alpha +...
\end{eqnarray*}
and $\mathbf{X}$ denotes the vector $\left( \mathbf{R,P}\right) $. In fact
in \cite{DIAGOEXACT} we show that $\mathbf{\emph{A}^{\mathbf{R}}}$ and $%
\mathbf{\emph{A}^{\mathbf{P}}}$ have to be corrected by a third order in $%
\hbar $ terms $\delta \mathbf{\emph{A}^{\mathbf{R}}}$ and $\delta \mathbf{%
\emph{A}^{\mathbf{P}}}$ but these terms will always be neglected here.

Therefore, the operator $T$ as a part of $O$ naturally leads to the
emergence of the covariant dynamical coordinates which turn out the be the
physical dynamical variables of particles as shown in several situations
\cite{PIERRESEMIDIAG}\cite{SERIESPIERRE}\cite{PIERREPHOTONGRAVIT}\cite%
{PIERREELECGRAVIT}.

Second, using the gauge setting to zero the anti-hermitian diagonal elements
of $U_{\alpha }$ \cite{DIAGOEXACT} we have
\begin{eqnarray}
N_{\alpha }U_{\alpha } &=&\left[ \left( \frac{\partial }{\partial \alpha }%
+\left\langle .\right\rangle \right) U_{\alpha }\right] =-\left[
.,\varepsilon _{\alpha }\right] ^{-1}.\left[ \mathcal{P}_{-}\left\{ \frac{1}{%
2}\left( \mathcal{A}_{\alpha }^{R_{l}}\nabla _{R_{l}}\varepsilon _{\alpha
}+\nabla _{R_{l}}\varepsilon _{\alpha }\mathcal{A}_{\alpha }^{R_{l}}+%
\mathcal{A}_{\alpha }^{P_{l}}\nabla _{P_{l}}\varepsilon _{\alpha }+\nabla
_{P_{l}}\varepsilon _{\alpha }\mathcal{A}_{\alpha }^{P_{l}}\right) \right\}
\right.  \notag \\
&&\left. +\frac{i}{4}\mathcal{P}_{-}\left\{ \left[ \varepsilon _{\alpha },%
\mathcal{A}_{\alpha }^{R_{l}}\right] \mathcal{A}_{\alpha }^{P_{l}}-\left[
\varepsilon _{\alpha },\mathcal{A}_{\alpha }^{P_{l}}\right] \mathcal{A}%
_{\alpha }^{R_{l}}\right\} +H.C.\right] -\frac{i}{4}\mathcal{P}_{-}\left\{ %
\left[ \mathcal{A}_{\alpha }^{R_{l}},\mathcal{A}_{\alpha }^{P_{l}}\right]
U_{\alpha }\right\}  \label{NU}
\end{eqnarray}
where the inverse of the commutator operation $\left[ ,\varepsilon _{\alpha }%
\right] $ has the following properties \
\begin{eqnarray}
\left[ \left[ .,\varepsilon _{\alpha }\right] ^{-1}.M,\varepsilon _{\alpha }%
\right] &=&\left[ .,\varepsilon _{\alpha }\right] ^{-1}.\left[ M,\varepsilon
_{\alpha }\right] =M\text{\ \ \ for \ \ }\left[ M,\varepsilon _{\alpha }%
\right] \neq 0\text{ }  \notag \\
\left[ .,\varepsilon _{\alpha }\right] ^{-1}.M &=&0\text{ \ \ if \ }\left[
M,\varepsilon _{\alpha }\right] =0  \label{NU2}
\end{eqnarray}
for an arbitrary matrix valued operator $M$. It means that the operator $%
\left[ .,\varepsilon _{\alpha }\right] ^{-1}$ acts on the space of
endomorphism of the state space in the following way : it is zero when
acting on the kernel of the operator $\left[ .,\varepsilon _{\alpha }\right]
$ whose action is to compute the commutator with $\varepsilon _{\alpha }$,
and is the inverse of $\left[ .,\varepsilon _{\alpha }\right] $ on the
complementary subspace of the kernel.

Note that, as it will appear clearly later on, having both $\varepsilon
\left( \mathbf{X}\right) $ and $U\left( \mathbf{X}\right) $ at order $n$ in $%
\hbar $, and reinserting in the exponential of Eqs. $\left( \ref{EE}\right)
\left( \ref{U}\right) $ allows us to find $\varepsilon \left( \mathbf{X}%
\right) $ and $U\left( \mathbf{X}\right) $ at order $n+1$ in $\hbar $. A
needed assumption for this procedure to work is that the diagonalization $%
\varepsilon _{0}\left( \mathbf{X}_{0}\right) =U_{0}H_{0}\left( \mathbf{X}%
_{0}\right) U_{0}^{+}$ is explicitly known when $\hbar =0$, i.e. when $%
\mathbf{P}$ and $\mathbf{R}$ are treated as commuting variables\textbf{. }

We end up this section with a technical remark that will be important for
the sequel. As shown in our formula $\left( \ref{U}\right) $ the solutions
for the diagonalization process depends on $\varepsilon _{0}\left( \mathbf{X}%
_{0}\right) $, $U_{0}\left( \mathbf{X}_{0}\right) $. While the final results
do not depend on the way variables are symmetrized (that is the order we
write the products of components of $\mathbf{X}$, one has to start with an
initial symmetrization for the diagonalized energy at the zeroth order as
well as for $U_{0}\left( \mathbf{X}_{0}\right) $ or equivalently the Berry
phase. Since this detail will be important only at the second order in $%
\hbar $ while considering the Bloch electron in an electromagnetic field, we
do not mention any choice for the moment.

Eqs. $\left( \ref{U}\right) \left( \ref{NU}\right) $ show that a diagonal
Hamiltonian representation can be found to any desired order in $\hbar $,
and in the following we will carry it out until the second order. Let start
first with the first order.

\subsection{Covariant dynamical operator algebra}

From equations Eq. $\left( \ref{RP1}\right) $ we readily deduce the
following non trivial algebra between the dynamical operators
\begin{eqnarray}
\left[ r_{i},r_{j}\right] &=&i\hbar ^{2}\Theta _{ij}^{rr}=i\hbar ^{2}\left(
\nabla _{P_{i}}\emph{A}_{R_{j}}-\nabla _{P_{j}}\emph{A}_{R_{i}}\right)
+\hbar ^{2}\left[ \emph{A}_{R_{j}},\emph{A}_{R_{i}}\right]  \notag \\
\left[ p_{i},p_{j}\right] &=&i\hbar ^{2}\Theta _{ij}^{pp}=-i\hbar ^{2}\left(
\nabla _{R_{i}}\emph{A}_{P_{j}}-\nabla _{R_{j}}\emph{A}_{P_{i}}\right)
+\hbar ^{2}\left[ \emph{A}_{P_{i}},\emph{A}_{P_{j}}\right]  \notag \\
\left[ p_{i},r_{j}\right] &=&-i\hbar \delta _{ij}+i\hbar ^{2}\Theta
_{ij}^{pr}=-i\hbar \delta _{ij}-i\hbar ^{2}\left( \nabla _{R_{i}}\emph{A}%
_{R_{j}}+\nabla _{P_{j}}\emph{A}_{P_{i}}\right) +\hbar ^{2}\left[ \emph{A}%
_{P_{i}},\emph{A}_{R_{j}}\right]  \label{commalgebra}
\end{eqnarray}
where the terms $\Theta _{ij}$ are Berry curvatures definitions. Of course
these non trivial commutation relations also give new contributions to the
equations of motion and thus lead to new phenomena \cite{PIERREEUROPHYS}\cite%
{PIERRESEMIDIAG}\cite{SERIESPIERRE}\cite{PIERREPHOTONGRAVIT}\cite%
{PIERREELECGRAVIT}. The commutation relations are valid to any order in $%
\hbar ,$ but in practice we can compute them as well as the energy $%
\varepsilon \left( \mathbf{X}\right) $ in a series expansion in $\hbar .$

\subsection{Diagonal representation at the first order in $\hbar $}

In this section we derive by a straightforward application of Eq.\ $\left( %
\ref{EE}\right) $ the semiclassical effective diagonal in-bands Hamiltonian\
$\varepsilon $ for one particle. Using the fact that at this level $%
\int_{0}^{\hbar }O_{\alpha }d\alpha \varepsilon _{0}\left( \mathbf{X}%
_{\alpha }\right) =\hbar \left( T_{\hbar }+M_{\hbar }\right) \varepsilon
_{0}\left( \mathbf{X}\right) $ we have
\begin{equation}
\varepsilon =\varepsilon _{0}\left( \mathbf{X}\right) +\hbar \left( T_{\hbar
}+M_{\hbar }\right) \varepsilon _{0}\left( \mathbf{X}\right) +O(\hbar ^{2})
\label{E}
\end{equation}
From the definition of $T$ and $M$ in Eq. $\left( \ref{TM}\right) $ we see
that we must determine the quantities $\mathcal{A}_{\hbar }^{\mathbf{X}%
}=\left( \mathcal{A}_{\hbar }^{\mathbf{R}},\mathcal{A}_{\hbar }^{\mathbf{P}%
}\right) $ Eq. $\left( \ref{Ber}\right) $ defined previously. At this level
of the approximation and due to the factor $\hbar $ in the previous equation
it is enough to know this quantity at the zeroth order in $\hbar $. Skipping
the $\hbar $ index, they are thus simply given by

\begin{eqnarray*}
\mathcal{A}_{\hbar }^{\mathbf{X}} &=&\mathcal{A}_{0}^{\mathbf{R}%
}=iU_{0}\left( \mathbf{X}\right) \mathbf{\nabla }_{\mathbf{P}%
}U_{0}^{+}\left( \mathbf{X}\right) \\
\mathcal{A}_{\hbar }^{\mathbf{P}} &=&\mathcal{A}_{0}^{\mathbf{P}%
}=-iU_{0}\left( \mathbf{X}\right) \mathbf{\nabla }_{\mathbf{R}%
}U_{0}^{+}\left( \mathbf{X}\right)
\end{eqnarray*}
where as before $U_{0}$ is the matrix that would diagonalize the Hamiltonian
if $\mathbf{R}$ and $\mathbf{P}$ where commuting variables (that is the
matrix $U_{0}$ which diagonalizes the classical energy when $\hbar =0$). It
means that, interestingly, at this first order in $\hbar $, the
diagonalizing matrix $U$\ is the matrix defined at the order $\hbar ^{0}$ in
which the classical variables $\mathbf{X}_{0}$ have been replaced by the
quantum operator $\mathbf{X.}$ This is the clue of the method: at each order
$\hbar ^{n}$ we just need to know the matrix $U$ at the order $\hbar ^{n-1}.$
Therefore, the semiclassical diagonalization is within the present approach
simple to achieve (and simpler than any other method we are aware for) as it
only requires the determination of the matrix diagonalizing the classical
energy.

Introducing also the covariant dynamical operators $\mathbf{x}=\mathbf{X+}%
\emph{A}^{\mathbf{X}}$ with the Berry connection $\emph{A}^{\mathbf{X}%
}=\hbar \mathcal{P}_{+}\left[ \mathcal{A}_{\hbar }^{\mathbf{X}}\right] $, as
well as the integrated non projected Berry connections $\mathcal{A}^{\mathbf{%
X}}=\int_{0}^{\hbar }\mathcal{A}_{\alpha }^{\mathbf{X}}d\alpha =\hbar
\mathcal{A}_{\hbar }^{\mathbf{X}}$ at this order, the diagonal energy Eq. $%
\left( \ref{E}\right) $ can be written as
\begin{equation}
\varepsilon =\varepsilon _{0}\left( \mathbf{x}\right) +\frac{i}{2\hbar }%
\mathcal{P}_{+}\left\{ \left[ \varepsilon _{0}\left( \mathbf{X}\right) ,%
\mathcal{A}^{R^{i}}\right] \mathcal{A}^{P_{i}}-\left[ \varepsilon _{0}\left(
\mathbf{X}\right) ,\mathcal{A}^{P_{i}}\right] \mathcal{A}^{R^{i}}\right\}
+O(\hbar ^{2})  \label{SE}
\end{equation}
where summation over $i$ is assumed. Here $\varepsilon _{0}\left( \mathbf{x}%
\right) $ corresponds to the classical diagonal energy operators in which
the classical variables $\mathbf{X}_{0}$ have been replaced by the full
noncomutative covariant operators which satisfy a non commutative algebra as
we will see later when considering the practical examples. The second
contribution which is of order $\hbar $, gives for instance for Bloch
electrons in a magnetic field $\mathbf{B}$ as we shall see later, the
coupling between $\mathbf{B}$ and the magnetic moment operator .

Note also that this general diagonal energy for an arbitrary one particle
system was already derived in previous works by different methods \cite%
{PIERRESEMIDIAG}\cite{SERIESPIERRE} and turns out to be very useful for the
study of several different physical systems \cite{PIERREEUROPHYS}\cite%
{PIERREELECGRAVIT}\cite{PIERREPHOTONGRAVIT}\cite{PIERREGRAPHENE}.

\subsection{Diagonal representation at the second order in $\hbar $}

The Hamiltonian diagonalization at this order requires the matrix $U$ at the
first order: $U\left( \mathbf{X}\right) =U_{0}\left( \mathbf{X}\right)
+\hbar U_{1}\left( \mathbf{X}\right) U_{0}\left( \mathbf{X}\right) $ where $%
U_{1}\left( \mathbf{X}\right) $ is determined from the relation $\hbar
U_{1}\left( \mathbf{X}\right) =\int_{0}^{\hbar }N_{\alpha }d\alpha
U_{0}\left( \mathbf{X}\right) $ as a consequence of Eq. $\left( \ref{U}%
\right) .$ With expression $\left( \ref{NU}\right) $ we readily obtain :

\begin{eqnarray}
U_{1}\left( \mathbf{X}\right) &=&\left[ .,\varepsilon _{0}\right] ^{-1}.%
\left[ \mathcal{P}_{-}\left\{ \frac{1}{2}\left( \mathcal{A}%
_{0}^{R_{l}}\nabla _{R_{l}}\varepsilon _{0}\left( \mathbf{X}\right) +\nabla
_{R_{l}}\varepsilon _{0}\left( \mathbf{X}\right) \mathcal{A}_{0}^{R_{l}}+%
\mathcal{A}_{0}^{P_{l}}\nabla _{P_{l}}\varepsilon _{0}\left( \mathbf{X}%
\right) +\nabla _{P_{l}}\varepsilon _{0}\left( \mathbf{X}\right) \mathcal{A}%
_{0}^{P_{l}}\right) \right\} \right.  \notag \\
&&\left. -\frac{i}{2}\left\{ \left[ \varepsilon _{0}\left( \mathbf{X}\right)
,\mathcal{A}_{0}^{R_{l}}\right] \mathcal{A}_{0}^{P_{l}}-\left[ \varepsilon
_{0}\left( \mathbf{X}\right) ,\mathcal{A}_{0}^{P_{l}}\right] \mathcal{A}%
_{0}^{R_{l}}\right\} \right] -\frac{i}{4}\left[ \mathcal{A}_{0}^{R_{l}},%
\mathcal{A}_{0}^{P_{l}}\right]  \label{U1}
\end{eqnarray}
At the same order the (non-diagonal) Berry connections $\mathcal{A}_{\alpha
}^{\mathbf{X}}=\left( \mathcal{A}_{\alpha }^{\mathbf{R}},\mathcal{A}_{\alpha
}^{\mathbf{P}}\right) $ are again given by
\begin{eqnarray}
\mathcal{A}_{\alpha }^{\mathbf{R}}\left( \mathbf{X}_{\alpha }\right) &=&i%
\left[ U_{\alpha }\left( \mathbf{X}_{\alpha }\right) \mathbf{\nabla }_{%
\mathbf{P}}U_{\alpha }^{+}\left( \mathbf{X}_{\alpha }\right) \right] =\frac{1%
}{\alpha }U_{\alpha }\left( \mathbf{X}_{\alpha }\right) \mathbf{R}_{\alpha
}U_{\alpha }^{+}\left( \mathbf{X}_{\alpha }\right) \\
\text{and }\mathcal{A}_{\alpha }^{\mathbf{P}}\left( \mathbf{X}_{\alpha
}\right) &=&-i\left[ U_{\alpha }\left( \mathbf{X}_{\alpha }\right) \mathbf{%
\nabla }_{\mathbf{R}}U_{\alpha }^{+}\left( \mathbf{X}_{\alpha }\right) %
\right] =\frac{1}{\alpha }U_{\alpha }\left( \mathbf{X}_{\alpha }\right)
\mathbf{P}_{\alpha }U_{\alpha }^{+}\left( \mathbf{X}_{\alpha }\right)  \notag
\end{eqnarray}
where now $U\left( \mathbf{X}_{\alpha }\right) $ is the transformation to
the first order in $\alpha $ i.e. $U_{0}\left( \mathbf{X}_{\alpha }\right)
+\alpha U_{1}\left( \mathbf{X}_{\alpha }\right) ,$ in which $\mathbf{X}$ is
replaced by the running operator $\mathbf{X}_{\alpha }$. Using the
hermiticity of $\mathcal{A}_{\alpha }^{\mathbf{X}}$, so that one has $%
\mathcal{A}_{\alpha }^{\mathbf{X}}=\mathcal{A}_{\alpha }^{\mathbf{X}}+\left(
\mathcal{A}_{\alpha }^{\mathbf{X}}\right) ^{+}$ we can expand $\mathcal{A}^{%
\mathbf{X}}$ as :
\begin{equation*}
\mathcal{A}_{\alpha }^{\mathbf{X}}=\left( \frac{1}{2}\left( \left[ 1+\alpha
U_{1}\left( \mathbf{X}_{\alpha }\right) \right] U_{0}\left( \mathbf{X}%
_{\alpha }\right) \right) \frac{\mathbf{X}_{\alpha }}{\alpha }\left(
U_{0}^{+}\left( \mathbf{X}_{\alpha }\right) \left[ 1+\alpha U_{1}^{+}\left(
\mathbf{X}_{\alpha }\right) \right] \right) +\frac{1}{2}H.C.\right) -\frac{%
\mathbf{X}_{\alpha }}{\alpha }
\end{equation*}
(the $\frac{1}{\alpha }$ factor reminds that in our definition of $\mathcal{A%
}_{\alpha }^{\mathbf{X}}$ the Gradient with respect to $\mathbf{X}_{\alpha }$
is normalized, i.e. divided by $\alpha $). After some recombinations, the
previous expression can be written in a more convenient form :
\begin{eqnarray*}
\mathcal{A}_{\alpha }^{\mathbf{X}} &=&\frac{1}{2\alpha }U_{0}\left( \mathbf{X%
}_{\alpha }\right) \left[ \mathbf{X}_{\alpha }\mathbf{,}U_{0}^{+}\left(
\mathbf{X}_{\alpha }\right) \right] +H.C. \\
&&+\frac{1}{2}\left[ U_{1}\left( \mathbf{X}_{\alpha }\right) \left[
U_{0}\left( \mathbf{X}_{\alpha }\right) \mathbf{X}_{\alpha }U_{0}^{+}\left(
\mathbf{X}_{\alpha }\right) \right] +\left[ U_{0}\left( \mathbf{X}_{\alpha
}\right) \mathbf{X}_{\alpha }U_{0}^{+}\left( \mathbf{X}_{\alpha }\right) %
\right] U_{1}^{+}\left( \mathbf{X}_{\alpha }\right) \right] +H.C.-\mathbf{X}%
_{\alpha }
\end{eqnarray*}
using now the fact that at the lowest order $U_{0}\left( \mathbf{X}_{\alpha
}\right) \left[ \frac{\mathbf{X}_{\alpha }}{\alpha },U_{0}^{+}\left( \mathbf{%
X}_{\alpha }\right) \right] =\mathcal{A}_{0}^{\mathbf{X}}\left( \mathbf{X}%
_{\alpha }\right) $, one has :
\begin{eqnarray*}
\mathcal{A}_{\alpha }^{\mathbf{X}} &=&\frac{1}{2\alpha }\left[ U_{0}\left(
\mathbf{X}_{\alpha }\right) \left[ \mathbf{X}_{\alpha }\mathbf{,}%
U_{0}^{+}\left( \mathbf{X}_{\alpha }\right) \right] +H.C.\right] \\
&&+\frac{1}{2}\left[ U_{1}\left( \mathbf{X}_{\alpha }\right) \mathcal{A}%
_{0}^{\mathbf{X}}\left( \mathbf{X}_{\alpha }\right) +\mathcal{A}_{0}^{%
\mathbf{X}}\left( \mathbf{X}_{\alpha }\right) U_{1}^{+}\left( \mathbf{X}%
_{\alpha }\right) +\left[ \mathbf{X}_{\alpha },U_{1}^{+}\left( \mathbf{X}%
_{\alpha }\right) \right] +H.C.\right]
\end{eqnarray*}
decomposing $U_{1}^{+}\left( \mathbf{X}_{\alpha }\right) $ into Hermitian
and antihermitian part we are thus led to :
\begin{eqnarray*}
\mathcal{A}_{\alpha }^{\mathbf{X}} &=&\frac{1}{2\alpha }\left( U_{0}\left(
\mathbf{X}_{\alpha }\right) \left[ \mathbf{X}_{\alpha }\mathbf{,}%
U_{0}^{+}\left( \mathbf{X}_{\alpha }\right) \right] +H.C.\right) +\left[
\mathbf{X}_{\alpha }\mathbf{+}\mathcal{A}_{0}^{\mathbf{X}}\mathbf{,}\mathit{%
ah}\left( U_{1}^{+}\left( \mathbf{X}_{\alpha }\right) \right) \right] \\
&&+\left( \mathcal{A}_{0}^{\mathbf{X}}H\left( U_{1}^{+}\left( \mathbf{X}%
_{\alpha }\right) \right) +\mathit{h}\left( U_{1}^{+}\left( \mathbf{X}%
_{\alpha }\right) \right) \mathcal{A}_{0}^{\mathbf{X}}\right)
\end{eqnarray*}
where $\mathit{ah}\left( Z\right) $ and $\mathit{ah}\left( Z\right) $ denote
the antihermitian and Hermitian part of an operator $Z$ respectively. Now
with Eq. $\left( \ref{U1}\right) $ and after the integration, we are led for
$\mathcal{A}^{\mathbf{X}}=\hbar \mathcal{A}_{0}^{\mathbf{X}}+\hbar ^{2}%
\mathcal{A}_{1}^{\mathbf{X}}$ to the following expression :
\begin{equation}
\mathcal{A}_{\alpha }^{\mathbf{X}}=\mathcal{A}_{0}^{\mathbf{X}}\left(
\mathbf{R+}\frac{\alpha }{2}\mathcal{A}_{0}^{R_{l}},\mathbf{P+}\frac{\alpha
}{2}\mathcal{A}_{0}^{P_{l}}\right) -\left[ \mathcal{B},\mathbf{X}_{\alpha }+%
\mathcal{A}_{0}^{\mathbf{X}}\right]  \label{Conn2}
\end{equation}
where we introduced the notations
\begin{equation}
\mathcal{A}_{0}^{\mathbf{X}}\left( \mathbf{R+}\frac{\alpha }{2}\mathcal{A}%
_{0}^{R_{l}},\mathbf{P+}\frac{\alpha }{2}\mathcal{A}_{0}^{P_{l}}\right)
\equiv \mathcal{A}_{0}^{\mathbf{X}}+\frac{\alpha }{4}\left\{ \mathcal{A}%
_{0}^{R_{l}}\nabla _{R_{l}}\mathcal{A}_{0}^{\mathbf{X}}+\mathcal{A}%
_{0}^{P_{l}}\nabla _{R_{l}}\mathcal{A}_{0}^{\mathbf{X}}+H.C.\right\}
\label{conn2A}
\end{equation}
and
\begin{eqnarray}
\mathcal{B} &=&\left[ .,\varepsilon _{0}\right] ^{-1}.\left( \mathcal{P}%
_{-}\left\{ \frac{1}{2}\mathcal{A}_{0}^{R_{l}}\nabla _{R_{l}}\varepsilon
_{0}\left( \mathbf{X}\right) +\frac{1}{2}\mathcal{A}_{0}^{P_{l}}\nabla
_{P_{l}}\varepsilon _{0}\left( \mathbf{X}\right) +H.C.\right\} \right.
\notag \\
&&\left. -\frac{i}{4}\left\{ \left[ \varepsilon _{0}\left( \mathbf{X}\right)
,\mathcal{A}_{0}^{R_{l}}\right] \mathcal{A}_{0}^{P_{l}}-\left[ \varepsilon
_{0}\left( \mathbf{X}\right) ,\mathcal{A}_{0}^{P_{l}}\right] \mathcal{A}%
_{0}^{R_{l}}+H.C.\right\} \right)  \notag \\
&=&\left[ .,\varepsilon _{0}\right] ^{-1}.\left( \mathcal{P}_{-}\left\{
\frac{1}{2}\mathcal{A}_{0}^{R_{l}}\nabla _{R_{l}}\varepsilon _{0}\left(
\mathbf{X}\right) +\frac{1}{2}\mathcal{A}_{0}^{P_{l}}\nabla
_{P_{l}}\varepsilon _{0}\left( \mathbf{X}\right) +H.C.\right\} \right.
\notag \\
&&\left. -\frac{i}{4}\left\{ \mathcal{P}_{-}\mathcal{A}_{0}^{R_{l}}\mathcal{P%
}_{+}\mathcal{A}_{0}^{P_{l}}-\mathcal{P}_{-}\mathcal{A}_{0}^{P_{l}}\mathcal{P%
}_{+}\mathcal{A}_{0}^{R_{l}}+H.C.\right\} \right)  \label{conn2B}
\end{eqnarray}
These formula, although abstruse will be useful in our next sections. Having
found the matrices $U\left( \mathbf{X}\right) $ and $\mathcal{A}_{\alpha }^{%
\mathbf{X}}$ at the required order we can now determine the series expansion
for the diagonalized Hamiltonian to the second order in $\hbar $. From
equation Eq. $\left( \ref{EE}\right) $ we can write

\begin{equation*}
\varepsilon \left( \mathbf{X}\right) =\varepsilon _{0}\left( \mathbf{X}%
\right) +\int_{0}^{\hbar }O_{\alpha }d\alpha \varepsilon _{0}\left( \mathbf{X%
}_{\alpha }\right) +\int_{0}^{\hbar }O_{\alpha _{1}}\int_{0}^{\alpha
_{1}}O_{\alpha _{2}}d\alpha _{2}d\alpha _{1}\varepsilon _{0}\left( \mathbf{X}%
_{\alpha _{2}}\right) -\frac{\hbar }{2}\left\langle \varepsilon _{0}\left(
\mathbf{X}\right) \right\rangle
\end{equation*}
where the last contribution is given by
\begin{equation*}
-\frac{1}{2}\hbar \left\langle \varepsilon _{0}\left( \mathbf{X}\right)
\right\rangle =\frac{i}{4}\hbar Asym\left\{ \nabla _{P_{l}}\nabla
_{R_{l}}\varepsilon _{0}\left( \mathbf{X}\right) \right\}
\end{equation*}
The first contribution $\int_{0}^{\hbar }O_{\alpha }d\alpha \varepsilon
_{0}\left( \mathbf{X}\right) $ can be expanded as before as:
\begin{eqnarray*}
\int_{0}^{\hbar }O_{\alpha }d\alpha \varepsilon _{0}\left( \mathbf{X}\right)
&=&\int_{0}^{\hbar }\mathcal{P}_{+}\left\{ \frac{1}{2}\left( \mathcal{A}%
_{\alpha }^{R_{l}}\nabla _{R_{l}}\varepsilon _{0}\left( \mathbf{X}_{\alpha
}\right) +\nabla _{R_{l}}\varepsilon _{0}\left( \mathbf{X}_{\alpha }\right)
\mathcal{A}_{\alpha }^{R_{l}}+\mathcal{A}_{\alpha }^{P_{l}}\nabla
_{P_{l}}\varepsilon _{0}\left( \mathbf{X}_{\alpha }\right) +\nabla
_{P_{l}}\varepsilon _{0}\left( \mathbf{X}_{\alpha }\right) \mathcal{A}%
_{\alpha }^{P_{l}}\right) \right\} d\alpha \\
&&+\int_{0}^{\hbar }\mathcal{P}_{+}\left\{ \frac{i}{4}\left\{ \left[
\varepsilon _{0}\left( \mathbf{X}_{\alpha }\right) ,\mathcal{A}_{\alpha
}^{R_{l}}\right] \mathcal{A}_{\alpha }^{P_{l}}-\left[ \varepsilon _{0}\left(
\mathbf{X}_{\alpha }\right) ,\mathcal{A}_{\alpha }^{P_{l}}\right] \mathcal{A}%
_{\alpha }^{R_{l}}\right\} +H.C.\right. \\
&&\left. +\left[ U_{\alpha }\left( \left( \frac{\partial }{\partial \alpha }%
+\left\langle .\right\rangle \right) H\left( \mathbf{X}_{\alpha }\right)
\right) U_{\alpha }^{+}\right] \right\} d\alpha
\end{eqnarray*}
where the Berry connections have to be expanded to the first order, whereas
the second order contribution
\begin{equation*}
\int_{0}^{\hbar }O_{\alpha _{1}}\int_{0}^{\alpha _{1}}O_{\alpha _{2}}d\alpha
_{2}d\alpha _{1}\varepsilon _{0}\left( \mathbf{X}\right) -\frac{\hbar }{2}%
\left\langle \varepsilon _{0}\left( \mathbf{X}\right) \right\rangle
\end{equation*}
has to be expanded to the zeroth order in the Berry connections. Notice that
due to the integration process, the squared terms in $\mathcal{A}_{0}^{%
\mathbf{R}}$, $\mathcal{A}_{0}^{\mathbf{P}}$ as well as the first order
terms in the Berry phase get a $\frac{1}{2}$ factor.

The consequence is that these contributions can be recombined to yield :
\begin{eqnarray}
\varepsilon \left( \mathbf{X}\right) &=&\varepsilon _{0}\left( \mathbf{x}%
\right) +\frac{i}{2}\mathcal{P}_{+}\left\{ \left[ \varepsilon _{0}\left(
\mathbf{x}\right) ,\mathcal{\hat{A}}^{R_{l}}\right] \mathcal{\hat{A}}%
^{P_{l}}-\left[ \varepsilon _{0}\left( \mathbf{x}\right) ,\mathcal{\hat{A}}%
^{P_{l}}\right] \mathcal{\hat{A}}^{R_{l}}-\left[ \varepsilon _{0}\left(
\mathbf{x}\right) ,\left[ \mathcal{\hat{A}}^{R_{l}},\mathcal{\hat{A}}^{P_{l}}%
\right] \right] \right\}  \notag \\
&&-\frac{1}{8}\mathcal{P}_{+}\left\{ \left[ \left[ \varepsilon _{0}\left(
\mathbf{x}\right) ,\mathcal{\hat{A}}^{R_{l}}\right] \mathcal{\hat{A}}%
^{P_{l}}-\left[ \varepsilon _{0}\left( \mathbf{x}\right) ,\mathcal{\hat{A}}%
^{P_{l}}\right] \mathcal{\hat{A}}^{R_{l}},\mathcal{\hat{A}}^{R_{l}}\right]
\mathcal{\hat{A}}^{P_{l}}\right.  \notag \\
&&\left. -\left[ \left[ \varepsilon _{0}\left( \mathbf{x}\right) ,\mathcal{%
\hat{A}}^{R_{l}}\right] \mathcal{\hat{A}}^{P_{l}}-\left[ \varepsilon
_{0}\left( \mathbf{x}\right) ,\mathcal{\hat{A}}^{P_{l}}\right] \mathcal{\hat{%
A}}^{R_{l}},\mathcal{\hat{A}}^{P_{l}}\right] \mathcal{\hat{A}}%
^{R_{l}}\right\} -\frac{\hbar }{2}\left\langle \varepsilon _{0}\left(
\mathbf{x}\right) \right\rangle  \label{E2}
\end{eqnarray}
with :
\begin{eqnarray*}
\mathcal{\hat{A}}^{R_{l}} &=&\frac{1}{2}\left[ 1-\frac{1}{2}\left( \mathcal{P%
}_{+}\mathcal{A}_{0}^{\mathbf{X}}.\mathbf{\nabla }_{\mathbf{X}}\right) %
\right] \mathcal{A}^{R_{l}}\left( \mathbf{x}\right) +H.C. \\
\mathcal{\hat{A}}^{P_{l}} &=&\frac{1}{2}\left[ 1-\frac{1}{2}\left( \mathcal{P%
}_{+}\mathcal{A}_{0}^{\mathbf{X}}.\mathbf{\nabla }_{\mathbf{X}}\right) %
\right] \mathcal{A}^{P_{l}}\left( \mathbf{x}\right) +H.C.
\end{eqnarray*}
and where we have denoted again $\mathcal{A}^{\mathbf{X}}\left( \mathbf{X}%
\right) =\int_{0}^{\hbar }\mathcal{A}_{\alpha }^{\mathbf{X}}\left( \mathbf{X}%
\right) d\alpha $. Remark that the last term $-\frac{\hbar }{2}\left\langle
\varepsilon _{0}\left( \mathbf{x}\right) \right\rangle $ in the expression
for $\varepsilon \left( \mathbf{X}\right) $ is of order $\hbar ^{2}$ since $%
\left\langle \varepsilon _{0}\left( \mathbf{x}\right) \right\rangle $
involves some commutators and is thus of order $\hbar $. As explained
above,\ one can, at each order of the expansion,\ safely replace the
canonical operators $\mathbf{X=}\left( \mathbf{R,P}\right) $\ by the
covariant ones $\mathbf{x=}\left( \mathbf{r,p}\right) $\ which are given by
the following expression to the second order (dropping once again the $\hbar
$ index) :

\begin{equation}
\mathbf{x}=\mathbf{X+}\emph{A}^{\mathbf{X}}\equiv \mathbf{X+}\hbar \emph{A}%
_{0}^{\mathbf{X}}+\frac{\hbar }{2}^{2}\emph{A}_{1}^{\mathbf{X}}
\label{rpnew}
\end{equation}
with : $\emph{A}^{\mathbf{X}}=\left( \emph{A}^{\mathbf{R}},\emph{A}^{\mathbf{%
P}}\right) $ and (as before we drop the index $\hbar $) $\mathbf{X}=\left(
\mathbf{R},\mathbf{P}\right) =\mathbf{X}_{\hbar }=\left( \mathbf{R}_{\hbar },%
\mathbf{P}_{\hbar }\right) $:
\begin{eqnarray*}
\mathbf{\emph{A}^{\mathbf{R}}} &=&\int\limits_{0<\alpha <\hbar }\mathcal{P}%
_{+}\left[ \mathcal{A}_{\alpha }^{\mathbf{R}}\right] d\alpha
+\int\limits_{0<\alpha <\hbar }\frac{1}{2}\left[ \left[ \frac{1}{2}\left(
\mathcal{P}_{+}\left[ \mathcal{A}_{\alpha }^{\mathbf{X}}\right] .\mathbf{%
\nabla }_{\mathbf{X}}+H.C.\right) \int\limits_{0<\alpha _{1}<\alpha }%
\mathcal{P}_{+}\left[ \mathcal{A}_{\alpha _{1}}^{\mathbf{R}}\right] \right]
+H.C.\right] d\alpha _{1}d\alpha \\
&=&\mathcal{P}_{+}\left[ \hbar \mathcal{A}_{0}^{\mathbf{R}}\left( \mathbf{R+}%
\frac{\hbar }{4}\mathcal{A}_{0}^{R_{l}},\mathbf{P+}\frac{\hbar }{4}\mathcal{A%
}_{0}^{P_{l}}\right) -\frac{\hbar ^{2}}{2}\left[ \mathcal{B},\mathbf{R}+%
\mathcal{A}_{0}^{\mathbf{R}}\right] \right] +\frac{\hbar ^{2}}{4}\left(
\mathcal{P}_{+}\left[ \mathcal{A}_{0}^{\mathbf{X}}\right] .\mathbf{\nabla }_{%
\mathbf{X}}\mathcal{P}_{+}\left[ \mathcal{A}_{0}^{\mathbf{R}}\right]
+H.C.\right) \\
\mathbf{\emph{A}^{\mathbf{P}}} &=&\int\limits_{0<\alpha <\hbar }\mathcal{P}%
_{+}\left[ \mathcal{A}_{\alpha }^{\mathbf{P}}\right] d\alpha
+\int\limits_{0<\alpha <\hbar }\frac{1}{2}\left[ \left[ \frac{1}{2}\left(
\mathcal{P}_{+}\left[ \mathcal{A}_{\alpha }^{\mathbf{X}}\right] .\mathbf{%
\nabla }_{\mathbf{X}}+H.C.\right) \int\limits_{0<\alpha _{1}<\alpha }%
\mathcal{P}_{+}\left[ \mathcal{A}_{\alpha _{1}}^{\mathbf{P}}\right] \right]
+H.C.\right] d\alpha _{1}d\alpha +... \\
&=&\mathcal{P}_{+}\left[ \hbar \mathcal{A}_{0}^{\mathbf{P}}\left( \mathbf{R+}%
\frac{\hbar }{4}\mathcal{A}_{0}^{R_{l}},\mathbf{P+}\frac{\hbar }{4}\mathcal{A%
}_{0}^{P_{l}}\right) -\frac{\hbar ^{2}}{2}\left[ \mathcal{B},\mathbf{P}+%
\mathcal{A}_{0}^{\mathbf{P}}\right] \right] +\frac{\hbar ^{2}}{4}\left(
\mathcal{P}_{+}\left[ \mathcal{A}_{0}^{\mathbf{X}}\right] .\mathbf{\nabla }_{%
\mathbf{X}}\mathcal{P}_{+}\left[ \mathcal{A}_{0}^{\mathbf{P}}\right]
+H.C.\right)
\end{eqnarray*}
Remark ultimately, that had we chosen the variables $\mathbf{X}$ rather than
$\mathbf{x}$ to express our Hamiltonian, we would have rather written at our
order of approximation :
\begin{eqnarray*}
\mathcal{\hat{A}}^{R_{l}} &=&\frac{1}{2}\left[ 1+\frac{1}{2}\left( \mathcal{P%
}_{+}\mathcal{A}_{0}^{\mathbf{X}}.\mathbf{\nabla }_{\mathbf{X}}\right) %
\right] \mathcal{A}^{R_{l}}\left( \mathbf{X}\right) +H.C. \\
\mathcal{\hat{A}}^{P_{l}} &=&\frac{1}{2}\left[ 1+\frac{1}{2}\left( \mathcal{P%
}_{+}\mathcal{A}_{0}^{\mathbf{X}}.\mathbf{\nabla }_{\mathbf{X}}\right) %
\right] \mathcal{A}^{P_{l}}\left( \mathbf{X}\right) +H.C.
\end{eqnarray*}
However, as explained before, the transformed variables $\mathbf{x}$ fit
better to write the Hamiltonian since they enter directly in $\varepsilon
_{0}\left( \mathbf{x}\right) $.

Eq $\left( \ref{E2}\right) $ is the desired series expansion to the second
order in $\hbar $ of the diagonal Hamiltonian. In the next section it will
be the angular stone for the computation of the effective in-bands
Hamiltonian of a Bloch electron in an external electromagnetic field. Once
this will be achieved the case of several Bloch electrons will be
investigated.

\section{Bloch electron in an electromagnetic field}

To start, an outline of the first order diagonalization for the special case
of an electron in an crystal lattice perturbated by the presence of an
external electromagnetic field considered in refs. \cite{PIERREEUROPHYS} and
\cite{PIERRESEMIDIAG} is given.

The Hamiltonian of an electron in an crystal lattice perturbated by the
presence of an external electromagnetic field is
\begin{equation*}
H=\frac{\left( \mathbf{P-}e\mathbf{A}\right) ^{2}}{2m}+V_{p}\left( \mathbf{R}%
\right) +eV\left( \mathbf{R}\right)
\end{equation*}
($e<0$) where $V_{p}\left( \mathbf{R}\right) $ the periodic potential, $%
\mathbf{A}$ and $V$ the vector and scalar potential respectively. Our
purpose is this section is to compute the diagonal in-bands energy
Hamiltonian for this system to the second order in a series expansion in $%
\hbar .$ This can be done by using the general results of the previous
section, in particular with Eqs. $\left( \ref{SE}\right) \left( \ref{E2}%
\right) $. The major difficulty to find the diagonal representation relies
on the fact that in presence of an electromagnetic field, the lattice
translation operators $\mathbf{T}$ do not commute any more (see \cite{NIU}
and references therein). To deal with this problem it is convenient to
express the total magnetic field as the sum of a constant field $\mathbf{B}%
_{0}$ and small nonuniform part $\delta \mathbf{B}(\mathbf{R})$. The
Hamiltonian can be written $H=H_{0}+eV(\mathbf{R})$, with $H_{0}$ the
magnetic contribution ($V$ being the electric potential) which reads
\begin{equation}
H_{0}=\frac{1}{2m}\left( \mathbf{P}-e\mathbf{A}(\mathbf{R})-e\delta \mathbf{A%
}(\mathbf{R})\right) ^{2}+V_{p}\left( \mathbf{R}\right)  \label{Hmagnetic}
\end{equation}
where $\mathbf{A}(\mathbf{R})$ and $\delta \mathbf{A}(\mathbf{R})$ are the
vectors potential of the homogeneous and inhomogeneous magnetic field,
respectively. $V_{p}\left( \mathbf{R}\right) $ is the periodic potential.
The large constant part $\mathbf{B}_{0}$ is chosen such that the magnetic
flux through a unit cell is a rational fraction of the flux quantum $h/e$.
The advantage of such a decomposition is that for $\delta \mathbf{A}(\mathbf{%
R})=0$ the magnetic translation operators are commuting quantities allowing
to exactly diagonalize the Hamiltonian and to treat $\delta \mathbf{A}(%
\mathbf{R})$ as a small perturbation. The state space of the Bloch electron
is spanned by the basis vector $\left\vert n,\mathbf{k}\right\rangle
=\left\vert \mathbf{k}\right\rangle \otimes \left\vert n\right\rangle $ with
$n$ corresponding to a band index and $\mathbf{k}$ a common eigenvalue of
the translation operators. In this representation $\mathbf{K}\left\vert n,%
\mathbf{k}\right\rangle =\mathbf{k}\left\vert n,\mathbf{k}\right\rangle $
and the position operator is $\mathbf{R=}i\partial /\partial \mathbf{K}$
which implies the canonical commutation relation $\left[ \mathbf{R}_{i}%
\mathbf{,K}_{j}\right] =i\delta _{ij}$. Note that here $\mathbf{R}$ and $%
\mathbf{K}$, rather than $\mathbf{R}$ and $\mathbf{P}$ will play the role of
canonical variables.

The diagonalization of the Hamiltonian in Eq.$\left( \ref{Hmagnetic}\right) $
is first derived for $\delta \mathbf{A}=0$ by diagonalizing simultaneously $%
H_{0}$ and the magnetic translation operators $\mathbf{T}$. Start with an
arbitrary basis of eigenvectors of $\mathbf{T}$. As explained in \cite%
{PIERRESEMIDIAG}, in this basis $H_{0}$ can be seen as a square matrix with
operators entries and is diagonalized through a unitary matrix $U(\mathbf{K}%
) $ which depends only on $\mathbf{K}$ (since $U$ should leave $\mathbf{K}$
invariant, i.e., $U\mathbf{K}U^{+}=\mathbf{K}$), such that in-bands energy
matrix is $\varepsilon =UHU^{+}=\varepsilon _{0}(\mathbf{K})+V(U\mathbf{R}%
U^{+})$, where $\varepsilon _{0}(\mathbf{K})$ is the unperturbated ($\delta
\mathbf{A=0}$) diagonal energy matrix made of the magnetic bands elements $%
\varepsilon _{0,n}(\mathbf{K})$ with $n$ the band index.

Now, to add a perturbation $\delta A(\mathbf{R)}$ that breaks the
translational symmetry, we have to replace $\mathbf{K}$ in all expressions
by
\begin{equation}
\mathbf{\Pi }=\mathbf{K}-\frac{e}{\hbar }\delta A(\mathbf{R)}
\end{equation}
the (band) electron momentum and as the flux $\mathbf{\delta B}$ on a
plaquette is not a rational multiple of the flux quantum, we cannot
diagonalize simultaneously its components $\Pi _{i}$ since they do not
commute anymore. Actually
\begin{equation}
\hbar \lbrack \Pi ^{i},\Pi ^{j}]=ie\varepsilon ^{ijk}\delta B_{k}(\mathbf{R})
\end{equation}
To deal with this non-commutativity, we adapt our method to diagonalize the
Hamiltonian perturbatively in $\hbar $. To start, an outline of the first
order diagonalization for the special case of an electron in an crystal
lattice perturbated by the presence of an external electromagnetic field
considered in refs. \cite{PIERREEUROPHYS} and \cite{PIERRESEMIDIAG} is now
given.

\subsection{Semiclassical diagonalization: Generalized Peierls substitution}

Following section II, the diagonalization at the lowest order is just
obtained by replacing $U(\mathbf{K})$ by $U\left( \mathbf{\Pi }\right) .$
This last matrix would actually diagonalize the Hamiltonian if $\mathbf{R}$
in $\delta A(\mathbf{R)}$ was a parameter commuting with $\mathbf{K}$). Note
that a subtlety arises here (and that we will find again later) in the
application in our method. We do not consider, at this level that $\mathbf{R}
$ and $\mathbf{K}$ commute, but only that the $\mathbf{R}$ in $\delta A(%
\mathbf{R)}$ and $V\left( \mathbf{R}\right) $ commutes with $\mathbf{K}$. In
other words we assume that $\mathbf{R}$ has been replaced by a parameter in
these potentials. The reason of this difference with our general set up
comes from the fact that the initial diagonalization is not performed for a
function of, say, $\mathbf{P}$ alone, but both of $\mathbf{P}$ and $\mathbf{R%
}$ through the periodic potential. However this difference does not alter
our method which allows to recover the contributions of the electromagnetic
potential as a series of $\hbar $.

As consequence of our procedure, the non projected Berry connections are $%
\mathcal{A}_{0}^{R_{i}}=iU\nabla _{K_{i}}U^{+}$ and $\mathcal{A}%
_{0}^{K_{l}}=e\nabla _{R_{l}}\delta A_{k}\mathbf{(R)}\mathcal{A}_{0}^{R_{k}}$%
. However, it turns out be more relevant to replace $\mathbf{K}$ by the
covariant momentum $\mathbf{\Pi }$ in the physical expressions so that
instead of $\mathcal{A}_{0}^{K_{l}}$ we better consider the quantity $%
\mathcal{A}_{0}^{\Pi _{l}}=e\mathcal{A}_{0}^{R_{k}}B^{m}\varepsilon _{kml}.$
Remark that the Berry connections just defined are non diagonal and are
matrices whose index correspond to interband transitions.

The physical dynamical variables for the $n$-th band dynamics to the first
order in $\hbar $ imply a projection on the $n$-th band. For the intraband
coordinate operator $\mathbf{\mathbf{r\equiv r}}_{n}=\mathcal{P}_{n}(U(%
\mathbf{\Pi })\mathbf{R}U^{+}(\mathbf{\Pi }))$ we obtain
\begin{equation}
\mathbf{\mathbf{r}}\simeq \mathbf{R}+\emph{A}_{0}^{\mathbf{R}}(\pi )+O(\hbar
)  \label{rn}
\end{equation}
and for covariant intraband momentum we obtain in the same manner $\mathbf{%
\pi \equiv \pi }_{n}=\mathbf{\Pi }+\mathcal{A}_{0}^{\mathbf{\Pi }}$ which
writes also
\begin{equation}
\mathbf{\pi }\simeq \mathbf{\Pi }+e\emph{A}_{0}^{\mathbf{R}}(\pi )\times
\delta \mathbf{B}(\mathbf{r})/\hbar +O(\hbar )  \label{pn}
\end{equation}
with $\emph{A}_{0}^{\mathbf{R}}=\hbar \mathcal{P}_{n}\left( \mathcal{A}_{0}^{%
\mathbf{R}}\right) $ and $\mathcal{P}_{n}$ the projection on the $n$-th
band. Remember that previously $\mathcal{P}_{+}$ was the projection on the
diagonal elements of a matrix. Using now the general expression for the
semiclassical Hamiltonian Eq.\ $\left( \ref{SE}\right) $, we obtain the
desired semiclassical $n$th-band Hamiltonian (dropping the index $n$) $%
\varepsilon \left( \mathbf{\pi ,r}\right) $ as :
\begin{equation}
\varepsilon \left( \mathbf{\pi ,r}\right) =\varepsilon _{0}\left( \mathbf{%
\pi }\right) +eV\left( \mathbf{\mathbf{r}}\right) -\mathcal{M}(\mathbf{\pi }%
).\delta \mathbf{B}(\mathbf{\mathbf{r})}  \label{E1bloch}
\end{equation}
where $\varepsilon _{0}\left( \mathbf{\pi }\right) $ is the unperturbated $n$%
th-magnetic band energy in which $\mathbf{K}$ has been replaced by $\mathbf{%
\pi },$ a procedure that we can adequately call the \textit{generalized}
Peierls substitution and which was introduced for the first time in \cite%
{PIERREEUROPHYS}. The second term in Eq. $\left( \ref{E1bloch}\right) $
constitutes the electrostatic potential, and the third is the coupling
between\ $\delta \mathbf{B}$ and the magnetic moment (or magnetization)
which is given:
\begin{equation*}
\mathcal{M}(\mathbf{\pi })=\mathcal{P}_{n}(\frac{-ie}{2\hbar }\left[
\varepsilon _{0}(\mathbf{\pi }),\hbar \mathcal{A}_{0}^{\mathbf{R}}(\mathbf{%
\pi })\right] \times \hbar \mathcal{A}_{0}^{\mathbf{R}}(\mathbf{\pi }))
\end{equation*}
It can explicitly check that this expression of the magnetization is the
same than previous expression found with different approaches \cite{NIU}\cite%
{Landau}.

\subsubsection{Dynamical operators algebra and equations of motion}

From the dynamical operators a new algebra has to be considered . Indeed we
have
\begin{eqnarray}
\left[ r^{i},r^{j}\right] &=&i\Theta (\mathbf{\pi })^{ij}  \notag \\
\hbar \left[ \pi ^{i},\pi ^{j}\right] &=&ie\varepsilon ^{ijk}\delta B_{k}(%
\mathbf{r})+ie^{2}\varepsilon ^{ipk}\delta B_{k}\varepsilon ^{jql}\delta
B_{l}\Theta ^{pq}/\hbar  \notag \\
\left[ r^{i},\hbar \pi ^{j}\right] &=&i\hbar \delta ^{ij}-ie\varepsilon
^{jlk}\delta B_{k}(\mathbf{r})\Theta ^{il}(\mathbf{\pi })  \label{CC}
\end{eqnarray}
with $\Theta ^{ij}(\mathbf{\pi })=\partial ^{i}\emph{A}^{j}(\mathbf{\pi }%
)-\partial ^{j}\emph{A}^{i}(\mathbf{\pi })$ the Berry curvature. Whereas the
term of order $\delta B^{2}$ in $\left[ \pi ^{i},\pi ^{j}\right] $ is
usually negligible, it turns out that all terms in $\left[ r^{i},\hbar \pi
^{j}\right] $ are essential for the correct computation of the semiclassical
equations of motion which are
\begin{eqnarray}
\overset{\cdot }{\mathbf{r}} &=&\partial \varepsilon /\hbar \partial \mathbf{%
\pi }-\dot{\mathbf{\pi }}\times \Theta (\mathbf{\pi })  \notag \\
\hbar \dot{\mathbf{\pi }} &=&e\mathbf{E}+e\dot{\mathbf{r}}\times \delta
\mathbf{B}(\mathbf{r})-\mathcal{M}.\partial \delta \mathbf{B}/\partial
\mathbf{r}  \label{EQM}
\end{eqnarray}
where we have defined the vector $\Theta ^{i}=\varepsilon ^{ijk}\Theta
_{jk}/2$. As shown in the following, the generalization Peierls
substitution, not only is essential for the correct determination of the
equations of motion, but also for the Bohr-Sommerfeld (BS) quantization
condition.

\subsubsection{Bohr-Sommerfeld quantization}

To underline the relevance of the generalized Peierls substitution at the
level of the semiclassical quantization of energy levels for an electron
motion in an external uniform magnetic, we adapt the arguments of ref. \cite%
{PIERREEBS}. For $V=0$ the equations of motion Eq. $\left( \ref{EQM}\right) $
become
\begin{equation}
\mathbf{\dot{r}}=D\left( \frac{\partial \varepsilon }{\hbar \partial \mathbf{%
\pi }}\right) \ \ \ \ \text{and \ }\ \hbar \overset{\cdot }{\mathbf{\pi }}%
=eD\left( \frac{\partial \varepsilon }{\hbar \partial \mathbf{\pi }}\mathbf{%
\times B}\right)  \label{EQBS}
\end{equation}
with $D^{-1}=1+\frac{e}{\hbar }\mathbf{B\Theta .}$ For convenience $\delta
\mathbf{B\equiv B}$ is chosen in the $z$-direction $\mathbf{B=}B\mathbf{k,}$
the energy reads $\varepsilon =\varepsilon _{0}\left( \pi \right)
-M_{z}\left( \pi \right) B.$ Consequently the orbits satisfies the
conditions $\varepsilon _{0}=$const and $\pi _{z}=$const. The semiclassical
quantization of energy levels can be done according to the Bohr-Sommerfeld
quantization rule
\begin{equation}
\oint K_{\perp }dR_{\perp }=2\mathcal{\pi }\left( n+1/2\right)  \label{BSS}
\end{equation}
where $K_{\perp }$ is the canonical pseudo-momentum in the plane
perpendicular to the axis $\pi _{z}=cte.$ The integration is taken over a
period of the motion and $n$ is a large integer. Now, it turns out to be
convenient to choose the gauge $\widetilde{A}_{y}=BX,$\ $\widetilde{A}_{x}=%
\widetilde{A}_{z}=0.$ In this gauge, one has $\pi _{z}=P_{z}=cte$, and the
usual covariant momentum $\Pi _{y}=K_{y}-\frac{eB}{\hbar }X.$ In this case
the BS condition reads $\oint \Pi _{x}d\Pi _{y}=\frac{-2\mathcal{\pi }eB}{%
\hbar }\left( n+1/2\right) .$ Assuming that the physically relevant
variables are instead the covariant ones, writing thus $BX=B(x-\emph{A}_{x})$
the generalized covariant momentum defined as $\pi _{y}=\Pi _{y}-\frac{eB}{%
\hbar }\emph{A}_{x}$ becomes
\begin{equation}
\pi _{y}=K_{y}-\frac{eB}{\hbar }x
\end{equation}
which is formally the same relation as the one between the canonical
variables, but now relating the new covariant generalized dynamical
operators. This relation with the help of the equations of motion Eq. $%
\left( \ref{EQBS}\right) $\ gives $\overset{\cdot }{K}_{y}=\overset{\cdot }{%
\pi }_{y}+\frac{eB}{\hbar }\overset{\cdot }{x}=0$ thus $K_{y}$ is a constant
of motion so that $\oint K_{Y}dY=K_{Y}\oint dY=0$ and Eq.$\left( \ref{BSS}%
\right) $ becomes simply $\oint K_{x}dX=2\mathcal{\pi }\left( n+1/2\right) .$
Now using the definition of the generalized momentum $K_{x}=\pi _{x}-\frac{e%
\emph{A}_{y}}{\hbar }B$ and the differential of the canonical position $%
dX=dx-d\emph{A}_{x}=\frac{\hbar d\pi _{y}}{-eB}-d\emph{A}_{x},$ the
Bohr-Sommerfeld condition Eq. $\left( \ref{BSS}\right) $ becomes
\begin{equation}
\oint \pi _{x}d\pi _{y}=-2\mathcal{\pi }eB\left( n+\frac{1}{2}-\frac{1}{2%
\mathcal{\pi }}\oint \emph{A}_{\bot }d\mathbf{\pi }_{\perp }\right)
\label{pipiA}
\end{equation}
where the integral is now taken along a closed trajectory $\Gamma $ in the $%
\pi $ space and $\frac{1}{2\mathcal{\pi }}\oint \emph{A}_{\bot }d\mathbf{\pi
}_{\perp }=\phi _{B}$ is the Berry phase for the orbit $\Gamma $.\
Integration in Eq. $\left( \ref{pipiA}\right) $ defines the cross-sectional
area $\oint \pi _{x}d\pi _{y}=S_{0}(\varepsilon _{0},K_{z})$ of the orbit $%
\Gamma $ which is the intersection of the constant energy surface $%
\varepsilon _{0}\left( \pi \right) =$const and the plane $\pi _{z}=K_{z}=$%
const. As shown in \cite{PIERREEBS} $S_{0}(\varepsilon =\varepsilon
_{0}-M_{z}B,\pi _{z})$ can then be written as
\begin{equation}
S_{0}(\mathcal{E},\pi _{z})=\frac{2\pi \left\vert e\right\vert B}{\hbar }%
\left( n+\frac{1}{2}-\phi _{B}-\frac{1}{2\mathcal{\pi }\left\vert
e\right\vert }\oint \frac{M_{z}\left( \mathbf{\pi }\right) d\kappa }{%
\left\vert \partial \varepsilon /\hbar \partial \mathbf{\pi }_{\perp
}\right\vert }\right)  \label{SO}
\end{equation}
here $d\kappa =\sqrt{d\pi _{x}^{2}+d\pi _{y}^{2}}$ is an elementary length
of the $\pi $ orbit. We have thus succeeded to deduce the required result
Eq. $\left( \ref{SO}\right) $ (first found by Roth \cite{ROTH} in a
different way and without mentioning the Berry phase), as resulting from the
generalized Peierls substitution in the BS condition.\ The importance in
this expression of the Berry's phase for electrons in metal in connection to
band degeneracy was later discussed by Mikitik and Sharlay \cite{MIKITIK}. \
Eq. $\left( \ref{SO}\right) $ implicitly determines the energy levels $%
\varepsilon _{n}\left( K_{z}\right) $. For instance for the case where the
Fermi surface is an ellipsoid of revolution characterized by two effective
masses, a transverse $m_{\perp }$ and a longitudinal $m_{l}$ one has
\begin{equation}
\varepsilon _{0}=\hbar ^{2}\left( \frac{\mathbf{\pi }_{\perp }^{2}}{%
2m_{\perp }}+\frac{K_{z}^{2}}{2m_{l}}\right)
\end{equation}
and $S(\varepsilon ,K_{z})$ is a disc of radius square $\mathbf{\pi }_{\perp
}^{2}$. Therefore for $M_{z}=$const we have the following generalized
relation for the Landau levels:
\begin{equation*}
\varepsilon _{n}=\frac{\left\vert e\right\vert B\hbar }{m_{\perp }}\left( n+%
\frac{1}{2}-\phi _{B}-\frac{m_{\perp }}{\left\vert e\right\vert }%
M_{z}\right) +\frac{\hbar ^{2}K_{z}^{2}}{2m_{l}}
\end{equation*}
which shows that both the magnetic moment and Berry's phase can influence
the energy levels. A nice illustration of this result is provided by
electrons in graphene with broken inversion symmetry \cite{PIERREGRAPHENE}.

\subsection{Second order diagonalization}

As explained in section 2, the Hamiltonian diagonalization at this order
requires the computation of the Berry connections and the covariant
dynamical variables. This will thus be our first task. Note that, from now
on, for the sake on simplicity, we will keep the notation introduced in the
previous section, $\delta \mathbf{B\equiv B}$.

\subsubsection{Computation of the Berry phases and dynamical variables to
the second order}

Starting from the general expressions for the non projected Berry phases at
the second order Eqs. \ref{Conn2}\ref{conn2A}\ref{conn2B} we have in the
present case for $\mathcal{A}^{\mathbf{X}}=\int_{0}^{\hbar }\mathcal{A}%
_{\alpha }^{\mathbf{X}}d\alpha $ :
\begin{eqnarray*}
\mathcal{A}^{\mathbf{X}} &=&\hbar \mathcal{A}_{0}^{\mathbf{X}}\left( \mathbf{%
R+}\frac{\hbar }{4}\mathcal{A}_{0}^{R_{l}},\mathbf{K+}\frac{\hbar }{4}%
\mathcal{A}_{0}^{P_{l}}\right) \\
&&-\frac{\hbar ^{2}}{2}\left\{ \left[ .,\varepsilon _{0}\right] ^{-1}.\left(
\mathcal{P}_{-}\left\{ \frac{1}{2}\mathcal{A}_{0}^{P_{l}}\nabla
_{K_{l}}\varepsilon _{0}+\frac{1}{2}\mathcal{A}_{0}^{R_{l}}\nabla
_{R_{l}}\varepsilon _{0}+H.C.\right\} \right) \right. \\
&&\left. -\frac{i}{4}\left\{ \mathcal{P}_{-}\mathcal{A}_{0}^{R_{k}}\mathcal{P%
}_{+}\mathcal{A}_{0}^{R_{l}}+H.C.\right\} \varepsilon _{klm}eB_{m},\mathbf{X}%
+\hbar \mathcal{A}_{0}^{\mathbf{X}}\right\}
\end{eqnarray*}
where now $\mathbf{X=}\left( \mathbf{R,K}\right) $ and the same for $\mathbf{%
Y}$. We aim now at writing the connection $\mathcal{A}^{\mathbf{X}}$ in a
more convenient form as an expansion in terms of the zero-order (actually
first order in $\hbar $) $\mathcal{A}_{0}^{\mathbf{X}}.$ To do so, we first
start to express the crystal momentum Berry phase $\mathcal{A}^{\mathbf{K}}$
as a function of the position Berry phase $\mathcal{A}^{\mathbf{R}}$. The
computation of $\mathcal{A}^{\mathbf{K}}$ involves a commutator with $%
\mathbf{K}+\mathcal{A}_{0}^{\mathbf{K}}$ that we compute first. Notice that
the first order energy operator $\varepsilon _{0}$ in Eq. $\left( \ref%
{E1bloch}\right) $ and the Berry connections at the zeroth order depend on $%
\mathbf{\Pi }=\mathbf{K-}\frac{e}{\hbar }\mathbf{A}^{\text{ }}$ as well as
on $\mathcal{A}_{0}^{K_{m}}=e\nabla _{R_{m}}A_{l}\mathcal{A}_{0}^{R_{l}}$,
which allows us to write the commutator of $K_{m}+\hbar \mathcal{A}%
_{0}^{K_{m}}$ with any such operator , e.g. $\left[ K_{m}+\hbar \mathcal{A}%
_{0}^{K_{m}},\cdot \right] $ in the following way :
\begin{equation*}
\left[ K_{m}+\hbar \mathcal{A}_{0}^{K_{m}},\cdot \right] =e\nabla
_{R_{m}}A_{l}\left[ R_{l}+\hbar \mathcal{A}_{0}^{R_{l}},\cdot \right]
\end{equation*}
Let us stress that this formula is not valid for the electric potential term
appearing in the Berry phase formula since it depends exclusively on $%
\mathbf{R}$. The action of $\left[ K_{m}+\hbar \mathcal{A}_{0}^{K_{m}},%
\right] $ on this term has thus to be computed independently. The additional
contribution due to the electric field and the magnetic field to the
momentum Berry phase is seen easily from the formula for the to be
\begin{equation*}
-i\frac{\hbar ^{2}}{2}\left[ \left[ .,\hat{\varepsilon}_{0}\right]
^{-1}.\left( \mathcal{A}_{0}^{R_{l}}\nabla _{R_{m}}\nabla _{R_{l}}V\left(
\mathbf{R}\right) \right) -\frac{ie}{2}\left\{ \mathcal{P}_{-}\mathcal{A}%
_{0}^{\mathbf{R}}\times \mathcal{P}_{+}\mathcal{A}_{0}^{\mathbf{R}}\right\}
.\nabla _{R_{m}}\mathbf{B}\right]
\end{equation*}

Now the first term and the last term in the crystal momentum Berry phase can
also be replaced as a function of the position Berry phase contribution. But
at this point we need to go back to the technicality mentioned in the first
section. While the final results do not depend on the way variables are
symmetrized, one has to choose an initial symmetrization for the
diagonalized energy at the zeroth order as well as for the Berry phase. For
the diagonalized energy, we will choose to write it as a series of symmetric
monomials in the momentum $\mathbf{\Pi }$. Concerning the Berry phase for
the crystal momentum the natural choice is to symmetrize initially the
variables such that at the lowest order in $\hbar $ one has $\mathcal{A}%
_{0}^{\mathbf{K}}=\frac{1}{2}e\nabla _{R_{m}}A_{l}\left( \mathbf{R}\right)
\mathcal{A}_{0}^{\mathbf{R}}+\frac{1}{2}e\mathcal{A}_{0}^{\mathbf{R}}\nabla
_{R_{m}}A_{l}\left( \mathbf{R}\right) $. This will introduce some technical
problem later, but it is the most simple choice for us. As a consequence, we
obtain readily for the terms of interest :
\begin{eqnarray*}
&&\hbar \mathcal{A}_{0}^{\mathbf{K}}\left( \mathbf{R+}\frac{\hbar }{4}%
\mathcal{A}_{0}^{R_{l}}\right) +\frac{\hbar ^{2}}{4}\left( \mathcal{P}_{+}%
\left[ \mathcal{A}_{0}^{\mathbf{Y}}\right] .\mathbf{\nabla }_{\mathbf{Y}}%
\mathcal{P}_{+}\left[ \mathcal{A}_{0}^{K_{m}}\right] +H.C.\right) \\
&=&\frac{1}{2}\hbar e\nabla _{R_{m}}A_{l}\left( \mathbf{R+}\frac{1}{4}%
\mathcal{A}_{0}^{\mathbf{R}}\right) \mathcal{A}_{0}^{\mathbf{R}}\left(
\mathbf{R+}\frac{\hbar }{4}\mathcal{A}_{0}^{R_{l}},\mathbf{K+}\frac{\hbar }{4%
}\mathcal{A}_{0}^{P_{l}}\right) \\
&&+\frac{1}{2}\hbar \mathcal{A}_{0}^{\mathbf{R}}\left( \mathbf{R+}\frac{%
\hbar }{4}\mathcal{A}_{0}^{R_{l}}\right) e\nabla _{R_{m}}A_{l}\left( \mathbf{%
R+}\frac{1}{4}\mathcal{A}_{0}^{\mathbf{R}},\mathbf{P+}\frac{1}{4}\mathcal{A}%
_{0}^{\mathbf{P}}\right) \\
&&+\frac{\hbar ^{2}}{4}e\nabla _{R_{m}}A_{l}\left( \mathcal{P}_{+}\left[
\mathcal{A}_{0}^{\mathbf{R}}\right] .\mathbf{\nabla }_{\mathbf{Y}}\mathcal{P}%
_{+}\left[ \mathcal{A}_{0}^{R_{m}}\right] +H.C.\right)
\end{eqnarray*}
so that can now gather all these terms and compute $\mathcal{A}^{K_{m}}$ as
a function of $\mathcal{A}^{R_{l}}$ :
\begin{eqnarray*}
\mathcal{A}^{K_{m}} &=&\frac{1}{2}\left( e\nabla _{R_{m}}A_{l}\left( \mathbf{%
R+}\frac{1}{4}\mathcal{A}_{0}^{\mathbf{R}},\mathbf{P+}\frac{1}{4}\mathcal{A}%
_{0}^{\mathbf{P}}\right) \mathcal{A}^{R_{l}}+\mathcal{A}^{R_{l}}e\nabla
_{R_{m}}A_{l}\left( \mathbf{R+}\frac{1}{4}\mathcal{A}_{0}^{\mathbf{R}},%
\mathbf{P+}\frac{1}{4}\mathcal{A}_{0}^{\mathbf{P}}\right) \right) \\
&&-i\frac{\hbar ^{2}}{2}\left[ \left[ .,\hat{\varepsilon}_{0}\right]
^{-1}.\left( \mathcal{A}_{0}^{R_{l}}\nabla _{R_{m}}\nabla _{R_{l}}V\left(
\mathbf{R}\right) \right) -\frac{ie}{2}\left\{ \mathcal{P}_{-}\mathcal{A}%
_{0}^{\mathbf{R}}\times \mathcal{P}_{+}\mathcal{A}_{0}^{\mathbf{R}}\right\}
.\nabla _{R_{m}}\mathbf{B}\right]
\end{eqnarray*}
The potential $A_{l}$ being developer to the first order in $\hbar $. As
expected, only the contribution due to the electric term is not rewritten as
a function of $\mathcal{A}^{R_{l}}$. The quantity $-i\frac{\hbar ^{2}}{2}%
\left[ \left[ .,\hat{\varepsilon}_{0}\right] ^{-1}.\left( \mathcal{A}%
_{0}^{R_{l}}\nabla _{R_{m}}\nabla _{R_{l}}V\left( \mathbf{R}\right) \right) -%
\frac{ie}{2}\left\{ \mathcal{P}_{-}\mathcal{A}_{0}^{\mathbf{R}}\times
\mathcal{P}_{+}\mathcal{A}_{0}^{\mathbf{R}}\right\} .\nabla _{R_{m}}\mathbf{B%
}\right] $ can be computed on the bands as being equal to :
\begin{eqnarray*}
&&-i\frac{\hbar ^{2}}{2}\left[ \left[ .,\hat{\varepsilon}_{0}\right]
^{-1}.\left( \mathcal{A}_{0}^{R_{l}}\nabla _{R_{m}}\nabla _{R_{l}}V\left(
\mathbf{R}\right) \right) -\frac{ie}{2}\left\{ \mathcal{P}_{-}\mathcal{A}%
_{0}^{\mathbf{R}}\times \mathcal{P}_{+}\mathcal{A}_{0}^{\mathbf{R}}\right\}
.\nabla _{R_{m}}\mathbf{B}\right] _{MN} \\
&=&i\frac{\hbar ^{2}}{2}\frac{\left( \mathcal{A}_{0}^{R_{l}}\right)
_{MN}\nabla _{R_{m}}\nabla _{R_{l}}V\left( R\right) }{\hat{\varepsilon}_{0M}-%
\hat{\varepsilon}_{0N}}-\frac{\hbar ^{2}}{4}\left\{ \mathcal{P}_{-}\mathcal{A%
}_{0}^{\mathbf{R}}\times \mathcal{P}_{+}\mathcal{A}_{0}^{\mathbf{R}}\right\}
_{MN}.\nabla _{R_{m}}\mathbf{B}\text{ if }M\neq N \\
&=&0\text{ if }M=N
\end{eqnarray*}
so that we have for the $(M,N)$ component :
\begin{eqnarray*}
\mathcal{A}_{MN}^{K_{m}} &=&\frac{1}{2}\left[ \left( e\nabla
_{R_{m}}A_{l}\left( \mathbf{R+}\frac{\hbar }{4}\mathcal{A}_{0}^{\mathbf{R}%
}\right) \mathcal{A}^{R_{l}}+\mathcal{A}^{R_{l}}e\nabla _{R_{m}}A_{l}\left(
\mathbf{R+}\frac{\hbar }{4}\mathcal{A}_{0}^{\mathbf{R}}\right) \right) %
\right] _{MN} \\
&&+\left[ i\frac{\hbar ^{2}}{2}\frac{\left( \mathcal{A}_{0}^{R_{l}}\right)
_{MN}\nabla _{R_{m}}\nabla _{R_{l}}V\left( R\right) }{\hat{\varepsilon}_{0M}-%
\hat{\varepsilon}_{0N}}-\frac{\hbar ^{2}}{4}\left\{ \mathcal{P}_{-}\mathcal{A%
}_{0}^{\mathbf{R}}\times \mathcal{P}_{+}\mathcal{A}_{0}^{\mathbf{R}}\right\}
_{MN}.\nabla _{R_{m}}\mathbf{B}\right] \hat{\delta}_{MN} \\
&=&\frac{1}{2}\left[ \left( e\nabla _{R_{m}}A_{l}\left( \mathbf{R}\right)
\mathcal{A}^{R_{l}}+\mathcal{A}^{R_{l}}e\nabla _{R_{m}}A_{l}\left( \mathbf{R}%
\right) \right) \right] _{MN} \\
&&\mathbf{+}\frac{\hbar ^{2}}{8}e\nabla _{R_{k}}\nabla _{R_{m}}A_{l}\left(
\mathbf{R},\mathbf{P}\right) \left( \mathcal{A}_{0}^{R_{l}}\mathcal{A}%
_{0}^{R_{k}}+\mathcal{A}_{0}^{R_{k}}\mathcal{A}_{0}^{R_{l}}\right) \\
&&+\left[ i\frac{\hbar ^{2}}{2}\frac{\left( \mathcal{A}_{0}^{R_{l}}\right)
_{MN}\nabla _{R_{m}}\nabla _{R_{l}}V\left( R\right) }{\hat{\varepsilon}_{0M}-%
\hat{\varepsilon}_{0N}}-\frac{\hbar ^{2}}{4}\left\{ \mathcal{P}_{-}\mathcal{A%
}_{0}^{\mathbf{R}}\times \mathcal{P}_{+}\mathcal{A}_{0}^{\mathbf{R}}\right\}
_{MN}.\nabla _{R_{m}}\mathbf{B}\right] \hat{\delta}_{MN}
\end{eqnarray*}
where we defined the notation $\hat{\delta}_{MN}=1-\delta _{MN}$,

As explained before, the relevant variables for the physical expressions are
rather the momentum variables $\mathbf{\Pi =K-A}$. Since at first order we
already had $\mathcal{A}_{0}^{\mathbf{\pi }}=e\mathcal{A}_{0}^{\mathbf{R}%
}\times \mathbf{B}$ we would expect a similar formula at the second order.
However, since our method has been designed to deal with quantities
symmetrized as functions of $\mathbf{R}$ and $\mathbf{K}$, it appears that
the momentum variable does not appear directly in our Hamiltonian. We thus
postpone its introduction till the computation of the diagonalized
Hamiltonian.

We can now, as a second step, concentrate on the position Berry phase. We
rewrite it as :
\begin{eqnarray*}
\mathcal{A}^{\mathbf{R}} &=&\hbar \mathcal{A}_{0}^{\mathbf{R}}+\frac{\hbar
^{2}}{8}\left\{ \mathcal{A}_{0}^{R_{l}}\nabla _{R_{l}}\mathcal{A}_{0}^{%
\mathbf{R}}+\mathcal{A}_{0}^{K_{l}}\nabla _{K_{l}}\mathcal{A}_{0}^{\mathbf{R}%
}+H.C.\right\} \\
&&+\left[ R+\hbar \mathcal{A}_{0}^{\mathbf{R}},\frac{\hbar ^{2}}{2}\left[ .,%
\hat{\varepsilon}_{0}\right] ^{-1}.\left[ \left( \mathcal{P}_{-}\left\{
\frac{1}{2}\mathcal{A}_{0}^{R_{l}}\nabla _{R_{l}}\varepsilon _{0}\left(
\mathbf{X}\right) +\frac{1}{2}\mathcal{A}_{0}^{K_{l}}\nabla
_{K_{l}}\varepsilon _{0}\left( \mathbf{X}\right) +H.C.\right\} \right) %
\right] \right] \\
&&+\left[ R+\hbar \mathcal{A}_{0}^{\mathbf{R}},\frac{e\hbar ^{2}}{4}\left(
\mathcal{P}_{-}\mathcal{A}_{0}^{R_{n}}\mathcal{P}_{+}\mathcal{A}%
_{0}^{R_{r}}+H.C.\right) \varepsilon ^{nru}.B_{u}\right]
\end{eqnarray*}
We can also rewrite the third term $\left( M,N\right) $ component\ as :
\begin{eqnarray*}
&&\left( \left[ .,\hat{\varepsilon}_{0}\right] ^{-1}.\left[ \left( \mathcal{P%
}_{-}\left\{ \frac{1}{2}\mathcal{A}_{0}^{R_{l}}\nabla _{R_{l}}\varepsilon
_{0}\left( \mathbf{X}\right) +\frac{1}{2}\mathcal{A}_{0}^{K_{l}}\nabla
_{K_{l}}\varepsilon _{0}\left( \mathbf{X}\right) +H.C.\right\} \right) %
\right] \right) _{MN} \\
&=&\frac{1}{2}\frac{\left( \mathcal{A}_{0}^{R_{l}}\right) _{MN}\nabla
_{R_{l}}\left( \varepsilon _{0M}+\varepsilon _{0N}\right) +\left( \mathcal{A}%
_{0}^{K_{l}}\right) _{MN}\nabla _{K_{l}}\left( \varepsilon _{0M}+\varepsilon
_{0N}\right) }{\hat{\varepsilon}_{0M}-\hat{\varepsilon}_{0N}}\text{ if }%
M\neq N \\
&=&0\text{ if }M=N
\end{eqnarray*}
Using also, for quantities depending on $\mathbf{\Pi =K-A}$ and $\mathbf{R}$
that $\left( \mathcal{A}_{0}^{R_{l}}\right) _{MP}\nabla _{R_{l}}+\left(
\mathcal{A}_{0}^{K_{l}}\right) _{MP}\nabla _{K_{l}}=-e\mathbf{B}.\left(
\left( \mathcal{A}_{0}^{\mathbf{R}}\right) _{MP}\times \nabla _{\mathbf{\Pi }%
}\right) $ at the lowest order in $\hbar $ we now give an expanded formula
for the \ position Berry phase at the second order :
\begin{eqnarray*}
\left( \mathcal{A}^{\mathbf{R}}\right) _{MN} &=&\left( \hbar \mathcal{A}%
_{0}^{\mathbf{R}}\right) _{MN}-\frac{\hbar ^{2}}{16}\sum_{P}\left\{ e\mathbf{%
B}.\left( \left( \mathcal{A}_{0}^{\mathbf{R}}\right) _{MP}\times \nabla _{%
\mathbf{\Pi }}\right) \left( \mathcal{A}_{0}^{\mathbf{R}}\right)
_{PN}+M\leftrightarrow N\right\} \\
&&+\frac{\hbar ^{2}}{2}i\mathbf{\nabla }_{\mathbf{\Pi }}\frac{\left(
\mathcal{A}_{0}^{R_{l}}\right) _{MN}\nabla _{R_{l}}V\left( \mathbf{R}\right)
+e\mathbf{B}.\left( \left( \mathcal{A}_{0}^{\mathbf{R}}\right) _{MN}\times
\nabla _{\mathbf{\Pi }}\right) \left( \hat{\varepsilon}_{0M}+\hat{\varepsilon%
}_{0N}\right) }{\hat{\varepsilon}_{0M}-\hat{\varepsilon}_{0N}}\hat{\delta}%
_{MN} \\
&&+\frac{\hbar ^{2}}{2}\sum_{P}\left( \mathcal{A}_{0}^{\mathbf{R}}\right)
_{MP}\hat{\delta}_{PM}\frac{\left( \mathcal{A}_{0}^{R_{l}}\right)
_{PN}\nabla _{R_{l}}V\left( \mathbf{R}\right) +e\mathbf{B}.\left( \left(
\mathcal{A}_{0}^{\mathbf{R}}\right) _{PN}\times \nabla _{\mathbf{\Pi }%
}\right) \left( \hat{\varepsilon}_{0P}+\hat{\varepsilon}_{0N}\right) }{\hat{%
\varepsilon}_{0P}-\hat{\varepsilon}_{0N}} \\
&&-\frac{\hbar ^{2}}{2}\sum_{P}\frac{\left( \mathcal{A}_{0}^{R_{l}}\right)
_{MP}\nabla _{R_{l}}V\left( \mathbf{R}\right) +e\mathbf{B}.\left( \left(
\mathcal{A}_{0}^{\mathbf{R}}\right) _{MP}\times \nabla _{\mathbf{\Pi }%
}\right) \left( \hat{\varepsilon}_{0M}+\hat{\varepsilon}_{0P}\right) }{\hat{%
\varepsilon}_{0M}-\hat{\varepsilon}_{0P}}\hat{\delta}_{MP}\left( \mathcal{A}%
_{0}^{\mathbf{R}}\right) _{PN} \\
&&-\frac{e\hbar ^{2}}{4}\mathbf{\nabla }_{\mathbf{\Pi }}\left( \left(
\mathcal{A}_{0}^{\mathbf{R}}\right) _{MN}\times \left( \mathcal{A}_{0}^{%
\mathbf{R}}\right) _{NN}\right) .\mathbf{B}\hat{\delta}_{MN} \\
&&+i\frac{e\hbar ^{2}}{8}\left( \sum_{P}\left( \mathcal{A}_{0}^{\mathbf{R}%
}\right) _{MP}\left( \left( \mathcal{A}_{0}^{\mathbf{R}}\right) _{PN}\times
\left( \mathcal{A}_{0}^{\mathbf{R}}\right) _{NN}\right) .\mathbf{B}\hat{%
\delta}_{PN}+H.C.\right) \\
&&-i\frac{e\hbar ^{2}}{8}\left( \sum_{P}\left( \left( \mathcal{A}_{0}^{%
\mathbf{R}}\right) _{MP}\times \left( \mathcal{A}_{0}^{\mathbf{R}}\right)
_{PP}\right) .\mathbf{B}\hat{\delta}_{MP}\left( \mathcal{A}_{0}^{\mathbf{R}%
}\right) _{PN}+H.C.\right)
\end{eqnarray*}
and its projection on Band $N$.
\begin{eqnarray*}
\mathcal{P}_{N}\left( \mathcal{A}_{NN}^{\mathbf{R}}\right) &=&\left(
\mathcal{A}^{\mathbf{R}}\right) _{NN}=\left( \hbar \mathcal{A}_{0}^{\mathbf{R%
}}\right) _{NN}-\frac{\hbar ^{2}}{8}\sum_{P}\left\{ e\mathbf{B}.\left(
\left( \mathcal{A}_{0}^{\mathbf{R}}\right) _{NP}\times \nabla _{\mathbf{\Pi }%
}\right) \left( \mathcal{A}_{0}^{\mathbf{R}}\right) _{PN}\right\} \\
&&+\frac{\hbar ^{2}}{2}\sum_{P}\left( \left( \mathcal{A}_{0}^{\mathbf{R}%
}\right) _{NP}\left( \mathcal{A}_{0}^{R_{l}}\right) _{PN}+\left( \mathcal{A}%
_{0}^{R_{l}}\right) _{NP}\left( \mathcal{A}_{0}^{\mathbf{R}}\right)
_{PN}\right) \frac{\nabla _{R_{l}}V\left( \mathbf{R}\right) }{\hat{%
\varepsilon}_{0P}-\hat{\varepsilon}_{0N}}\hat{\delta}_{PN} \\
&&-\frac{\hbar ^{2}}{2}\left( \sum_{P}\left( e\mathbf{B}.\left( \left(
\mathcal{A}_{0}^{\mathbf{R}}\right) _{NP}\times \nabla _{\mathbf{\Pi }%
}\right) \right) \frac{\left( \hat{\varepsilon}_{0P}+\hat{\varepsilon}%
_{0N}\right) }{\hat{\varepsilon}_{0P}-\hat{\varepsilon}_{0N}}\hat{\delta}%
_{PN}\left( \mathcal{A}_{0}^{\mathbf{R}}\right) _{PN}-H.C.\right) \\
&&+i\frac{e\hbar ^{2}}{8}\sum_{P}\left( \left( \mathcal{A}_{0}^{\mathbf{R}%
}\right) _{NP}\left( \left( \mathcal{A}_{0}^{\mathbf{R}}\right) _{PN}\times
\left( \mathcal{A}_{0}^{\mathbf{R}}\right) _{NN}\right) -\left( \left(
\mathcal{A}_{0}^{\mathbf{R}}\right) _{NP}\times \left( \mathcal{A}_{0}^{%
\mathbf{R}}\right) _{PP}\left( \mathcal{A}_{0}^{\mathbf{R}}\right)
_{PN}\right) .\mathbf{B}\right) \hat{\delta}_{PN} \\
&&+H.C.
\end{eqnarray*}
For later purpose we will need also the modified Berry phase $\emph{A}_{N}^{%
\mathbf{R}}$ appearing in the dynamical Band variable $\mathbf{r}_{N}$. A
direct application of our general set up of the previous section yields.
\begin{eqnarray*}
\emph{A}_{N}^{\mathbf{R}} &=&\mathcal{A}_{NN}^{\mathbf{R}}+\frac{\hbar ^{2}}{%
4}\left( \mathcal{P}_{N}\left[ \mathcal{A}_{0}^{\mathbf{Y}}\right] .\mathbf{%
\nabla }_{\mathbf{Y}}\mathcal{P}_{N}\left[ \mathcal{A}_{0}^{\mathbf{X}}%
\right] +H.C.\right) \\
&=&\mathcal{A}_{NN}^{\mathbf{R}}+\frac{\hbar ^{2}}{4}\left( \mathcal{A}%
_{0NN}^{\mathbf{R}}.\mathbf{\nabla }_{\mathbf{R}}\mathcal{A}_{0NN}^{\mathbf{R%
}}+\mathcal{A}_{0NN}^{\mathbf{P}}.\mathbf{\nabla }_{\mathbf{P}}\mathcal{A}%
_{0NN}^{\mathbf{R}}+H.C.\right) \\
&=&\mathcal{A}_{NN}^{\mathbf{R}}-e\frac{\hbar ^{2}}{4}\left( \mathbf{B}%
.\left( \left( \mathcal{A}_{0}^{\mathbf{R}}\right) _{NN}\times \nabla _{%
\mathbf{\Pi }}\right) \mathcal{A}_{0NN}^{\mathbf{R}}+H.C.\right)
\end{eqnarray*}
Ultimately, the projected dynamical variables follow directly from our
general method of diagonalization. They are given by the canonical variables
shifted by the modified projected Berry phases :
\begin{eqnarray*}
\mathbf{r}_{N} &=&\mathbf{R}+\emph{A}_{N}^{\mathbf{R}} \\
\mathbf{k}_{N} &=&\mathbf{K}+\emph{A}_{N}^{\mathbf{K}}\equiv \mathbf{K}+%
\mathcal{A}_{NN}^{\mathbf{K}} \\
\mathbf{\pi }_{N} &=&\mathbf{\Pi }+\mathcal{A}^{\mathbf{\pi }}=\mathbf{\Pi }+%
\frac{1}{2}\left[ e\mathcal{A}\mathbf{\times B}\left( \mathbf{R+}\frac{\hbar
}{4}\mathcal{A}_{0}^{\mathbf{R}},\mathbf{P+}\frac{\hbar }{4}\mathcal{A}_{0}^{%
\mathbf{P}}\right) +H.C.\right]
\end{eqnarray*}
Note that since $-i\nabla _{R_{m}}\frac{\hbar ^{2}}{4}\left[ .,\hat{%
\varepsilon}_{0}\right] ^{-1}.\left( \mathcal{A}_{0}^{R_{l}}\nabla
_{R_{l}}V\left( \mathbf{R}\right) +H.C.\right) $ has no diagonal element
\begin{eqnarray*}
\emph{A}_{N}^{K_{m}} &=&\frac{1}{2}\left[ \left( e\nabla _{R_{m}}A_{l}\left(
\mathbf{R}\right) \emph{A}_{N}^{R_{l}}+\emph{A}_{N}^{R_{l}}e\nabla
_{R_{m}}A_{l}\left( \mathbf{R}\right) \right) \right] _{NN} \\
&&\mathbf{+}\frac{\hbar ^{2}}{8}e\nabla _{R_{k}}\nabla _{R_{m}}A_{l}\left(
\mathbf{R},\mathbf{P}\right) \mathcal{P}_{N}\left( \mathcal{A}_{0}^{R_{l}}%
\mathcal{A}_{0}^{R_{k}}+\mathcal{A}_{0}^{R_{k}}\mathcal{A}_{0}^{R_{l}}\right)
\\
&&+\frac{\hbar ^{2}}{4}e\nabla _{R_{k}}\nabla _{R_{m}}A_{l}\left( \mathbf{R},%
\mathbf{P}\right) \left( \mathcal{P}_{N}\left[ \mathcal{A}_{0}^{R_{l}}\right]
\mathcal{P}_{N}\left[ \mathcal{A}_{0}^{R_{k}}\right] +\mathcal{P}_{N}\left[
\mathcal{A}_{0}^{R_{k}}\right] \mathcal{P}_{N}\left[ \mathcal{A}_{0}^{R_{l}}%
\right] \right)
\end{eqnarray*}

\subsubsection{Band Hamiltonian}

We now turn to the Problem of deriving the diagonalized Hamiltonian. As
explained in section II, the effective diagonal Hamiltonian for the $Nth$
band can be written (see Eq. $\left( \ref{E2}\right) $ )
\begin{eqnarray}
H_{N}\left( \mathbf{X}\right) &=&\varepsilon _{0N}\left( \mathbf{x}\right) +%
\frac{i}{2}\mathcal{P}_{N}\left\{ \left[ \varepsilon _{0}\left( \mathbf{x}%
\right) ,\mathcal{\hat{A}}^{R_{l}}\right] \mathcal{\hat{A}}^{P_{l}}-\left[
\varepsilon _{0}\left( \mathbf{x}\right) ,\mathcal{\hat{A}}^{P_{l}}\right]
\mathcal{\hat{A}}^{R_{l}}\right\}  \notag \\
&&-\frac{1}{8}\mathcal{P}\left\{ _{N}\left[ \mathcal{P}_{+}\left[ \left[
\varepsilon _{0}\left( \mathbf{x}\right) ,\mathcal{\hat{A}}^{R_{l}}\right]
\mathcal{\hat{A}}^{P_{l}}-\left[ \varepsilon _{0}\left( \mathbf{x}\right) ,%
\mathcal{\hat{A}}^{P_{l}}\right] \mathcal{\hat{A}}^{R_{l}}\right] ,\mathcal{%
\hat{A}}^{R_{l}}\right] \mathcal{\hat{A}}^{P_{l}}\right.  \notag \\
&&\left. -\left[ \mathcal{P}_{+}\left[ \left[ \varepsilon _{0}\left( \mathbf{%
x}\right) ,\mathcal{\hat{A}}^{R_{l}}\right] \mathcal{\hat{A}}^{P_{l}}-\left[
\varepsilon _{0}\left( \mathbf{x}\right) ,\mathcal{\hat{A}}^{P_{l}}\right] %
\right] \mathcal{\hat{A}}^{R_{l}},\mathcal{\hat{A}}^{P_{l}}\right] \mathcal{%
\hat{A}}^{R_{l}}\right\} -\frac{\hbar }{2}\left\langle \varepsilon
_{0}\left( \mathbf{x}\right) \right\rangle
\end{eqnarray}
Where $\mathcal{P}_{N}$ is the projection on this band. As explained in \cite%
{DIAGOEXACT} this Hamiltonian can be rewritten in a more enlightening way as
:
\begin{equation*}
H_{N}=\hat{\varepsilon}_{0}\left( \mathbf{\pi }\right) +M\left[ \varepsilon
_{0}\right] +\frac{1}{2}M\left[ M\left[ \varepsilon _{0}\right] \right]
+V\left( \mathbf{r}\right) -\frac{\hbar }{2}\left\langle \varepsilon
_{0}\left( \mathbf{x}\right) \right\rangle
\end{equation*}
Where the magnetization operator $M$ acts on the following way :
\begin{equation*}
M\left[ X\right] =\frac{i}{4\hbar }\mathcal{P}_{N}\left\{ \left[ X,\mathcal{%
\hat{A}}^{R_{l}}\right] \mathcal{\hat{A}}^{K_{l}}-\left[ X,\mathcal{\hat{A}}%
^{K_{l}}\right] \mathcal{\hat{A}}^{R_{l}}\right\} +H.C.
\end{equation*}
As consequence, the Hamiltonian is given by a series of magnetization terms
acting on the \textquotedblright classical\textquotedblright\ Hamiltonian,
that is the Hamiltonian obtained for classical commuting dynamical
variables. We will more give detailed formula for the various terms as
functions of the first order Berry phases, but we turn first to the problem
of choice of variables in writing the energy operator.

\paragraph{Momentum variables. Second order generalized Peierls substitution}

For the moment, the pair of canonical variables implied by our method are
the \textquotedblright non physical one\textquotedblright\ $x=\left( \mathbf{%
r}_{N},\mathbf{k}_{N}\right) $
\begin{eqnarray*}
\mathbf{r}_{N} &=&\mathbf{R}+\emph{A}_{N}^{\mathbf{R}} \\
\mathbf{k}_{N} &=&\mathbf{K}+\emph{A}_{N}^{\mathbf{K}}\equiv \mathbf{K}+%
\mathcal{A}_{NN}^{\mathbf{K}}
\end{eqnarray*}
One aims at replacing $\mathbf{k}_{N}$ by a \textquotedblright
physical\textquotedblright\ momentum variable which reduces to $\mathbf{\Pi }%
=\mathbf{K-}\frac{e}{\hbar }\mathbf{A}$ at the lowest order. To do so assume
the electric potential is set to $0$ since it plays no role here, and
rewrite the quantity $\varepsilon _{0N}\left( \mathbf{x}\right) $ that
arises in the diagonalized Hamiltonian as :
\begin{equation*}
\varepsilon _{0N}\left( \mathbf{x}\right) =\varepsilon _{0N}\left( \mathbf{k}%
_{N}-\mathbf{A}\left( \mathbf{r}_{N}\right) \right)
\end{equation*}
so that one would be tempted to choose $\mathbf{k}_{N}-\mathbf{A}\left(
\mathbf{r}_{N}\right) $ as the right momentum variable. However a
computation of this quantity yields :
\begin{eqnarray*}
\mathbf{k}_{N}-e\mathbf{A}\left( \mathbf{r}_{N}\right) &=&\mathbf{K}+\emph{A}%
_{N}^{\mathbf{K}}-e\mathbf{A}\left( \mathbf{R}+\emph{A}_{N}^{\mathbf{R}%
}\right) \\
&=&\mathbf{K}-\mathbf{A}\left( \mathbf{R}\right) +\frac{1}{2}e\left[ \mathbf{%
\nabla }A_{l}\left( \mathbf{R}\right) \emph{A}_{N}^{R_{l}}+\emph{A}%
_{N}^{R_{l}}\mathbf{\nabla }A_{l}\left( \mathbf{R}\right) \right] \\
&&\mathbf{+}\frac{\hbar ^{2}}{8}e\nabla _{R_{k}}\mathbf{\nabla }A_{l}\left(
\mathbf{R},\mathbf{P}\right) \mathcal{P}_{N}\left( \mathcal{A}_{0}^{R_{l}}%
\mathcal{A}_{0}^{R_{k}}+\mathcal{A}_{0}^{R_{k}}\mathcal{A}_{0}^{R_{l}}\right)
\\
&&+\frac{\hbar ^{2}}{4}e\mathbf{\nabla }\nabla _{R_{k}}A_{l}\left( \mathbf{R}%
,\mathbf{P}\right) \left( \mathcal{P}_{N}\left[ \mathcal{A}_{0}^{R_{l}}%
\right] \mathcal{P}_{N}\left[ \mathcal{A}_{0}^{R_{k}}\right] +\mathcal{P}_{N}%
\left[ \mathcal{A}_{0}^{R_{k}}\right] \mathcal{P}_{N}\left[ \mathcal{A}%
_{0}^{R_{l}}\right] \right) \\
&&-e\left[ \nabla _{R_{k}}\mathbf{A}\left( \mathbf{R}\right) \emph{A}%
_{N}^{R_{k}}+\emph{A}_{N}^{R_{k}}\nabla _{R_{k}}\mathbf{A}\left( \mathbf{R}%
\right) \right] \\
&&-\frac{\hbar ^{2}}{4}e\nabla _{R_{k}}\nabla _{R_{l}}\mathbf{A}\left(
\mathbf{R},\mathbf{P}\right) \left( \mathcal{P}_{N}\left[ \mathcal{A}%
_{0}^{R_{l}}\right] \mathcal{P}_{N}\left[ \mathcal{A}_{0}^{R_{k}}\right] +%
\mathcal{P}_{N}\left[ \mathcal{A}_{0}^{R_{k}}\right] \mathcal{P}_{N}\left[
\mathcal{A}_{0}^{R_{l}}\right] \right) \\
&=&\mathbf{\Pi }+e\emph{A}_{N}^{\mathbf{R}}\times \mathbf{B}\left( \mathbf{R}%
\right) +e\frac{\hbar }{4}\left[ \mathcal{P}_{N}\left[ \mathcal{A}%
_{0}^{R_{k}}\right] \left( \emph{A}_{N}^{\mathbf{R}}\times \nabla _{R_{k}}%
\mathbf{B}\left( \mathbf{R}\right) \right) +H.C.\right] \\
&&+\frac{\hbar ^{2}}{8}e\nabla _{R_{k}}\mathbf{\nabla }A_{l}\left( \mathbf{R}%
,\mathbf{P}\right) \mathcal{P}_{N}\left( \mathcal{A}_{0}^{R_{l}}\mathcal{A}%
_{0}^{R_{k}}+\mathcal{A}_{0}^{R_{k}}\mathcal{A}_{0}^{R_{l}}\right)
\end{eqnarray*}
While the first three terms are physical since they involve the momentum $%
\mathbf{\Pi }$ and the magnetic field, the last one is problematic since it
will involve in the Hamiltonian some symmetric combinations of the vector
potential of the kind $\nabla _{R_{k}}A_{l}+\nabla _{R_{l}}A_{k}$. The
appearance of such terms is non physical but has nothing surprising here
since we ordered all our expressions in $\mathbf{R}$ and $\mathbf{K}$, not
in $\mathbf{R}$ and $\mathbf{\Pi }$. As a consequence, we cannot expect in
one individual expression to have only contributions of the momentum and
magnetic field operators. An other way to understand this problem is to
remember our choice of symmetrization for the crystal momentum Berry phase
that was easy to handle but does not fit when aiming at rewriting the
Hamiltonian in terms of $\mathbf{\Pi }$. Actually, having let the
derivatives of the electromagnetic potential half on the left and half on
the right automatically induces commutators terms proportional to $\nabla
_{R_{k}}A_{l}+\nabla _{R_{l}}A_{k}$. However, we know that the overall
process of diagonalization of the initial Hamiltonian has to depend only on $%
\mathbf{\Pi }$, $\mathbf{R}$, $\mathbf{B}\left( \mathbf{R}\right) $ so that
we know that the symmetric contributions $\nabla _{R_{k}}A_{l}+\nabla
_{R_{l}}A_{k}$ to the Hamiltonian have to cancel.

This assertion will be checked explicitly for the above problematic term
whose contribution (developing $\varepsilon _{0N}\left( \mathbf{x}\right) $
to the required order) is $\frac{\hbar ^{2}}{8}e\nabla _{R_{k}}\nabla
_{R_{n}}A_{l}\left( \mathbf{R},\mathbf{P}\right) \mathcal{P}_{N}\left(
\mathcal{A}_{0}^{R_{l}}\mathcal{A}_{0}^{R_{k}}+\mathcal{A}_{0}^{R_{k}}%
\mathcal{A}_{0}^{R_{l}}\right) \nabla _{\Pi _{n}}\varepsilon _{0N}\left(
\mathbf{x}\right) $, when diagonalizing the Hamiltonian. Actually we will
find a counterpart to it. However, it is unnecessary in general to check the
cancellation. These have to vanish by construction, so that we can discard
them directly. Their local appearance in individual terms will be discussed
briefly while computing the Bracket term for $\varepsilon _{0N}\left(
\mathbf{x}\right) $.

As a consequence of this discussion, our relevant variable for the momentum
in the Hamiltonian will be :
\begin{equation*}
\mathbf{\pi =\Pi }+e\emph{A}_{N}^{\mathbf{R}}\times \mathbf{B}\left( \mathbf{%
R}\right) +e\frac{\hbar }{4}\left[ \mathcal{P}_{N}\left[ \mathcal{A}%
_{0}^{R_{k}}\right] \left( \emph{A}_{N}^{\mathbf{R}}\times \nabla _{R_{k}}%
\mathbf{B}\left( \mathbf{R}\right) \right) +H.C.\right]
\end{equation*}
and in the sequel, the relevant couple of canonical variables will be :
\begin{equation*}
\left( \mathbf{x}\right) =\left( \mathbf{r}_{N},\mathbf{\pi }_{N}\right)
=\left( \mathbf{r,\pi }\right)
\end{equation*}
(the index $N$ being understood).

\paragraph{Magnetization terms}

At this order of approximation, and using our expressions for the Berry
phases, one can directly develop the expression for $M\left[ \varepsilon _{0}%
\right] $ to obtain the following decomposition :
\begin{eqnarray*}
M\left[ \varepsilon _{0}\right] &=&-\frac{1}{2}\left( \mathbf{\mu }\left(
\mathbf{x}\right) \mathbf{.B}\left( \mathbf{x}\right) +\mathbf{B}\left(
\mathbf{x}\right) \mathbf{.\mu }\left( \mathbf{x}\right) \right) -\mathbf{%
\hat{\mu}}^{k}\left( \mathbf{x}\right) \mathbf{.}\nabla _{R_{k}}\mathbf{B}%
\left( \mathbf{x}\right) \\
+ &&\text{Darwin term}
\end{eqnarray*}
where the Darwin term refers to an expression which reveals to be of the
same kind as the Darwin term in the Dirac Hamiltonian as we will see soon.

In this expression we have defined
\begin{eqnarray*}
\mu _{n} &=&-\frac{i}{2\hbar }\mathcal{P}_{N}\left\{ \left[ \varepsilon
_{0}\left( \mathbf{x}\right) ,\mathcal{\hat{A}}^{R_{l}}\right] \mathcal{\hat{%
A}}^{R_{m}}\right\} \varepsilon ^{lmn} \\
\hat{\mu}_{l}^{k} &=&-\frac{1}{16}\mathcal{P}_{N}\left[ \left[ \left[
\varepsilon _{0}\left( \mathbf{x}\right) ,\mathcal{\hat{A}}^{R_{m}}\right]
,\nabla _{\Pi _{k}}\mathcal{\hat{A}}^{R_{n}}\right] \right] \varepsilon
^{lmn} \\
&&-\frac{i\hbar ^{2}}{8}\left[ \varepsilon _{0}\left( \mathbf{x}\right) ,%
\mathcal{A}_{0}^{R_{k}}\right] _{NM}\left\{ \left( \mathcal{P}_{-}\mathcal{A}%
_{0}^{\mathbf{R}}\times \mathcal{P}_{+}\mathcal{A}_{0}^{\mathbf{R}}\right)
_{l}\right\} _{MN}+\frac{i\hbar ^{2}}{8}\left[ \varepsilon _{0}\left(
\mathbf{x}\right) ,\left\{ \left( \mathcal{P}_{-}\mathcal{A}_{0}^{\mathbf{R}%
}\times \mathcal{P}_{+}\mathcal{A}_{0}^{\mathbf{R}}\right) _{l}\right\} %
\right] _{NM}\mathcal{A}_{0MN}^{R_{k}}
\end{eqnarray*}
These expressions are directly obtained by expanding the magnetization term $%
M\left[ \varepsilon _{0}\right] =\frac{i}{4\hbar }\mathcal{P}_{N}\left\{ %
\left[ \varepsilon _{0},\mathcal{\hat{A}}^{R_{l}}\right] \mathcal{\hat{A}}%
^{K_{l}}-\left[ \varepsilon _{0},\mathcal{\hat{A}}^{K_{l}}\right] \mathcal{%
\hat{A}}^{R_{l}}\right\} $ and computing the commutators that appear when
letting the magnetic field half on the right and half of the left of the
relevant expressions. Note at this point that, doing so, some contributions
involving $\nabla _{R_{k}}A_{l}+\nabla _{R_{l}}A_{k}$ appear that cancel the
corresponding contribution arising from the second order in $\hbar $
development of $\varepsilon _{0N}\left( \mathbf{x}\right) $ as announced
before.

More precisely, the derivation of $\mathbf{\mu }\left( \mathbf{x}\right) $
and $\mathbf{\hat{\mu}}^{k}\left( \mathbf{x}\right) $ is as follows :
starting with the term $\frac{i}{2}\mathcal{P}_{N}\left\{ \left[ \varepsilon
_{0}\left( \mathbf{x}\right) ,\mathcal{\hat{A}}^{R_{l}}\right] \mathcal{\hat{%
A}}^{P_{l}}-\left[ \varepsilon _{0}\left( \mathbf{x}\right) ,\mathcal{\hat{A}%
}^{P_{l}}\right] \mathcal{\hat{A}}^{R_{l}}\right\} $ we have to put the
magnetic field half to the left and half to the right in the following way
\begin{eqnarray*}
&&\frac{i}{4}e\mathcal{P}_{N}\left\{ \left[ \varepsilon _{0}\left( \mathbf{x}%
\right) ,\mathcal{\hat{A}}^{R_{l}}\right] \mathcal{\hat{A}}^{K_{l}}-\left[
\varepsilon _{0}\left( \mathbf{x}\right) ,\mathcal{\hat{A}}^{K_{l}}\right]
\mathcal{\hat{A}}^{R_{l}}\right\} +H.C. \\
&=&\frac{i}{8}e\mathcal{P}_{N}\left\{ \left[ \varepsilon _{0}\left( \mathbf{x%
}\right) ,\mathcal{\hat{A}}^{R_{l}}\right] \left( \mathcal{\hat{A}}%
^{R_{n}}\nabla _{l}A_{n}+\nabla _{l}A_{n}\mathcal{\hat{A}}^{R_{n}}\right) -%
\left[ \varepsilon _{0}\left( \mathbf{x}\right) ,\left( \mathcal{\hat{A}}%
^{R_{n}}\nabla _{l}A_{n}+\nabla _{l}A_{n}\mathcal{\hat{A}}^{R_{n}}\right) %
\right] \mathcal{\hat{A}}^{R_{l}}\right\} +H.C.
\end{eqnarray*}
Letting the gradient of the potential to the left or to the right yields the
required contribution for $\mathbf{\mu }\left( \mathbf{x}\right) $ and $%
\mathbf{\hat{\mu}}^{k}\left( \mathbf{x}\right) $ is obtained by computing
the commutators involving the gauge field. These terms are (we skip $%
\mathcal{P}_{N}$ for convenience and introduced a $\frac{1}{2}$ to take into
account the integral over $\alpha $) :
\begin{eqnarray*}
&&\frac{i}{16}e\left( \left[ \varepsilon _{0}\left( \mathbf{x}\right) ,%
\mathcal{\hat{A}}^{R_{l}}\right] \left[ \nabla _{l}A_{n}-\nabla _{n}A_{l},%
\mathcal{\hat{A}}^{R_{n}}\right] \right) -\frac{ie}{16}\left[ \nabla
_{l}A_{n},\left[ \varepsilon _{0}\left( \mathbf{x}\right) ,\mathcal{\hat{A}}%
^{R_{n}}\right] \mathcal{\hat{A}}^{R_{l}}\right] \\
&&-\frac{i}{16}e\left[ \left( \mathcal{\hat{A}}^{R_{n}}\left[ \varepsilon
_{0}\left( \mathbf{x}\right) ,\nabla _{l}A_{n}\right] \mathcal{\hat{A}}%
^{R_{l}}+\left[ \varepsilon _{0}\left( \mathbf{x}\right) ,\nabla _{l}A_{n}%
\right] \mathcal{\hat{A}}^{R_{n}}\mathcal{\hat{A}}^{R_{l}}\right) \right]
+H.C. \\
&=&-\frac{\hbar }{16}e\varepsilon ^{nlk}\left( \left[ \varepsilon _{0}\left(
\mathbf{x}\right) ,\mathcal{\hat{A}}^{R_{l}}\right] \nabla _{\Pi _{p}}%
\mathcal{\hat{A}}^{R_{n}}\nabla _{R_{p}}B_{k}\right) -\frac{ie}{16}\left[
\nabla _{l}A_{n},\left[ \varepsilon _{0}\left( \mathbf{x}\right) ,\mathcal{%
\hat{A}}^{R_{n}}\right] \mathcal{\hat{A}}^{R_{l}}\right] \\
&&-\frac{i}{16}e\left( \left[ \mathcal{\hat{A}}^{R_{n}},\left[ \varepsilon
_{0}\left( \mathbf{x}\right) ,\nabla _{l}A_{n}\right] \mathcal{\hat{A}}%
^{R_{l}}\right] +\left[ \varepsilon _{0}\left( \mathbf{x}\right) ,\nabla
_{l}A_{n}\right] \left( \mathcal{\hat{A}}^{R_{n}}\mathcal{\hat{A}}^{R_{l}}+%
\mathcal{\hat{A}}^{R_{l}}\mathcal{\hat{A}}^{R_{n}}\right) \right) +H.C. \\
&=&-\frac{\hbar }{16}e\varepsilon ^{nlk}\left( \left[ \varepsilon _{0}\left(
\mathbf{x}\right) ,\mathcal{\hat{A}}^{R_{l}}\right] \nabla _{\Pi _{p}}%
\mathcal{\hat{A}}^{R_{n}}\nabla _{R_{p}}B_{k}\right) -\frac{i}{16}\left( e%
\left[ \nabla _{l}A_{n},\left[ \varepsilon _{0}\left( \mathbf{x}\right) ,%
\mathcal{\hat{A}}^{R_{n}}\right] \mathcal{\hat{A}}^{R_{l}}\right] \right) \\
&&-\frac{i}{16}e\left( \left[ \mathcal{\hat{A}}^{R_{n}},\left[ \varepsilon
_{0}\left( \mathbf{x}\right) ,\nabla _{l}A_{n}\right] \right] \mathcal{\hat{A%
}}^{R_{l}}+\left[ \varepsilon _{0}\left( \mathbf{x}\right) ,\nabla _{l}A_{n}%
\right] \left[ \mathcal{\hat{A}}^{R_{n}},\mathcal{\hat{A}}^{R_{l}}\right]
\right. \\
&&\left. +\left[ \varepsilon _{0}\left( \mathbf{x}\right) ,\nabla _{l}A_{n}%
\right] \left( \mathcal{\hat{A}}^{R_{n}}\mathcal{\hat{A}}^{R_{l}}+\mathcal{%
\hat{A}}^{R_{l}}\mathcal{\hat{A}}^{R_{n}}\right) \right) +H.C.
\end{eqnarray*}
the projection of the last term
\begin{eqnarray*}
&&\frac{-i}{16}e\left[ \varepsilon _{0}\left( \mathbf{x}\right) ,\nabla
_{l}A_{n}\right] \mathcal{P}_{N}\left( \mathcal{\hat{A}}^{R_{n}}\mathcal{%
\hat{A}}^{R_{l}}+\mathcal{\hat{A}}^{R_{l}}\mathcal{\hat{A}}^{R_{n}}\right)
+H.C. \\
&=&-\frac{1}{8}e\nabla _{l}\nabla _{m}A_{n}\mathcal{P}_{N}\left( \mathcal{%
\hat{A}}^{R_{n}}\mathcal{\hat{A}}^{R_{l}}+\mathcal{\hat{A}}^{R_{l}}\mathcal{%
\hat{A}}^{R_{n}}\right) \nabla _{\Pi _{m}}\varepsilon _{0N}\left( \mathbf{x}%
\right)
\end{eqnarray*}
gives a contribution that cancel the one we have neglected in the momentum
term as announced before. Moreover
\begin{equation*}
-\frac{i}{16}e\left[ \varepsilon _{0}\left( \mathbf{x}\right) ,\nabla
_{l}A_{n}\right] \left[ \mathcal{\hat{A}}^{R_{n}},\mathcal{\hat{A}}^{R_{l}}%
\right] +H.C.=-\left[ \frac{i}{16}e\left[ \varepsilon _{0}\left( \mathbf{x}%
\right) ,\nabla _{l}A_{n}\right] ,\left[ \mathcal{\hat{A}}^{R_{n}},\mathcal{%
\hat{A}}^{R_{l}}\right] \right]
\end{equation*}
and this term does not contribute to the diagonal part at our order of
approximation.

The last term we are left to compute is then :
\begin{eqnarray*}
&&-\frac{\hbar }{16}e\varepsilon ^{nlk}\left( \left[ \varepsilon _{0}\left(
\mathbf{x}\right) ,\mathcal{\hat{A}}^{R_{l}}\right] \nabla _{\Pi _{p}}%
\mathcal{\hat{A}}^{R_{n}}\nabla _{R_{p}}B_{k}\right) -\left( \frac{i}{16}e%
\left[ \nabla _{l}A_{n},\left[ \varepsilon _{0}\left( \mathbf{x}\right) ,%
\mathcal{\hat{A}}^{R_{n}}\right] \mathcal{\hat{A}}^{R_{l}}\right]
+H.C.\right) \\
&&-\left( \frac{i}{16}e\left[ \mathcal{\hat{A}}^{R_{n}},\left[ \varepsilon
_{0}\left( \mathbf{x}\right) ,\nabla _{l}A_{n}\right] \right] \mathcal{\hat{A%
}}^{R_{l}}+H.C.\right) \\
&=&-\frac{\hbar }{16}e\varepsilon ^{nlk}\left( \left[ \left[ \varepsilon
_{0}\left( \mathbf{x}\right) ,\mathcal{\hat{A}}^{R_{l}}\right] ,\nabla _{\Pi
_{p}}\mathcal{\hat{A}}^{R_{n}}\right] \nabla _{R_{p}}B_{k}\right) -\frac{i}{%
16}e\left[ \nabla _{l}A_{n},\left[ \left[ \varepsilon _{0}\left( \mathbf{x}%
\right) ,\mathcal{\hat{A}}^{R_{n}}\right] ,\mathcal{\hat{A}}^{R_{l}}\right] %
\right] \\
&&-\frac{i}{16}e\left[ \left[ \mathcal{\hat{A}}^{R_{n}},\left[ \varepsilon
_{0}\left( \mathbf{x}\right) ,\nabla _{l}A_{n}\right] \right] ,\mathcal{\hat{%
A}}^{R_{l}}\right] \\
&=&-\frac{\hbar }{16}e\varepsilon ^{nlk}\left( \left[ \varepsilon _{0}\left(
\mathbf{x}\right) ,\mathcal{\hat{A}}^{R_{l}}\right] \nabla _{\Pi _{p}}%
\mathcal{\hat{A}}^{R_{n}}\nabla _{R_{p}}B_{k}\right) +\frac{\hbar }{16}%
e\nabla _{p}\nabla _{l}A_{n},\nabla _{\Pi _{p}}\left[ \left[ \varepsilon
_{0}\left( \mathbf{x}\right) ,\mathcal{\hat{A}}^{R_{n}}\right] ,\mathcal{%
\hat{A}}^{R_{l}}\right] \\
&&-\frac{i}{16}e\left[ \mathcal{\hat{A}}^{R_{n}},\left[ \varepsilon
_{0}\left( \mathbf{x}\right) ,\nabla _{l}A_{n}\right] ,\mathcal{\hat{A}}%
^{R_{l}}\right] \\
&=&-\frac{\hbar }{16}e\varepsilon ^{nlk}\left( \left[ \varepsilon _{0}\left(
\mathbf{x}\right) ,\mathcal{\hat{A}}^{R_{l}}\right] \nabla _{\Pi _{p}}%
\mathcal{\hat{A}}^{R_{n}}\nabla _{R_{p}}B_{k}\right) \\
&&+\frac{\hbar }{32}e\nabla _{p}\nabla _{l}A_{n},\nabla _{\Pi _{p}}\left( %
\left[ \left[ \varepsilon _{0}\left( \mathbf{x}\right) ,\mathcal{\hat{A}}%
^{R_{n}}\right] ,\mathcal{\hat{A}}^{R_{l}}\right] +\left[ \left[ \varepsilon
_{0}\left( \mathbf{x}\right) ,\mathcal{\hat{A}}^{R_{l}}\right] ,\mathcal{%
\hat{A}}^{R_{n}}\right] +\left[ \left[ \mathcal{\hat{A}}^{R_{l}},\mathcal{%
\hat{A}}^{R_{n}}\right] ,\varepsilon _{0}\left( \mathbf{x}\right) \right]
\right) \\
&&-\frac{i}{32}e\left[ \left[ \mathcal{\hat{A}}^{R_{n}},\left[ \varepsilon
_{0}\left( \mathbf{x}\right) ,\nabla _{l}A_{n}\right] \right] ,\mathcal{\hat{%
A}}^{R_{l}}\right] +\left[ \left[ \mathcal{\hat{A}}^{R_{l}},\left[
\varepsilon _{0}\left( \mathbf{x}\right) ,\nabla _{l}A_{n}\right] \right] ,%
\mathcal{\hat{A}}^{R_{n}}\right] \\
&&+\left[ \left[ \mathcal{\hat{A}}^{R_{n}},\mathcal{\hat{A}}^{R_{l}}\right] ,%
\left[ \varepsilon _{0}\left( \mathbf{x}\right) ,\nabla _{l}A_{n}\right] %
\right]
\end{eqnarray*}
where we used twice the Jacobi in the last identity to rewrite $\frac{\hbar
}{16}e\nabla _{p}\nabla _{l}A_{n},\nabla _{\Pi _{p}}\left[ \left[
\varepsilon _{0}\left( \mathbf{x}\right) ,\mathcal{\hat{A}}^{R_{l}}\right] ,%
\mathcal{\hat{A}}^{R_{n}}\right] $ and $-\frac{i}{16}e\left[ \mathcal{\hat{A}%
}^{R_{n}},\left[ \varepsilon _{0}\left( \mathbf{x}\right) ,\nabla _{l}A_{n}%
\right] ,\mathcal{\hat{A}}^{R_{l}}\right] $. While isolating the
contribution in the magnetic field, that is proportional to $\nabla
_{R_{k}}A_{l}-\nabla _{R_{l}}A_{k}$ and introducing $\mathcal{P}_{N}$ only
the first term in the last identity remains, which yields the first term in
the magnetization $\hat{\mu}_{l}^{k}$. The contribution $\nabla
_{R_{k}}A_{l}+\nabla _{R_{l}}A_{k}$ can be disgarded as discussed in the
next subsection.

The two last contributions for $\hat{\mu}_{l}^{k}$ arise from the term $-%
\frac{\hbar ^{2}}{4}\left\{ \mathcal{P}_{-}\mathcal{A}_{0}^{\mathbf{R}%
}\times \mathcal{P}_{+}\mathcal{A}_{0}^{\mathbf{R}}\right\} _{MN}.\nabla
_{R_{m}}\mathbf{B}$ derived before in the expansion of $\mathcal{\hat{A}}%
^{K_{l}}$. Inserted in $\frac{i}{2}e\mathcal{P}_{N}\left\{ \left[
\varepsilon _{0}\left( \mathbf{x}\right) ,\mathcal{\hat{A}}^{R_{l}}\right]
\mathcal{\hat{A}}^{K_{l}}-\left[ \varepsilon _{0}\left( \mathbf{x}\right) ,%
\mathcal{\hat{A}}^{K_{l}}\right] \mathcal{\hat{A}}^{R_{l}}\right\} $ yields
directly the terms announced.

Now, concerning the Darwin contribution, it is given by :
\begin{eqnarray*}
\text{Darwin term} &=&\frac{\hbar ^{2}}{4}\left[ \sum_{M}\left( \mathcal{A}%
_{0}^{R_{l}}\right) _{NM}\left( \mathcal{A}_{0}^{R_{m}}\right) _{MN}\right]
\nabla _{R_{m}}\nabla _{R_{l}}V\left( R\right) \\
&&+\frac{\hbar ^{2}}{4}\left[ \varepsilon _{0}\left( \mathbf{x}\right) ,%
\mathcal{A}_{0}^{R_{m}}\right] _{NM}\frac{\left( \mathcal{A}%
_{0}^{R_{l}}\right) _{MN}\nabla _{R_{m}}\nabla _{R_{l}}V\left( R\right) }{%
\hat{\varepsilon}_{0M}-\hat{\varepsilon}_{0N}}
\end{eqnarray*}
Its derivation is straightforward and corresponds to isolate in $\frac{i}{2}e%
\mathcal{P}_{N}\left\{ \left[ \varepsilon _{0}\left( \mathbf{x}\right) ,%
\mathcal{\hat{A}}^{R_{l}}\right] \mathcal{\hat{A}}^{K_{l}}-\left[
\varepsilon _{0}\left( \mathbf{x}\right) ,\mathcal{\hat{A}}^{K_{l}}\right]
\mathcal{\hat{A}}^{R_{l}}\right\} $ the contribution of $\mathcal{\hat{A}}%
^{K_{l}}$ involving the potential term (see above). At our order of
approximation, the $\mathcal{\hat{A}}^{\mathbf{X}}$ terms reduce to $%
\mathcal{A}_{0}^{\mathbf{X}}$.

The last important magnetization contribution in the Hamiltonian is due to
the squared action of the magnetization operator $M$. This is of second
order in field and can be written :
\begin{eqnarray*}
M\left[ M\left[ \varepsilon _{0}\right] \right] &=&\left( \frac{i\hbar }{4}%
\right) ^{2}\mathcal{P}_{N}\left\{ \left[ \mathcal{P}_{+}\left( \left[ \hat{%
\varepsilon}_{0},\mathcal{\hat{A}}_{0}^{R_{l}}\right] \mathcal{\hat{A}}%
_{0}^{P_{l}}-\left[ \hat{\varepsilon}_{0},\mathcal{\hat{A}}_{0}^{R_{l}}%
\right] \mathcal{\hat{A}}_{0}^{P_{l}}\right) ,\mathcal{\hat{A}}_{0}^{R_{m}}%
\right] \mathcal{\hat{A}}_{0}^{P_{m}}\right. \\
&&\left. -\left[ \mathcal{P}_{+}\left( \left[ \hat{\varepsilon}_{0},\mathcal{%
\hat{A}}_{0}^{R_{l}}\right] \mathcal{\hat{A}}_{0}^{P_{l}}-\left[ \hat{%
\varepsilon}_{0},\mathcal{\hat{A}}_{0}^{R_{l}}\right] \mathcal{\hat{A}}%
_{0}^{P_{m}}\right) ,\mathcal{\hat{A}}_{0}^{P_{m}}\right] \mathcal{\hat{A}}%
_{0}^{R_{m}}\right\} +H.C. \\
&=&\left( \frac{i\hbar }{2}\right) ^{2}\mathcal{P}_{N}\left\{ \left[
\mathcal{P}_{+}\left( \left[ \hat{\varepsilon}_{0},\mathcal{P}_{-}\mathcal{%
\hat{A}}_{0}^{R_{k}}\right] \mathcal{P}_{-}\mathcal{\hat{A}}%
_{0}^{R_{l}}\right) ,\mathcal{P}_{-}\mathcal{\hat{A}}_{0}^{R_{m}}\right]
\mathcal{P}_{-}\mathcal{\hat{A}}_{0}^{R_{n}}\varepsilon
_{klp}B_{p}\varepsilon _{mnq}B_{q}\right\} \\
&\equiv &\tilde{\mu}_{pq}B_{p}B_{q}
\end{eqnarray*}

\paragraph{Computation of $-\hbar \left\langle \protect\varepsilon %
_{0}\left( \mathbf{x}\right) \right\rangle $}

The form $\ $for the diagonalized Hamiltonian given in (first equat) is
taken from \cite{DIAGOEXACT}. As explained in this paper the computation of
the energy and in particular of the term $-\frac{\hbar }{2}\left\langle
\varepsilon _{0}\left( \mathbf{x}\right) \right\rangle $ has to be performed
by assuming a certain kind of symmetrization for the various expressions,
that is a way to order the powers of $\mathbf{R}$ and $\mathbf{K}$ in
expressions as $\varepsilon _{0}$. Its value is in fact the track of the
initial choice of symmetrization in the diagonalization process. Let us
remark first, that since at the lowest order (that is zeroth order in $\hbar
$), only $\hat{\varepsilon}_{0}\left( \mathbf{x}\right) $ mixes $\mathbf{R}$
and $\mathbf{K}$, the term $-\frac{\hbar }{2}\left\langle \varepsilon
_{0}\left( \mathbf{x}\right) \right\rangle $ will reduce to $-\frac{\hbar }{2%
}\left\langle \hat{\varepsilon}_{0}\left( \mathbf{x}\right) \right\rangle $.
Second, and more importantly, let us stress that the final Hamiltonian is
independent of this ordering, but this last one is necessary to give a
precise meaning to each expression. The bracket term $-\frac{\hbar }{2}%
\left\langle \hat{\varepsilon}_{0}\left( \mathbf{x}\right) \right\rangle $
has in fact to counterbalance different choices of symmetrization and
provide ultimately the same diagonalized Hamiltonian.

Concerning the choice of symmetrization between the canonical variables $%
\mathbf{R}$ and $\mathbf{K}$ a symmetric form in $\mathbf{R}$ and $\mathbf{K}
$ is in general chosen, such as putting the powers of $\mathbf{K}$ half on
the left and half to the right. An other choice is a complete symmetrization
in $\mathbf{R}$ and $\mathbf{K}$, that is permuting this to variables in all
monomials, all permutations being equally weighted. However, both this type
of symmetrization does not fit here since the relevant variable is $\mathbf{%
\Pi }\mathbf{=K}-eA\left( \mathbf{R}\right) $. We thus have rather to
consider a symmetrization in the gauge invariant variable $\mathbf{\Pi }$. \
We will compute $-\frac{\hbar }{2}\left\langle \hat{\varepsilon}_{0}\left(
\mathbf{x}\right) \right\rangle $ in two cases that might be relevant for
the applications, depending on the problem at stake.

Assume first for $\hat{\varepsilon}_{0}\left( \mathbf{x}\right) $ a
completely symmetrized form in the powers of components of $\mathbf{\Pi }$.
That is, we will consider $\hat{\varepsilon}_{0}\left( \mathbf{x}\right) $
written as a series of terms of the kind \ $\Psi ^{i_{1}i_{2}i_{3}...}\Pi
_{i_{1}}\Pi _{i_{2}}\Pi _{i_{3}}...$ symmetrized in the indices $i_{j}$.
This form corresponds in fact to the most general one, not taking into
account any spatial symmetry for the problem

The $\left\langle .\right\rangle $ operation defined in \cite{DIAGOEXACT}
has been recalled in the first section and can be computed explicitly for $%
\hat{\varepsilon}_{0}$, in the following way :

Start with $\Psi ^{i_{1}i_{2}i_{3}...i_{n}}\Pi _{i_{1}}\Pi _{i_{2}}\Pi
_{i_{3}}...\Pi _{i_{n}}$ , $\Psi ^{i_{1}i_{2}i_{3}...}$ being a completely
symmetrized tensor.

Applying the rule given in the first section, the contribution of this term
to $-\frac{\hbar }{2}\left\langle \hat{\varepsilon}_{0}\left( \mathbf{x}%
\right) \right\rangle $ is obtained by deriving (that is removing) by some
momentum components $\Pi _{i_{k}}$ $\Pi _{i_{l}}$ and inserting $\nabla
_{R_{i_{l}}}A^{i_{k}}$. One gets for this contribution :
\begin{equation*}
-ie\frac{\hbar }{4}\sum_{kl}\Psi ^{i_{1}i_{2}...i_{n}}\left[ \Pi
_{i_{1}}...\Pi _{i_{k-1}}\nabla _{R_{i_{l}}}A^{i_{k}}\Pi _{i_{k+1}}...\hat{%
\Pi}_{i_{l}}...\Pi _{i_{n}}-\Pi _{i_{1}}...\hat{\Pi}_{i_{k}}...\Pi
_{i_{l-1}}\nabla _{R_{i_{l}}}A^{i_{k}}\Pi _{i_{l+1}}...\Pi _{i_{n}}\right]
\end{equation*}
where the hat recalls that the variable is omitted from the series. Let move
the gradient of the potential half on the left and half on the right to
gather them by pair. We thus have :
\begin{eqnarray*}
&&-ie\frac{\hbar }{4}\sum_{kl}\Psi ^{i_{1}i_{2}...i_{n}}\left[ \Pi
_{i_{1}}...\Pi _{i_{k-1}}\nabla _{R_{i_{l}}}A^{i_{k}}\Pi _{i_{k+1}}...\hat{%
\Pi}_{i_{l}}...\Pi _{i_{n}}-\Pi _{i_{1}}...\hat{\Pi}_{i_{k}}...\Pi
_{i_{l-1}}\nabla _{R_{i_{l}}}A^{i_{k}}\Pi _{i_{l+1}}...\Pi _{i_{n}}\right] \\
&=&-ie\frac{\hbar }{8}\sum_{kl}\Psi ^{i_{1}i_{2}...i_{n}}\left[ \Pi
_{i_{1}}...\Pi _{i_{k-1}}\left[ \nabla _{R_{i_{l}}}A^{i_{k}}-\nabla
_{R_{i_{l}}}A^{i_{k}}\right] \Pi _{i_{k+1}}...\hat{\Pi}_{i_{l}}...\Pi
_{i_{n}}\right] \\
&&+ie\frac{\hbar }{8}\sum_{kl}\Psi ^{i_{1}i_{2}...i_{n}}\left[ \Pi
_{i_{1}}...\hat{\Pi}_{i_{k}}...\Pi _{i_{l-1}}\left[ \nabla
_{R_{i_{l}}}A^{i_{k}}-\nabla _{R_{i_{l}}}A^{i_{k}}\right] \Pi
_{i_{l+1}}...\Pi _{i_{n}}\right] \\
&&+e\frac{\hbar ^{2}}{8}\sum_{k<p<l}\nabla _{R_{i_{p}}}\left[ \nabla
_{R_{i_{l}}}A^{i_{k}}+\nabla _{R_{i_{l}}}A^{i_{k}}\right] \Pi _{i_{1}}...%
\hat{\Pi}_{i_{k}}...\hat{\Pi}_{i_{p}}...\hat{\Pi}_{i_{l}}...\Pi _{i_{n}} \\
&=&+e\frac{\hbar ^{2}}{8}\sum_{k<p<l}\nabla _{R_{i_{p}}}\left[ \nabla
_{R_{i_{l}}}A^{i_{k}}+\nabla _{R_{i_{l}}}A^{i_{k}}\right] \Pi _{i_{1}}...%
\hat{\Pi}_{i_{k}}...\hat{\Pi}_{i_{p}}...\hat{\Pi}_{i_{l}}...\Pi _{i_{n}}
\end{eqnarray*}
the last equality is obtained since $\Psi ^{i_{1}i_{2}...i_{n}}$ is
completely symmetric so that the contribution of $\nabla
_{R_{i_{l}}}A^{i_{k}}-\nabla _{R_{i_{l}}}A^{i_{k}}$ vanishes.

As a consequence one has :
\begin{equation*}
-\frac{\hbar }{2}\left\langle \hat{\varepsilon}_{0}\left( \mathbf{x}\right)
\right\rangle =\frac{ie}{8}\hbar \left( B_{k}\varepsilon ^{jik}\right)
\nabla _{\Pi _{i}}\nabla _{\Pi _{j}}\hat{\varepsilon}_{0}+\text{%
contributions proportional to }\mathbf{\nabla }\left( \nabla
_{j}A^{i}+\nabla _{i}A^{j}\right)
\end{equation*}

As a consequence, with our initial choice of symmetrization, $-\frac{\hbar }{%
2}\left\langle \hat{\varepsilon}_{0}\left( \mathbf{x}\right) \right\rangle $
contains only symmetric contributions. However, as we explained in the
previous section, since by construction ultimately $\varepsilon \left(
\mathbf{X}\right) $ has to depend only on $\mathbf{\Pi }$ and $\mathbf{B}$
(since $\mathbf{B}$ appears as the commutators of the components of $\mathbf{%
\Pi }$), this symmetric contribution has to find its counterpart in our \
previous development of the energy operator. This counterpart is
automatically the contributions we discarded before.

However, we can at this point explain a bit more how this works. Roughly, in
the diagonalization process, the Bracket of the energy enters in two ways in
our method since basically the trick to obtain the diagonalized Hamiltonian
is to add and subtract $\left\langle \hat{\varepsilon}_{0}\left( \mathbf{x}%
\right) \right\rangle $ (see \cite{DIAGOEXACT}, or more explicitly the
differential equation in \cite{SERIESPIERRE}) to two different quantities
that do not contain any symmetric term .

The first way comes from developing
\begin{eqnarray*}
&&\varepsilon _{\alpha }\left( \mathbf{X}_{\alpha +d\alpha }+d\mathbf{X}%
_{\alpha }\right) -U_{\alpha }\left( \mathbf{X}_{\alpha }\right) H_{0}\left(
\mathbf{X}_{\alpha }\right) U_{\alpha }^{+}\left( \mathbf{X}_{\alpha }\right)
\\
&=&U_{\alpha }\left( \mathbf{X}_{\alpha +d\alpha }+d\mathbf{X}_{\alpha
}\right) H_{0}\left( \mathbf{X}_{\alpha +d\alpha }+d\mathbf{X}_{\alpha
}\right) U_{\alpha }^{+}\left( \mathbf{X}_{\alpha +d\alpha }+d\mathbf{X}%
_{\alpha }\right) -U_{\alpha }\left( \mathbf{X}_{\alpha }\right) H_{0}\left(
\mathbf{X}_{\alpha }\right) U_{\alpha }^{+}\left( \mathbf{X}_{\alpha }\right)
\end{eqnarray*}
This term, yields all the relevant contributions to the diagonalized
Hamiltonian we derived before. However, due to our initial choice of
canonical variables $\mathbf{R}$, $\mathbf{K}$, the series expansion of $%
\left\langle UH_{0}U^{+}\right\rangle $ led us also to some contributions
proportional to $\mathbf{\nabla }\left( \nabla _{j}A^{i}+\nabla
_{i}A^{j}\right) $. They arise because the Bracket operation isolates non
physical terms such $\nabla _{j}A^{i}$. When we recombined them to get
magnetic field contributions the $\mathbf{\nabla }\left( \nabla
_{j}A^{i}+\nabla _{i}A^{j}\right) $ appeared.

The second way comes from developing
\begin{equation*}
-\left[ \varepsilon _{\alpha }\left( \mathbf{X}_{\alpha +d\alpha }+d\mathbf{X%
}_{\alpha }\right) -\varepsilon _{\alpha }\left( \mathbf{X}_{\alpha +d\alpha
}\right) \right]
\end{equation*}
that will yield the contribution $-\left\langle \hat{\varepsilon}_{0}\left(
\mathbf{x}\right) \right\rangle $ including its symmetric contributions. But
since $U_{\alpha }\left( \mathbf{X}_{\alpha }\right) H_{0}\left( \mathbf{X}%
_{\alpha }\right) U_{\alpha }^{+}\left( \mathbf{X}_{\alpha }\right) $ and $%
\varepsilon _{\alpha }\left( \mathbf{X}_{\alpha +d\alpha }\right) $ can be
written in terms including only physical variables $\mathbf{R}$ and $\mathbf{%
\Pi }$ (both the energy operator and diagonalization process are assumed to
depend on these variables), the symmetric contributions of these two terms
come only from $U_{\alpha }\left( \mathbf{X}_{\alpha +d\alpha }+d\mathbf{X}%
_{\alpha }\right) H_{0}\left( \mathbf{X}_{\alpha +d\alpha }+d\mathbf{X}%
_{\alpha }\right) U_{\alpha }^{+}\left( \mathbf{X}_{\alpha +d\alpha }+d%
\mathbf{X}_{\alpha }\right) -\varepsilon _{\alpha }\left( \mathbf{X}_{\alpha
+d\alpha }+d\mathbf{X}_{\alpha }\right) $, and as such cancel by
construction.

As a check, one can see that, typically, the contribution arising from $%
\left\langle UH_{0}U^{+}\right\rangle $ were proportional to $\nabla
_{i}\nabla _{l}\nabla _{j}U^{+}\nabla _{l}\left( \nabla _{j}A^{i}+\nabla
_{i}A^{j}\right) \hat{\varepsilon}_{0}\left( \mathbf{x}\right) \propto
\mathcal{A}_{0}^{R_{i}}\nabla _{l}\mathcal{A}_{0}^{R_{j}}\nabla _{l}\left(
\nabla _{j}A^{i}+\nabla _{i}A^{j}\right) $ which is exactly the kind of
symmetric terms we were left with while computing the magnetization term.
But precisely these contributions are by construction equal to those
produced by $\frac{\hbar }{2}\left\langle \hat{\varepsilon}_{0}\left(
\mathbf{x}\right) \right\rangle $ (since $UH_{0}U^{+}=\varepsilon _{0}\left(
\mathbf{x}\right) $). As a consequence, $-\frac{\hbar }{2}\left\langle \hat{%
\varepsilon}_{0}\left( \mathbf{x}\right) \right\rangle $ acts in fact\ as a
compensation term, and the overall terms in $\mathbf{\nabla }\left( \nabla
_{j}A^{i}+\nabla _{i}A^{j}\right) $ is constrained to cancel whatever the
form of the Hamiltonian or the Berry phases.

As a consequence of this discussion, we can neglect the symmetric term and
keep definitely for the choice of full symmetrization on the momentum
components :
\begin{equation*}
-\frac{\hbar }{2}\left\langle \hat{\varepsilon}_{0}\left( \mathbf{x}\right)
\right\rangle =0
\end{equation*}
However, the choice of symmetrizing all powers of the momentum $\mathbf{\Pi }
$ does not fit when the system presents some symmetries. The Basic example
is the Dirac Hamiltonian satisfying rotational invariance, but one can also
consider fermi surfaces having some ellipsoidal form. In that case a natural
choice of symmetrization can be done with respect to some or several
quadratic forms in the momentum of the kind $\Pi _{i}A^{ij}\Pi _{j}$
(typically for the Dirac case $A^{ij}=\delta ^{ij}$). To inspect this case,
we will thus compute $-\frac{\hbar }{2}\left\langle \hat{\varepsilon}%
_{0}\left( \mathbf{x}\right) \right\rangle $ \ when $\hat{\varepsilon}%
_{0}\left( \mathbf{x}\right) $ is written as a series expansion of products
and powers of such quadratic forms $\Pi _{i}A^{ij}\Pi _{j}$. To do so, we
develop $\hat{\varepsilon}_{0}\left( \mathbf{x}\right) $ as a symmetrized
series of monomial terms of the kind $C\left( X\right) ^{i}\Pi _{i}\left(
Q\right) \Pi _{j}\left( Y\right) ^{j}D$. The $\left( X\right) ^{i}$\ and $%
\left( Y\right) ^{j}$\ being shortcut for $\Pi _{j}A^{ij}$ and $\Pi
_{i}B^{ij}$ with $A^{ij}$ and $B^{ij}$ some quadratic forms arising in the
expansion. $C$, $D$, $\left( Q\right) $ are arbitrary. Note at this point
that implicitly, $\left( X\right) ^{i}\Pi _{i}\left( Q\right) \Pi _{j}\left(
Y\right) ^{j}$ has to be considered as part of a sum of four symmetric terms
obtained by permuting $\left( X\right) ^{i}$ and $\Pi _{i}$, $\Pi _{j}$ and $%
\left( Y\right) ^{j}$, that is $\frac{1}{4}\left[ \left( X\right) ^{i}\Pi
_{i}\left( Q\right) \Pi _{j}\left( Y\right) ^{j}+\Pi _{i}\left( X\right)
^{i}\left( Q\right) \Pi _{j}\left( Y\right) ^{j}+\left( X\right) ^{i}\Pi
_{i}\left( Q\right) \left( Y\right) ^{j}\Pi _{j}+\Pi _{i}\left( X\right)
^{i}\left( Q\right) \left( Y\right) ^{j}\Pi _{j}\right] $. For the sake of
simplicity we only keep $\left( X\right) ^{i}\Pi _{i}\left( Q\right) \Pi
_{j}\left( Y\right) ^{j}$, the symmetrization being implicit and we skip the
terms $C$ and $D$ that play no role in the sequel. $\left( X\right) ^{i}\Pi
_{i}\left( Q\right) \Pi _{j}\left( Y\right) ^{j}$ contributes to $-\frac{%
\hbar }{2}\left\langle \hat{\varepsilon}_{0}\left( \mathbf{x}\right)
\right\rangle $ by deriving (that is removing) $\Pi _{i}$ $\Pi _{j}$ and
inserting $\nabla _{j}A^{i}$ (the sums over the indices $i$,$j$ is
understood):
\begin{eqnarray*}
&&\frac{i}{4}e\hbar \left( \left( X\right) ^{i}\nabla _{j}A^{i}\left(
Q\right) \left( Y\right) ^{j}-\left( X\right) ^{j}\left( Q\right) \nabla
_{j}A^{i}\left( Y\right) ^{i}\right) \\
&=&\frac{i}{4}e\hbar \left( \left( X\right) ^{i}\nabla _{j}A^{i}\left(
Q\right) \left( Y\right) ^{j}-\left( X\right) ^{i}\left( Q\right) \nabla
_{i}A^{j}\left( Y\right) ^{j}\right)
\end{eqnarray*}
We can gather the potential terms by putting them half to the left of $Q$
and half to the right of $Q$. We can thus write this term at order $\hbar
^{2}$ (which is enough for us here given our order of approximation)
\begin{eqnarray*}
&&\frac{i}{4}e\hbar \left( \left( X\right) ^{i}\nabla _{j}A^{i}\left(
Q\right) \left( Y\right) ^{j}-\left( X\right) ^{i}\left( Q\right) \nabla
_{i}A^{j}\left( Y\right) ^{j}\right) \\
&=&\frac{ie}{8}\hbar \left( X\right) ^{i}\left( Q\right) \varepsilon
^{jik}B_{k}\left( Y\right) ^{j}+\frac{ie}{8}\hbar \left( X\right)
^{i}\varepsilon ^{jik}B_{k}\left( Q\right) \left( Y\right) ^{j} \\
&&-\frac{1}{8}e\hbar ^{2}\nabla _{l}\left( \nabla _{j}A^{i}+\nabla
_{i}A^{j}\right) \left( X\right) ^{i}\nabla _{l}\left( Q\right) \left(
Y\right) ^{j}+\frac{1}{4}e\hbar ^{2}\left( X\right) ^{i}\nabla _{l}\left(
Q\right) \left( Y\right) ^{j}\nabla _{l}\left( \nabla _{j}A^{i}+\nabla
_{i}A^{j}\right)
\end{eqnarray*}
To the order $\hbar ^{2}$, the terms $\frac{ie}{8}\hbar \left( X\right)
^{i}\left( Q\right) \varepsilon ^{jik}B_{k}\left( Y\right) ^{j}+\frac{ie}{8}%
\hbar \left( X\right) ^{i}\varepsilon ^{jik}B_{k}\left( Q\right) \left(
Y\right) ^{j}$, once summed over the symmetric monomials, involve all the
possible permutations of $\varepsilon ^{jik}B_{k}$ inside the series
expansion of the second derivative of $\hat{\varepsilon}_{0}$. As a
consequence, we can write :
\begin{equation*}
-\frac{\hbar }{2}\left\langle \hat{\varepsilon}_{0}\left( \mathbf{x}\right)
\right\rangle =\frac{ie}{8}\hbar \left( B_{k}\varepsilon ^{jik}\right)
\nabla _{\Pi _{i}}\nabla _{\Pi _{j}}\hat{\varepsilon}_{0}+\text{%
contributions proportional to }\mathbf{\nabla }\left( \nabla
_{j}A^{i}+\nabla _{i}A^{j}\right)
\end{equation*}
As explained just before, it is understood that $B_{k}$ is inserted in a
completely symmetric way inside the series expansion of $\nabla _{\Pi
_{i}}\nabla _{\Pi _{j}}\hat{\varepsilon}_{0}$ (that is cyclically permuted
inside the series).

Moreover, we can again discard the symmetric terms $\mathbf{\nabla }\left(
\nabla _{j}A^{i}+\nabla _{i}A^{j}\right) $.

We thus have with this choice of symmetrization :
\begin{equation*}
-\frac{\hbar }{2}\left\langle \hat{\varepsilon}_{0}\left( \mathbf{x}\right)
\right\rangle =\frac{ie}{8}\hbar \left( B_{k}\varepsilon ^{jik}\right)
\nabla _{\Pi _{i}}\nabla _{\Pi _{j}}\hat{\varepsilon}_{0}
\end{equation*}

We will conclude this section by obtaining a developer form for $\frac{ie}{8}%
\hbar \left( B_{k}\varepsilon ^{jik}\right) \nabla _{\Pi _{i}}\nabla _{\Pi
_{j}}\hat{\varepsilon}_{0}$. It is performed by going back to its series
expansion. Let us assume in a first time that $\hat{\varepsilon}_{0}$ is a
function of $\mathbf{\Pi }^{2}$ only (rotational invariance), so that we can
choose a natural symmetrization to express $\hat{\varepsilon}_{0}\left(
\mathbf{x}\right) $ as a power series of $\mathbf{\Pi }^{2}$.

The sum of monomials of the kind
\begin{equation*}
\frac{ie}{8}\hbar C\left( X\right) ^{i}\left( Q\right) \varepsilon
^{jik}B_{k}\left( Y\right) ^{j}D+\frac{ie}{8}\hbar C\left( X\right)
^{i}\varepsilon ^{jik}B_{k}\left( Q\right) \left( Y\right) ^{j}D
\end{equation*}
can be written (we now reintroduce our implicit permutations that let $B_{k}$
to be half on the left and half on the right of $\left( X\right) ^{i}$ and $%
\left( Y\right) ^{j}$) :
\begin{eqnarray*}
&&\frac{ie}{16}\hbar C\Pi _{i}\varepsilon ^{jik}B_{k}\left( Q\right) \Pi
_{j}D+\frac{ie}{16}\hbar C\Pi _{i}\left( Q\right) \varepsilon ^{jik}B_{k}\Pi
_{j}D \\
&&+\frac{ie}{16}\varepsilon ^{jik}B_{k}\hbar C\Pi _{i}\left( Q\right) \Pi
_{j}D+\frac{ie}{16}\hbar C\Pi _{i}\left( Q\right) \Pi _{j}\varepsilon
^{jik}B_{k}D
\end{eqnarray*}
Letting move the magnetic field half to the left and half to the right
yields to the order $\hbar ^{2}$ yields :
\begin{eqnarray*}
&&\frac{ie}{16}\hbar C\Pi _{i}\varepsilon ^{jik}B_{k}\left( Q\right) \Pi
_{j}D+\frac{ie}{16}\hbar C\Pi _{i}\left( Q\right) \varepsilon ^{jik}B_{k}\Pi
_{j}D \\
&&+\frac{ie}{16}\varepsilon ^{jik}B_{k}\hbar C\Pi _{i}\left( Q\right) \Pi
_{j}D+\frac{ie}{16}\hbar C\Pi _{i}\left( Q\right) \Pi _{j}\varepsilon
^{jik}B_{k}D \\
&=&\varepsilon ^{jik}B_{k}\frac{ie}{8}\hbar \left[ C\Pi _{i}\left( Q\right)
\Pi _{j}D+C\Pi _{i}\left( Q\right) \Pi _{j}D\right] +\frac{ie}{8}\hbar \left[
C\Pi _{i}\left( Q\right) \Pi _{j}D+C\Pi _{i}\left( Q\right) \Pi _{j}D\right]
\varepsilon ^{jik}B_{k} \\
&&+\left( \mathbf{\nabla }\times \mathbf{B}\right) .\mathbf{\Pi }\frac{e}{8}%
\hbar ^{2}C\left( Q\right) D
\end{eqnarray*}
Concentrate on the first term. One has
\begin{eqnarray*}
&&\varepsilon ^{jik}B_{k}\frac{ie}{8}\hbar \left[ C\Pi _{i}\left( Q\right)
\Pi _{j}D+C\Pi _{i}\left( Q\right) \Pi _{j}D\right] +\frac{ie}{8}\hbar \left[
C\Pi _{i}\left( Q\right) \Pi _{j}D+C\Pi _{i}\left( Q\right) \Pi _{j}D\right]
\varepsilon ^{jik}B_{k} \\
&=&\varepsilon ^{jik}B_{k}\frac{ie}{8}\hbar \left[ C\Pi _{i}\left( Q\right)
\Pi _{j}D-C\Pi _{j}\left( Q\right) \Pi _{i}D\right]
\end{eqnarray*}
to the order $\hbar ^{2}$. Now,
\begin{eqnarray*}
&&\varepsilon ^{jik}B_{k}\frac{ie}{8}\hbar \left[ C\Pi _{i}\left( Q\right)
\Pi _{j}D-C\Pi _{j}\left( Q\right) \Pi _{i}D\right] \\
&=&\frac{ie}{8}\hbar C\left( Q\right) \left[ \Pi _{i},\Pi _{j}\right]
D\varepsilon ^{jik}B_{k} \\
&&+\frac{ie}{8}\hbar C\Pi _{i}\left[ \left( Q\right) ,\Pi _{j}\right]
D\varepsilon ^{jik}B_{k}-\frac{ie}{8}\hbar C\Pi _{j}\left[ \left( Q\right)
,\Pi _{i}\right] D\varepsilon ^{jik}B_{k}
\end{eqnarray*}
Summing over the symmetric monomials of the series expansion of $\hat{%
\varepsilon}_{0}$ , each of this commutator will yield a corresponding
contribution to $-\frac{\hbar }{2}\left\langle \hat{\varepsilon}_{0}\left(
\mathbf{x}\right) \right\rangle $. The first commutator will yield the
contribution :
\begin{eqnarray*}
\frac{ie}{8}\hbar C\left( Q\right) \left[ \Pi _{i},\Pi _{j}\right]
D\varepsilon ^{jik}B_{k} &=&\frac{e}{8}\hbar ^{2}C\left( Q\right) D\left(
\varepsilon ^{ijk}B_{k}\right) ^{2} \\
&\rightarrow &\frac{1}{2}\frac{e^{2}}{8}\hbar ^{2}\left( \varepsilon
^{ijk}B_{k}\right) ^{2}4\left( \nabla _{\mathbf{\Pi }^{2}}\right) ^{2}\hat{%
\varepsilon}_{0} \\
&=&\frac{e^{2}}{2}\hbar ^{2}\left( B^{2}\right) \left( \nabla _{\mathbf{\Pi }%
^{2}}\right) ^{2}\hat{\varepsilon}_{0}
\end{eqnarray*}
the global $\frac{1}{2}$ amounts for the fact that given our conventions the
derivative with respect to $\Pi _{i}$ are always on the left of the
derivatives with respect to $\Pi _{j}$. This implies a $\frac{1}{2}$ factor
each $\nabla _{\Pi _{i}}\nabla _{\Pi _{j}}$ term. We have also used the fact
that given the rotational invariance $\left( \nabla _{\Pi _{i}}\right) ^{2}$
and $\left( \nabla _{\Pi _{i}}\right) ^{2}$ can be replaced by $4\left(
\nabla _{\mathbf{\Pi }^{2}}\right) ^{2}$. Actually the commutator $\left[
\Pi _{i},\Pi _{j}\right] $ amounts for taking twice the derivative with
respect to $\Pi _{i}$ and $\Pi _{j}$ (times $\varepsilon ^{ijk}B_{k}$) and
twice the derivative with respect to $\Pi _{i}$ and $\Pi _{j}$ yield each a
contribution proportional to $\nabla _{\mathbf{\Pi }^{2}}$. On the other
hand the second commutator will lead to the contribution :
\begin{eqnarray*}
+\frac{ie}{8}\hbar C\Pi _{j}\left[ \Pi _{i},\left( Q\right) \right]
D\varepsilon ^{ijk}B_{k} &\rightarrow &\frac{1}{2}\frac{e^{2}}{24}\hbar
^{2}\left( \nabla _{\Pi _{l}}^{\prime }\nabla _{\Pi _{j}}^{\prime }2\left(
\nabla _{\mathbf{\Pi }^{2}}\right) \hat{\varepsilon}_{0}\right) \varepsilon
^{jik}B_{k}\varepsilon ^{ilm}B_{m} \\
&=&\frac{e^{2}}{48}\hbar ^{2}\left( \left( B^{2}\right) 2\nabla _{\Pi
_{j}}^{\prime }\nabla _{\Pi _{j}}^{\prime }\left( \nabla _{\mathbf{\Pi }%
^{2}}\right) ^{2}\hat{\varepsilon}_{0}-B_{k}\nabla _{\Pi _{k}}^{\prime
}B_{m}\nabla _{\Pi _{m}}^{\prime }2\left( \nabla _{\mathbf{\Pi }^{2}}\right)
\hat{\varepsilon}_{0}\right)
\end{eqnarray*}
with the implicit convention, recalled by the $^{\prime }$, and implied by
construction and the definitions of $\left( Q\right) $, $\Pi _{i}$, $\Pi
_{j} $), that a factor $\mathbf{\Pi }^{2}$ that has been derived by $\nabla
_{\Pi _{m}}$ will not be derived again by $\nabla _{\Pi _{k}}$ (the same
convention applying of course for $\nabla _{\Pi _{l}}^{\prime }\nabla _{\Pi
_{j}}^{\prime }$). Similarly $\nabla _{\Pi _{j}}^{\prime }\nabla _{\Pi
_{j}}^{\prime }$ means that a power of $\mathbf{\Pi }^{2}$ is not derived
twice by $\Pi _{j}$.

The reason for the global $\frac{1}{2}$ factor is the same as before, the
derivatives with respect to $\Pi _{i}$ being on the left of the derivatives
with respect to $\Pi _{j}$. A $\frac{1}{3}$ factor has also arisen from the
fact that the derivative with respect to $\Pi _{l}$ has to be taken between
the derivatives with respect to $\Pi _{i}$ and $\Pi _{j}$. Due to the
symmetrization of the variables, it amounts for only one third of the
derivative with respect to $\Pi _{l}$.

Similarly, the last term will give :
\begin{eqnarray*}
-\frac{ie}{8}\hbar C\Pi _{j}\left[ \left( Q\right) ,\Pi _{i}\right]
D\varepsilon ^{jik}B_{k} &\rightarrow &\frac{e^{2}}{12}\hbar ^{2}\left(
\nabla _{\Pi _{l}}^{\prime }\nabla _{\Pi _{i}}^{\prime }\left( \nabla _{%
\mathbf{\Pi }^{2}}\right) \hat{\varepsilon}_{0}\right) \varepsilon
^{ijk}B_{k}\varepsilon ^{ljm}B_{m} \\
&=&\frac{e^{2}}{48}\hbar ^{2}\left( \left( B^{2}\right) 2\nabla _{\Pi
_{j}}^{\prime }\nabla _{\Pi _{j}}^{\prime }\left( \nabla _{\mathbf{\Pi }%
^{2}}\right) ^{2}\hat{\varepsilon}_{0}-B_{k}\nabla _{\Pi _{k}}^{\prime
}B_{m}\nabla _{\Pi _{m}}^{\prime }2\left( \nabla _{\mathbf{\Pi }^{2}}\right)
\hat{\varepsilon}_{0}\right)
\end{eqnarray*}
with the same convention as before for the derivatives.

Ultimately, we compute similarly the contribution :
\begin{eqnarray*}
\left( \mathbf{\nabla }\times \mathbf{B}\right) .\mathbf{\Pi }\frac{e}{8}%
\hbar ^{2}C\left( Q\right) D &\rightarrow &\frac{1}{2}\frac{e}{8}\hbar
^{2}\left( \mathbf{\nabla }\times \mathbf{B}\right) .\mathbf{\Pi }4\left(
\nabla _{\mathbf{\Pi }^{2}}\right) ^{2}\hat{\varepsilon}_{0} \\
&=&\frac{e}{4}\hbar ^{2}\left( \mathbf{\nabla }\times \mathbf{B}\right) .%
\mathbf{\Pi }\left( \nabla _{\mathbf{\Pi }^{2}}\right) ^{2}\hat{\varepsilon}%
_{0}
\end{eqnarray*}
Gathering all the relevant terms yields in the end :
\begin{eqnarray*}
-\frac{\hbar }{2}\left\langle \hat{\varepsilon}_{0}\left( \mathbf{x}\right)
\right\rangle &=&\frac{e^{2}}{2}\hbar \left( B^{2}\right) \left( \nabla _{%
\mathbf{\Pi }^{2}}\right) ^{2}\hat{\varepsilon}_{0}+\frac{e^{2}}{12}\hbar
^{2}\left( \left( B^{2}\right) \nabla _{\Pi _{j}}^{\prime }\nabla _{\Pi
_{j}}^{\prime }\left( \nabla _{\mathbf{\Pi }^{2}}\right) ^{2}\hat{\varepsilon%
}_{0}-B_{k}\nabla _{\Pi _{k}}^{\prime }B_{m}\nabla _{\Pi _{m}}^{\prime
}\left( \nabla _{\mathbf{\Pi }^{2}}\right) \hat{\varepsilon}_{0}\right) \\
&&+\frac{e}{4}\hbar ^{2}\left( \mathbf{\nabla }\times \mathbf{B}\right) .%
\mathbf{\Pi }\left( \nabla _{\mathbf{\Pi }^{2}}\right) ^{2}\hat{\varepsilon}%
_{0}
\end{eqnarray*}
A direct computation shows that specializing to the case of the Dirac
Hamiltonian in an electromagnetic field, will be given in \cite{PIERREDIRAC}.

The more general case can now be treated in a very similar way. Coming back
to the general form for the monomials, leads to consider again the monomials
$-\frac{ie}{8}\hbar \left( X\right) ^{i}\varepsilon ^{ijk}B_{k}\left(
Q\right) \left( Y\right) ^{j}-\frac{ie}{8}\hbar \left( X\right) ^{j}\left(
Q\right) \varepsilon ^{jik}B_{k}\left( Y\right) ^{i}$. Here to alleviate the
notation we assume again implicitly that $B_{k}$ is put\ half on the left
and half on the right of $\left( X\right) ^{i}$ and $\left( Y\right) ^{j}$).
the contribution to move them on the left of $\left( X\right) ^{i}$ and the
right of $\left( Y\right) ^{j}$ leads to a similar contribution to $-\frac{%
\hbar }{2}\left\langle \hat{\varepsilon}_{0}\left( \mathbf{x}\right)
\right\rangle $ as before : $\frac{e}{8}\hbar ^{2}\left( \mathbf{\nabla }%
\times \mathbf{B}\right) .\mathbf{\nabla }_{\mathbf{\Pi }}\left( \nabla _{%
\mathbf{\Pi }^{2}}\right) ^{2}\hat{\varepsilon}_{0}$ Concerning the other
terms, we recombine the monomial $-\frac{ie}{8}\hbar \left( X\right)
^{i}\left( Q\right) \left( Y\right) ^{j}\varepsilon ^{ijk}B_{k}$ with the
corresponding contribution $+\frac{ie}{8}\hbar \left( Y\right) ^{j}\left(
Q\right) \left( X\right) ^{i}\varepsilon ^{ijk}B_{k}$ (the $\varepsilon
^{jik}B_{k}$ being implicitly half on the left and half on the right of the
expressions):
\begin{eqnarray*}
&&-\frac{ie}{8}\hbar \left( X\right) ^{i}\left( Q\right) \left( Y\right)
^{j}\varepsilon ^{ijk}B_{k}+\frac{ie}{8}\hbar \left( Y\right) ^{j}\left(
Q\right) \left( X\right) ^{i}\varepsilon ^{ijk}B_{k} \\
&=&-\frac{ie}{8}\hbar \left[ \left( X\right) ^{i}\left( Q\right) \left(
Y\right) ^{j}-\left( Y\right) ^{j}\left( Q\right) \left( X\right) ^{i}\right]
\varepsilon ^{ijk}B_{k} \\
&=&-\frac{ie}{8}\hbar \left( \left( X\right) ^{i}\left( Y\right) ^{j}-\left(
Y\right) ^{j}\left( X\right) ^{i}\right) \left( Q\right) \varepsilon
^{ijk}B_{k} \\
&&-\frac{ie}{8}\hbar \left[ \left( X\right) ^{i},\left( Q\right) \right]
\left( Y\right) ^{j}\varepsilon ^{ijk}B_{k}+\frac{ie}{8}\hbar \left[ \left(
Y\right) ^{j},\left( Q\right) \right] \left( X\right) ^{i}\varepsilon
^{ijk}B_{k}
\end{eqnarray*}
As before, summing over the symmetric monomials of the series expansion of $%
\hat{\varepsilon}_{0}$ , each of this commutator will yield a corresponding
contribution in derivative of $\hat{\varepsilon}_{0}$. The first commutator
in $\left[ \left( X\right) ^{i}\left( Y\right) ^{j}-\left( X\right)
^{j}\left( Y\right) ^{i}\right] $ is computed in the following way. The
rotational invariance does not exist anymore now, but $\left( X\right) ^{i}$
and $\left( Y\right) ^{j}$ are still functions of $\mathbf{\Pi }$. As such
they will yield the contribution :
\begin{eqnarray*}
&&\frac{ie}{8}\hbar \left( Q\right) \left[ \left( X\right) ^{i}\left(
Y\right) ^{j}-\left( X\right) ^{j}\left( Y\right) ^{i}\right] \varepsilon
^{ijk}B_{k} \\
&\rightarrow &\frac{1}{2}\frac{e^{2}}{8}\hbar ^{2}\left( \varepsilon
^{ijk}B_{k}\right) \left( \varepsilon ^{lmn}B_{n}\right) \left( \nabla _{\Pi
_{l}}\right) \left( \nabla _{\Pi _{i}}\right) \left( \nabla _{\Pi
_{m}}\right) \left( \nabla _{\Pi _{j}}\right) \hat{\varepsilon}_{0}
\end{eqnarray*}
Once again, the global $\frac{1}{2}$ amounts for the fact that given our
conventions the derivative with respect to $\Pi _{i}$ are always on the left
of the derivatives with respect to $\Pi _{j}$.

Similarly, the second and third commutators will lead to the contribution :
\begin{eqnarray*}
&&-\frac{ie}{8}\hbar \left( X\right) ^{i}\left[ \left( Q\right) ,\left(
Y\right) ^{j}\right] \varepsilon ^{ijk}B_{k}+\frac{ie}{8}\hbar \left(
Y\right) ^{j}\left[ \left( Q\right) ,\left( X\right) ^{i}\right] \varepsilon
^{ijk}B_{k} \\
&\rightarrow &\frac{e^{2}}{48}\hbar ^{2}\left( \nabla _{\Pi _{l}}^{\prime
}\nabla _{\Pi _{i}}^{\prime }\left[ \left( \nabla _{\Pi _{n}}\right) .\left(
\nabla _{\Pi _{j}}\right) \right] \hat{\varepsilon}_{0}\right) \varepsilon
^{ijk}B_{k}\varepsilon ^{lnm}B_{m} \\
&&-\frac{e^{2}}{48}\hbar ^{2}\left( \nabla _{\Pi _{l}}^{\prime }\nabla _{\Pi
_{j}}^{\prime }\left[ \left( \nabla _{\Pi _{n}}\right) .\left( \nabla _{\Pi
_{i}}\right) \right] \hat{\varepsilon}_{0}\right) \varepsilon
^{ijk}B_{k}\varepsilon ^{lnm}B_{m} \\
&=&\frac{e^{2}}{24}\hbar ^{2}\left( \nabla _{\Pi _{l}}^{\prime }\nabla _{\Pi
_{i}}^{\prime }\left[ \left( \nabla _{\Pi _{n}}\right) .\left( \nabla _{\Pi
_{j}}\right) \right] \hat{\varepsilon}_{0}\right) \varepsilon
^{ijk}B_{k}\varepsilon ^{lnm}B_{m}
\end{eqnarray*}
with again the implicit convention, recalled by the $^{\prime }$ and implied
by construction and the definitions of $\left( Q\right) $, $\left( X\right)
^{i}$, $\left( Y\right) ^{j}$), that the derivatives $\nabla _{\Pi
_{l}}^{\prime }$ and $\nabla _{\Pi _{i}}^{\prime }$ are not applied on the
same quadratic term $\Pi _{i}A^{ij}\Pi _{j}$ in the series expansion of $%
\hat{\varepsilon}_{0}$. $\left[ \left( \nabla _{\Pi _{n}}\right) .\left(
\nabla _{\Pi _{j}}\right) \right] $ is a notation to recall that $\left(
\nabla _{\Pi _{n}}\right) \left( \nabla _{\Pi _{j}}\right) $ act both on the
same quadratic term,

Gathering all the relevant terms yields ultimately for the general case :
\begin{eqnarray*}
-\frac{\hbar }{2}\left\langle \hat{\varepsilon}_{0}\left( \mathbf{x}\right)
\right\rangle &=&\frac{e^{2}}{16}\hbar ^{2}\left( \varepsilon
^{ijk}B_{k}\right) \left( \varepsilon ^{lmn}B_{n}\right) \left( \nabla _{\Pi
_{l}}\right) \left( \nabla _{\Pi _{i}}\right) \left( \nabla _{\Pi
_{m}}\right) \left( \nabla _{\Pi _{j}}\right) \hat{\varepsilon}_{0} \\
&&+\frac{e^{2}}{24}\hbar ^{2}\left( \nabla _{\Pi _{l}}^{\prime }\nabla _{\Pi
_{i}}^{\prime }\left[ \left( \nabla _{\Pi _{n}}\right) .\left( \nabla _{\Pi
_{j}}\right) \right] \hat{\varepsilon}_{0}\right) \varepsilon
^{ijk}B_{k}\varepsilon ^{lnm}B_{m}+\frac{e}{8}\hbar ^{2}\left( \mathbf{%
\nabla }\times \mathbf{B}\right) .\mathbf{\nabla }_{\mathbf{\Pi }}\left(
\nabla _{\mathbf{\Pi }^{2}}\right) ^{2}\hat{\varepsilon}_{0}
\end{eqnarray*}

\paragraph{Final form for the Band Hamiltonian}

Gathering all the previous terms, and writing $\hat{\varepsilon}_{0}\left(
\mathbf{\pi }\right) $ as a completely symmetrized series in the powers of
the momentum (in the lack of any a priory symmetry), we have ultimately the
diagonalized Hamiltonian :
\begin{eqnarray}
H_{d} &=&\hat{\varepsilon}_{0}\left( \mathbf{\pi }\right) +V\left( \mathbf{r}%
\right) -\frac{1}{2}\left( \mathbf{\mu }\left( \mathbf{x}\right) \mathbf{.B}%
\left( \mathbf{x}\right) +\mathbf{B}\left( \mathbf{x}\right) \mathbf{.\mu }%
\left( \mathbf{x}\right) \right) -\mathbf{\hat{\mu}}^{k}\left( \mathbf{x}%
\right) \mathbf{.}\nabla _{R_{k}}\mathbf{B}\left( \mathbf{x}\right) +\mathbf{%
B.\tilde{\mu}.B}  \notag \\
&&+\frac{\hbar ^{2}}{4}\left[ \sum_{M}\left( \mathcal{A}_{0}^{R_{l}}\right)
_{NM}\left( \mathcal{A}_{0}^{R_{m}}\right) _{MN}\right] \nabla
_{R_{m}}\nabla _{R_{l}}V\left( R\right)  \notag \\
&&+\frac{\hbar ^{2}}{4}\left[ \varepsilon _{0}\left( \mathbf{x}\right) ,%
\mathcal{A}_{0}^{R_{m}}\right] _{NM}\frac{\left( \mathcal{A}%
_{0}^{R_{l}}\right) _{MN}\nabla _{R_{m}}\nabla _{R_{l}}V\left( R\right) }{%
\hat{\varepsilon}_{0M}-\hat{\varepsilon}_{0N}}  \notag \\
&&  \label{H1P}
\end{eqnarray}
The double scalar product $\mathbf{B.\tilde{\mu}.B}$ meaning that the two
index tensor $\mathbf{\tilde{\mu}}$ being contracted twice with $\mathbf{B}$.

\paragraph{Example : Darwin term for the Bloch electron}

We now consider more specifically what we have called the Darwin term. Its
interpretation turns out to be more transparent when $B=0,$ therefore we
keep here only the electrostatic potential. In this case, we have
\begin{equation*}
\mathcal{A}^{\mathbf{R}}=\hbar \mathcal{A}_{0}^{\mathbf{R}}\left( \mathbf{R+}%
\frac{\hbar }{4}\mathcal{A}_{0}^{R_{l}},\mathbf{K}\right) -e\frac{\hbar ^{2}%
}{2}\left[ \hbar \left[ .,\hat{\varepsilon}_{0}\right] ^{-1}.\left( \mathcal{%
P}_{-}\left\{ \frac{1}{2}\mathcal{A}_{0}^{R_{l}}\nabla _{R_{l}}V\left(
\mathbf{R}\right) +H.C.\right\} \right) ,\mathbf{R}+\mathcal{A}_{0}^{\mathbf{%
R}}\right]
\end{equation*}
and $\mathcal{A}_{0}^{K_{l}}=0$ as shown in \cite{PIERREEUROPHYS} but at the
second order we have the following contributions and
\begin{equation*}
\left( \mathcal{A}^{\mathbf{K}}\right) _{MN}=-\frac{i}{2}\hbar ^{2}\nabla
_{R_{m}}\frac{\left( \mathcal{A}_{0}^{R_{l}}\right) _{MN}\nabla
_{R_{l}}V\left( R\right) }{\hat{\varepsilon}_{0M}-\hat{\varepsilon}_{0N}}%
\hat{\delta}_{MN}
\end{equation*}
showing that the non diagonal part of $\mathcal{A}^{\mathbf{K}}$ is non
null. It will lead to a magnetization term : the Darwin term. We can write :
\begin{eqnarray*}
\left( \mathcal{A}^{\mathbf{R}}\right) _{MN} &=&\left( \hbar \mathcal{A}%
_{0}^{\mathbf{R}}\right) _{MN}+\frac{\hbar ^{2}}{16}\left\{ \left( \mathcal{A%
}_{0}^{R_{l}}\right) _{MP}\nabla _{R_{l}}\left( \mathcal{A}_{0}^{\mathbf{R}%
}\right) _{PN}+M\leftrightarrow N\right\} +\frac{\hbar ^{2}}{2}i\mathbf{%
\nabla }_{\mathbf{K}}\frac{\left( \mathcal{A}_{0}^{R_{l}}\right) _{MN}\nabla
_{R_{l}}V\left( \mathbf{R}\right) }{\hat{\varepsilon}_{0M}-\hat{\varepsilon}%
_{0N}}\hat{\delta}_{MN} \\
&&+\frac{\hbar ^{2}}{2}\left( \mathcal{A}_{0}^{\mathbf{R}}\right) _{MP}\hat{%
\delta}_{PM}\frac{\left( \mathcal{A}_{0}^{R_{l}}\right) _{PN}\nabla
_{R_{l}}V\left( \mathbf{R}\right) }{\hat{\varepsilon}_{0P}-\hat{\varepsilon}%
_{0N}}-\frac{\hbar ^{2}}{2}\frac{\left( \mathcal{A}_{0}^{R_{l}}\right)
_{MP}\nabla _{R_{l}}V\left( \mathbf{R}\right) }{\hat{\varepsilon}_{0M}-\hat{%
\varepsilon}_{0P}}\hat{\delta}_{MP}\left( \mathcal{A}_{0}^{\mathbf{R}%
}\right) _{PN}+H.C.
\end{eqnarray*}
and the projected variables :
\begin{eqnarray*}
\left( \emph{A}^{\mathbf{R}}\right) _{N} &=&\left( \mathcal{A}^{\mathbf{R}%
}\right) _{NN}=\left( \hbar \mathcal{A}_{0}^{\mathbf{R}}\right) _{MN}+\frac{%
\hbar ^{2}}{8}\left( \mathcal{A}_{0}^{R_{l}}\right) _{MP}\nabla
_{R_{l}}\left( \mathcal{A}_{0}^{\mathbf{R}}\right) _{PN} \\
&&+\frac{\hbar ^{2}}{2}\left( \left( \mathcal{A}_{0}^{\mathbf{R}}\right)
_{NP}\left( \mathcal{A}_{0}^{R_{l}}\right) _{PN}+\left( \mathcal{A}%
_{0}^{R_{l}}\right) _{NP}\left( \mathcal{A}_{0}^{\mathbf{R}}\right)
_{PN}\right) \hat{\delta}_{PN}\frac{\nabla _{R_{l}}V\left( \mathbf{R}\right)
}{\hat{\varepsilon}_{0P}-\hat{\varepsilon}_{0N}} \\
&&+H.C. \\
\left( \emph{A}^{\mathbf{K}}\right) _{N} &=&0
\end{eqnarray*}
Therefore the Hamiltonian is :
\begin{eqnarray*}
H_{d} &=&\hat{\varepsilon}_{0}\left( \mathbf{k}\right) +V\left( \mathbf{r}%
\right) +M\left[ \varepsilon _{0}\right] \\
&=&\hat{\varepsilon}_{0}+V\left( \mathbf{r}\right) -\frac{i\hbar ^{2}}{8}%
\mathcal{P}_{+}\left\{ \left[ \left[ \hat{\varepsilon}_{0},\mathcal{A}%
_{0}^{R_{m}}\right] \left[ X\mathcal{,}K_{m}\right] \right] -\left[ \left[
\hat{\varepsilon}_{0},\left[ X\mathcal{,}K_{m}\right] \right] \mathcal{A}%
_{0}^{R_{m}}\right] \right\} +H.C.
\end{eqnarray*}
with :
\begin{equation*}
X=\left[ .,\hat{\varepsilon}_{0}\right] ^{-1}.\left( \mathcal{P}_{-}\left\{
\frac{1}{2}\mathcal{A}_{0}^{R_{l}}\nabla _{R_{l}}V\left( \mathbf{R}\right)
+H.C.\right\} \right)
\end{equation*}
which reduces to
\begin{eqnarray}
H_{N} &=&\hat{\varepsilon}_{0}+V\left( \mathbf{r}\right) +\frac{\hbar ^{2}}{8%
}\left[ \sum_{M}\left( \mathcal{A}_{0}^{R_{l}}\right) _{NM}\left( \mathcal{A}%
_{0}^{R_{m}}\right) _{MN}\right] \nabla _{R_{m}}\nabla _{R_{l}}V\left(
R\right)  \notag \\
&&+\frac{\hbar ^{2}}{8}\left[ \varepsilon _{0}\left( \mathbf{x}\right) ,%
\mathcal{A}_{0}^{R_{m}}\right] _{NM}\frac{\left( \mathcal{A}%
_{0}^{R_{l}}\right) _{MN}\nabla _{R_{m}}\nabla _{R_{l}}V\left( R\right) }{%
\hat{\varepsilon}_{0M}-\hat{\varepsilon}_{0N}}+H.C.  \label{H1Darwin}
\end{eqnarray}
This is an interaction in second derivative of the potential, Darwin type.

The effective in-bands Hamiltonian Eq. $\left( \ref{H1P}\right) $ is the
desired result concerning one Bloch electron in an external electromagnetic
potential and will be very useful for the latter computation of the
effective Hamiltonian of several Bloch electrons in self-interaction. But at
this point it necessary to compare if it is possible with previous
approaches in particular with Blount's one.

\paragraph{Comparison with previous results}

In a powerful series of paper, Blount \cite{Blount}\cite{Blount1} designed a
method to derive the diagonalized Hamiltonian both for a Bloch electron in a
constant magnetic field and the Dirac electron in an arbitrary magnetic
field as series expansion in powers of the fields. In both cases, at the
first order in $\hbar $ our results coincide with his ones. At the second
order, the direct comparison for the Bloch electron is difficult to do,
since Blount's choice of variables (the canonical ones) differs from ours.
However, the comparison in the particular case of the Dirac electron (2
bands Hamiltonian) can be performed and will be explained in detail in \cite%
{PIERREDIRAC}. The result is that, despite some important similarities, our
results differ slightly From Blount's ones. \ Actually some mistake arise in
his results due to his choice of variables, which induces wrong expressions
for the Berry curvature, and as a consequence, in the Hamiltonian. As a
consistency check we have shown in \cite{PIERREDIRAC}, that at order $\frac{1%
}{m^{3}}$, we recover the usual Foldy-Wouthuysen formula for diagonalized
Hamiltonian of the Dirac electron in an electromagnetic field. This is
apparently not the case in \cite{Blount1} where the coefficient of the
squared magnetic field is apparently incorrect.

\section{Interacting Bloch electrons}

Having found the diagonalized Hamiltonian of a one particle system, we can
now focus on the multiparticles Bloch electrons in interaction. In this
context we must consider the interaction mediated by the Coulomb
interaction. Although this last one dominates the magnetic interactions, it
is known that, for material whose electrons have a non vanishing magnetic
moment, new effects such as interaction between the moments mediated by the
magnetic field can have important effects. For this reason, we are now
interested in investigating the diagonalization of the Hamiltonian of
several electrons in self-interaction through the full electromagnetic
field. However, for practical reasons and to deal with tractable formula, we
will restrict ourselves to the case of relatively small interaction.

\subsection{Derivation of the microscopic Hamiltonian}

To start , we will derive the microscopic classical Hamiltonian for this
system, before going to the quantum version. Note that the electromagnetic
field will always be considered as classical and only the particles will be
treated at the quantum level. Let us introduce the non relativistic
Lagrangian of $P$ particles self-interacting through the electromagnetic
field and moving in a periodic potential $V_{p}$\ which is the only external
potential:
\begin{equation*}
L=\frac{1}{2}\sum_{\alpha }m^{\left( \alpha \right) }\left( \mathbf{\dot{R}}%
^{\left( \alpha \right) }\right) ^{2}-V_{p}\left( \mathbf{R}^{\left( \alpha
\right) }\right) -\frac{1}{2}F_{\mu \nu }F^{\mu \nu }+\int d^{3}x\mathbf{J}.%
\mathbf{A-}\int d^{3}xeA_{0}J_{0}
\end{equation*}
Here $F_{\mu \nu }$ is a classical electromagnetic field whose potential is $%
\left( \mathbf{A,}A_{0}\right) $, $e$ are the particles charges. The current
$\mathbf{J}$ is given by $\mathbf{J}\left( \mathbf{x}\right) \mathbf{=}%
\sum_{\alpha }e\delta \left( \mathbf{x-R}^{\left( \alpha \right) }\right)
\mathbf{\dot{R}}^{\left( \alpha \right) }$ and $J_{0}$ is the density of
charges, $J_{0}=\sum_{\alpha }e\delta \left( \mathbf{x-R}^{\left( \alpha
\right) }\right) $.

Choosing a gauge for the electromagnetic field, for example the Coulomb
gauge (see \cite{Weinberg} ), some computations lead to write the particles
and field Hamiltonian as
\begin{eqnarray*}
H &=&\sum_{\alpha =1}^{P}\mathbf{P}^{\left( \alpha \right) }\mathbf{\dot{R}}%
^{\left( \alpha \right) }-\frac{1}{2}\sum_{\alpha }m^{\left( \alpha \right)
}\left( \mathbf{\dot{R}}^{\left( \alpha \right) }\right) ^{2}-\int d^{3}x%
\mathbf{J}.\mathbf{A+}\int d^{3}xeA_{0}J_{0}+\Pi _{em}^{2}+\frac{1}{2}\left(
\mathbf{\nabla \times A}\right) ^{2} \\
&&-\frac{1}{2}\left( \Pi +\mathbf{\nabla }A_{0}\right) ^{2}+V_{p}\left(
\mathbf{R}^{\left( \alpha \right) }\right)
\end{eqnarray*}
with $\Pi _{em}$ the solenoidal part of the electromagnetic momentum defined
by $\Pi _{em}=\dot{A}$ satisfying the constraint $\mathbf{\nabla }.\Pi
_{em}=0$ \cite{Weinberg}. $\mathbf{P}^{\left( \alpha \right) }$ is the usual
canonical particle momentum. Using also our definition for $\mathbf{J}$, we
are thus led to :
\begin{eqnarray}
H &=&\sum_{\alpha =1}^{P}\frac{\left( \mathbf{P}^{\left( \alpha \right) }-e%
\mathbf{A}^{\left( \alpha \right) }\left( \mathbf{R}^{\left( \alpha \right)
}\right) \right) ^{2}}{2m^{\left( \alpha \right) }}\mathbf{+}\frac{1}{2}\int
d^{3}\mathbf{x}e\left[ A_{0}\left( \mathbf{x}\right) J_{0}\left( \mathbf{x}%
\right) +\frac{1}{2}\Pi _{em}^{2}+\frac{1}{2}\left( \mathbf{\nabla \times A}%
\right) ^{2}-\left( \mathbf{\nabla }A_{0}\right) ^{2}\right]  \notag \\
&&+V_{p}\left( \mathbf{R}^{\left( \alpha \right) }\right)  \notag \\
&=&\sum_{\alpha =1}^{P}\frac{\left( \mathbf{P}^{\left( \alpha \right) }-e%
\mathbf{A}^{\left( \alpha \right) }\left( \mathbf{R}^{\left( \alpha \right)
}\right) \right) ^{2}}{2m^{\left( \alpha \right) }}\mathbf{+}\frac{1}{2}\int
d^{3}\mathbf{x}eA_{0}\left( \mathbf{x}\right) \sum_{\alpha }e\delta \left(
\mathbf{x-R}^{\left( \alpha \right) }\right)  \notag \\
&&+\frac{1}{2}\int d^{3}x\left[ \Pi _{em}^{2}+\frac{1}{2}\int d^{3}x\left(
\mathbf{\nabla \times A}\right) ^{2}-\left( \mathbf{\nabla }A_{0}\right) ^{2}%
\right] +V_{p}\left( \mathbf{R}^{\left( \alpha \right) }\right)  \label{HP}
\end{eqnarray}
Recall that $V_{p}\left( \mathbf{R}^{\left( \alpha \right) }\right) $ stands
for the periodic potential. Note also that, given that the particle density
is a sum of delta function centered around the particles positions, the
potential $\frac{1}{2}\int d^{3}\mathbf{x}eA_{0}\left( \mathbf{x}\right)
\sum_{\alpha }e\delta \left( \mathbf{x-R}^{\left( \alpha \right) }\right) $
reduces to :
\begin{equation*}
\frac{1}{2}\int d^{3}\mathbf{x}eA_{0}\left( \mathbf{x}\right) \sum_{\alpha
}e\delta \left( \mathbf{x-R}^{\left( \alpha \right) }\right) =\sum_{\alpha
\neq \beta }\frac{e^{2}}{8\pi \left\vert \mathbf{R}^{\left( \alpha \right) }-%
\mathbf{R}^{\left( \beta \right) }\right\vert }
\end{equation*}
However, we will not use this \textquotedblright solved
form\textquotedblright\ right now and rather keep the electromagnetic field $%
A_{0}\left( \mathbf{x}\right) $ (see below).

The previous expression is desired Hamiltonian for $P$ classical electrons
in interaction and in a periodic potential. It will be the starting point
for the quantification and for the computation of the effective in-bands
Hamiltonian to the second order in $\hbar .$ This order is required to
determine the magnetization-magnetization interaction in the same way that
the spin-spin interaction in the Breit Hamiltonian for Dirac particles \cite%
{BREIT}. The consideration of interacting electrons, each of them being
constrained to be in a single band, will require the Hamiltonian
diagonalization and the projection on each respective band. To do so we just
have to apply the method developer for the one particle scheme.

Remark, at this point that the Hamiltonian could be further simplified by
using the Coulomb gauge condition which allows to replace directly the
potential $A_{0}$. Actually, as a consequence of the Coulomb gauge
constraint one has $-\mathbf{\nabla }^{2}A_{0}=J_{0}$. However, for the sake
of symmetry we aim at treating the electrostatic potential as an external
field, as we do for the vector potential, and to replace it after
transformation. We will explain later why this is in fact simpler and
innocuous for the final result.

\subsection{Diagonalization process}

We will now diagonalize our Hamiltonian to the second order in $\hbar $ as
before. The space of states acting is the tensor product of $P$ copies of
the individual spaces $\mathcal{V}=\otimes _{\alpha =1,...,P}\mathcal{V}%
_{\alpha }$ with the Hamiltonian Eq. $\left( \ref{HP}\right) $ in which $%
V_{p}$ is the periodic potential, $\mathbf{A}^{\left( \alpha \right) }$ is
the potential created by the other charges on charge $\alpha $ and $v\left(
\mathbf{R}^{\left( \alpha \right) }\right) =\int d^{3}\mathbf{x}eA_{0}\left(
\mathbf{x}\right) e\delta \left( \mathbf{x-R}^{\left( \alpha \right)
}\right) $ the electrostatic potential involving the particle $\alpha $.

The Hamiltonian $H$ is diagonalized at the second order in $\hbar $ in a
straightforward way. Using our previous experience with the one electron,
consider :
\begin{equation*}
U_{0}=\otimes _{\alpha =1...P}U^{\left( \alpha \right) }
\end{equation*}
Where $U^{\left( \alpha \right) }$ are the individual particle
diagonalization matrix found in the previous section to the first order in $%
\hbar $. Note that this matrix do no act on the electromagnetic field, whose
integration variable $x$ is independent from the quantum mechanical
operators $\mathbf{R}^{\left( \alpha \right) }$, $\mathbf{P}^{\left( \alpha
\right) }$. This is the usual procedure for this kind of system. We first
diagonalize the part of the Hamiltonian which describes the particles
dynamics and leave the electromagnetic field untouched (see for instance
Feynman's book for this procedure \cite{FEYNMANBOOK}). This is essential in
order to have each particle living in a particular energy band. It means
that we assume here an adiabatic process in which the electromagnetic
interaction does not cause interband jump. If we were eliminating straight
the electromagnetic field in the initial Hamiltonian Eq. $\left( \ref{HP}%
\right) $ with the help of the Maxwell equations, we would of course get an
equivalent Hamiltonian for particles only, but it would automatically mix
all the energy bands. Then the possibility to assume an adiabatic process
and to project each individual particles Hamiltonian on a specific band
would be lost.

Therefore we will choose the first route and after the diagonalization one
will solve the Maxwell equations which will determine the exact form of the
particle currents. These last ones will then be further injected into the
Hamiltonian to get the final in-bands effective Hamiltonian of the
particles. Let us show how this procedure works.

From Eqs. $\left( \ref{H1P}\right) \left( \ref{HP}\right) $ the Hamiltonian
in the diagonal representation of the particles part is thus directly :
\begin{eqnarray*}
H_{N} &=&\sum_{\alpha }\hat{\varepsilon}_{0N^{\left( \alpha \right)
}}^{\left( \alpha \right) }\left( \mathbf{\pi }^{\left( \alpha \right)
}\right) +\sum_{\alpha }v\left( \mathbf{r}^{\left( \alpha \right) }\right) -%
\frac{1}{2}\sum_{\alpha }\left( \mathbf{\mu }\left( \mathbf{x}^{\left(
\alpha \right) }\right) \mathbf{.B}^{\left( \alpha \right) }\left( \mathbf{x}%
^{\left( \alpha \right) }\right) +\mathbf{B}^{\left( \alpha \right) }\left(
\mathbf{x}^{\left( \alpha \right) }\right) \mathbf{.\mu }\left( \mathbf{x}%
\right) \right) \\
&&-\mathbf{\hat{\mu}}^{k}\left( \mathbf{x}\right) \mathbf{.}\nabla _{R_{k}}%
\mathbf{B}^{\left( \alpha \right) }\left( \mathbf{x}^{\left( \alpha \right)
}\right) +\mathbf{B}^{\left( \alpha \right) }\mathbf{.\tilde{\mu}.B}^{\left(
\alpha \right) } \\
&&+\frac{\hbar ^{2}}{4}\sum_{\alpha }\left[ \sum_{M^{\left( \alpha \right)
}}\left( \mathcal{A}_{0}^{R_{l}^{\left( \alpha \right) }}\right) _{N^{\left(
\alpha \right) }M^{\left( \alpha \right) }}\left( \mathcal{A}%
_{0}^{R_{m}^{\left( \alpha \right) }}\right) _{M^{\left( \alpha \right)
}N^{\left( \alpha \right) }}\right] \nabla _{R_{m}^{\left( \alpha \right)
}}\nabla _{R_{l}^{\left( \alpha \right) }}v\left( \mathbf{r}^{\left( \alpha
\right) }\right) \\
&&+\frac{\hbar ^{2}}{4}\sum_{\alpha }\frac{\left[ \left[ \varepsilon
_{0}^{\left( \alpha \right) }\left( \mathbf{x}\right) ,\mathcal{A}%
_{0}^{R_{m}^{\left( \alpha \right) }}\right] _{N^{\left( \alpha \right)
}M^{\left( \alpha \right) }}\left( \mathcal{A}_{0}^{R_{l}^{\left( \alpha
\right) }}\right) _{M^{\left( \alpha \right) }N^{\left( \alpha \right)
}}\nabla _{R_{m}^{\left( \alpha \right) }}\nabla _{R_{l}^{\left( \alpha
\right) }}\right] v\left( \mathbf{r}^{\left( \alpha \right) }\right) }{\hat{%
\varepsilon}_{0M^{\left( \alpha \right) }}^{\left( \alpha \right) }-\hat{%
\varepsilon}_{0N^{\left( \alpha \right) }}} \\
&&+\frac{1}{2}\int d^{3}x\left[ \Pi _{em}^{2}+\frac{1}{2}\int d^{3}x\left(
\mathbf{\nabla \times A}\right) ^{2}-\left( \mathbf{\nabla }A_{0}\right) ^{2}%
\right]
\end{eqnarray*}
The notations of the previous section have been kept, adding only the
superscript $\left( \alpha \right) $ to label the particles in Eq. $\left( %
\ref{H1P}\right) $. For example, $\mathbf{x}^{\left( \alpha \right) }=\left(
\mathbf{r}^{\left( \alpha \right) },\mathbf{k}^{\left( \alpha \right)
}\right) $ is the couple of dynamical variables for the particle $\left(
\alpha \right) $, $\mathbf{x}^{\left( \alpha \right) }=\left( \mathbf{R}%
^{\left( \alpha \right) }+\emph{A}_{N^{\left( \alpha \right) }}^{\mathbf{R}%
\left( ^{\alpha }\right) },\mathbf{K}^{\left( \alpha \right) }+\emph{A}%
_{N^{\left( \alpha \right) }}^{\mathbf{K}^{\left( \alpha \right) }}\right) $
and the Berry phases $\emph{A}_{N^{\left( \alpha \right) }}^{\mathbf{R}%
\left( ^{\alpha }\right) }$, $\emph{A}_{N^{\left( \alpha \right) }}^{\mathbf{%
K}^{\left( \alpha \right) }}$ involved in these operators have been computed
in the single particle section. Notice also that $N$ is a shorthand for the
multi-index $N=\left( N^{\left( \alpha \right) }\right) _{\alpha =1,...P}$
recalling that the Hamiltonian describes $P$ particles respectively in the
bands $N^{\left( 1\right) }...N^{\left( P\right) }$. Ultimately $\mathbf{B}%
^{\left( \alpha \right) }$ is the field acting on particle $\alpha $.

\subsection{Effective Hamiltonian for $P$ particles and particles currents:}

Our aim is now to \ replace the electromagnetic field in the Hamiltonian as
a function of the dynamical variables of the particles system. We first do
so by solving the Hamiltonian equations for the electromagnetic field in the
limit of relatively weak interactions.

\subsubsection{Dynamics for the Electromagnetic Field}

To get rid of the electromagnetic field and writing an interaction
Hamiltonian for the particles, we first focus on the vector potential and
start with the Hamiltonian equation for the electromagnetic field in the
Coulomb gauge \cite{Weinberg} :

\begin{eqnarray*}
\mathbf{\dot{A}}\left( \mathbf{x},t\right) &=&\Pi _{em} \\
\dot{\Pi}_{em}^{i}\left( \mathbf{x},t\right) &=&-\int d^{3}y\left[ \delta
^{ij}\delta ^{3}\left( \mathbf{x-y}\right) +\frac{\partial ^{2}}{\partial
x^{i}\partial x^{j}}\frac{1}{4\pi \left\vert \mathbf{x-y}\right\vert }\right]
\mathbf{\nabla }\times \mathbf{\nabla }A^{j}\left( \mathbf{y},t\right) \\
&&+\frac{i}{\hbar }\left[ H_{mat},\dot{\Pi}_{em}^{i}\left( \mathbf{x}%
,t\right) \right] \\
&=&-\int d^{3}y\left[ \delta ^{ij}\delta ^{3}\left( \mathbf{x-y}\right) +%
\frac{\partial ^{2}}{\partial x^{i}\partial x^{j}}\frac{1}{4\pi \left\vert
\mathbf{x-y}\right\vert }\right] \mathbf{\nabla }\times \mathbf{\nabla }%
A^{j}\left( \mathbf{y},t\right) \\
&&-\int d^{3}y\left[ \delta ^{ij}\delta ^{3}\left( \mathbf{x-y}\right) +%
\frac{\partial ^{2}}{\partial x^{i}\partial x^{j}}\frac{1}{4\pi \left\vert
\mathbf{x-y}\right\vert }\right] \frac{\delta }{\delta A^{j}\left( \mathbf{%
x,t}\right) }H_{mat}
\end{eqnarray*}
where $H_{mat}$ denotes the matter part (that is excluding the free part of
the electromagnetic field Hamiltonian). Using the gauge constraints : $%
\mathbf{\nabla }.\mathbf{\Pi }_{em}=\mathbf{\nabla }.\mathbf{A}=0$. we are
led to :
\begin{equation*}
\square \mathbf{A}\left( \mathbf{\left( \mathbf{x},t\right) }\right) =-\int
d^{3}y\left[ \delta ^{ij}\delta ^{3}\left( \mathbf{x-y}\right) +\frac{%
\partial ^{2}}{\partial x^{i}\partial x^{j}}\frac{1}{4\pi \left\vert \mathbf{%
x-y}\right\vert }\right] \frac{\delta }{\delta A^{j}\left( \mathbf{y,t}%
\right) }H_{mat}
\end{equation*}
where $\frac{\delta }{\delta A^{i}\left( \mathbf{y,t}\right) }$ stands for
the functional derivative with respect to $A^{i}\left( \mathbf{y,t}\right) $
.

\subsubsection{Explicit expression for the current}

We aim now at writing a more explicit formula for the effective current $%
j^{j}\mathbf{\left( \mathbf{y},t\right) =-}\frac{\delta }{\delta A^{j}\left(
\mathbf{x,t}\right) }H_{mat}$. To do so, we decompose $H_{mat}$ as :
\begin{eqnarray*}
H_{mat} &=&\sum_{\alpha }\hat{\varepsilon}_{0N^{\left( \alpha \right)
}}^{\left( \alpha \right) }\left( \mathbf{\pi }^{\left( \alpha \right)
}\right) +\sum_{\alpha }v\left( \mathbf{r}^{\left( \alpha \right) }\right) \\
&&-\frac{1}{2}\sum_{\alpha }\left( \mathbf{\mu }\left( \mathbf{x}^{\left(
\alpha \right) }\right) \mathbf{.B}^{\left( \alpha \right) }\left( \mathbf{x}%
^{\left( \alpha \right) }\right) +\mathbf{B}^{\left( \alpha \right) }\left(
\mathbf{x}^{\left( \alpha \right) }\right) \mathbf{.\mu }\left( \mathbf{x}%
\right) \right) -\sum_{\alpha }\mathbf{\hat{\mu}}^{k}\left( \mathbf{x}%
\right) \mathbf{.}\nabla _{R_{k}}\mathbf{B}^{\left( \alpha \right) }\left(
\mathbf{x}^{\left( \alpha \right) }\right) \\
&&+\mathbf{B}^{\left( \alpha \right) }\mathbf{.\tilde{\mu}.B}^{\left( \alpha
\right) }+D^{\left( \alpha \right) }
\end{eqnarray*}
where $D^{\left( \alpha \right) }$ stands for the Darwin term :
\begin{eqnarray*}
D^{\left( \alpha \right) } &=&\frac{\hbar ^{2}}{4}\left[ \sum_{M^{\left(
\alpha \right) }}\left( \mathcal{A}_{0}^{R_{l}^{\left( \alpha \right)
}}\right) _{N^{\left( \alpha \right) }M^{\left( \alpha \right) }}\left(
\mathcal{A}_{0}^{R_{m}^{\alpha }}\right) _{M^{\left( \alpha \right)
}N^{\left( \alpha \right) }}\right] \nabla _{R_{m}^{\left( \alpha \right)
}}\nabla _{R_{l}^{\left( \alpha \right) }}v\left( \mathbf{r}^{\left( \alpha
\right) }\right) \\
&&+\frac{\hbar ^{2}}{4}\sum_{\alpha }\frac{\left[ \left[ \varepsilon
_{0}^{\left( \alpha \right) }\left( \mathbf{x}\right) ,\mathcal{A}%
_{0}^{R_{m}^{\left( \alpha \right) }}\right] _{N^{\left( \alpha \right)
}M^{\left( \alpha \right) }}\left( \mathcal{A}_{0}^{R_{l}^{\left( \alpha
\right) }}\right) _{M^{\left( \alpha \right) }N^{\left( \alpha \right)
}}\nabla _{R_{m}^{\left( \alpha \right) }}\nabla _{R_{l}^{\left( \alpha
\right) }}\right] v\left( \mathbf{r}^{\left( \alpha \right) }\right) }{\hat{%
\varepsilon}_{0M^{\left( \alpha \right) }}^{\left( \alpha \right) }-\hat{%
\varepsilon}_{0N^{\left( \alpha \right) }}}
\end{eqnarray*}
The computation of $j^{j}\mathbf{\left( \mathbf{y},t\right) =-}\frac{\delta
}{\delta A^{j}\left( \mathbf{x,t}\right) }H_{mat}$ has to be performed
carefully since the dependence in the gauge field in $H_{mat}$ is intricate.
We aim at computing it as a function of the \textquotedblright
physical\textquotedblright\ variables $\mathbf{r}^{\left( \alpha \right) }$,
$\mathbf{\pi }^{\left( \alpha \right) }$ and not as a function of the
initial microscopic canonical variables. However, the physical variables $%
\mathbf{r}^{\left( \alpha \right) }$, $\mathbf{\pi }^{\left( \alpha \right)
} $ depend on the value of the fields evaluated at the microscopic positions
and momentum.
\begin{eqnarray*}
\mathbf{r}^{\left( \alpha \right) } &=&\mathbf{R}^{\left( \alpha \right) }%
\mathbf{+}\emph{A}_{N}^{\mathbf{R}^{\left( \alpha \right) }}\left( \mathbf{R}%
^{\left( \alpha \right) }\mathbf{,\Pi }^{\left( \alpha \right) }\right) \\
\mathbf{\pi }^{\left( \alpha \right) } &=&\mathbf{\Pi }^{\left( \alpha
\right) }+e\emph{A}_{N}^{\mathbf{R}^{\left( \alpha \right) }}\left( \mathbf{R%
}^{\left( \alpha \right) }\mathbf{,\Pi }^{\left( \alpha \right) }\right)
\times \mathbf{B}\left( \mathbf{R}^{\left( \alpha \right) }\right) \\
&&+e\frac{\hbar }{4}\mathcal{P}_{N}\left[ \mathcal{A}_{0}^{R_{k}^{\left(
\alpha \right) }}\right] \left( \emph{A}_{N}^{\mathbf{R}^{\left( \alpha
\right) }}\left( \mathbf{R}^{\left( \alpha \right) }\mathbf{,\Pi }^{\left(
\alpha \right) }\right) \times \nabla _{R_{k}^{\left( \alpha \right) }}%
\mathbf{B}\left( \mathbf{R}^{\left( \alpha \right) }\right) +H.C.\right)
\end{eqnarray*}
to turn this difficulty we reintroduce the variables $\mathbf{r}^{\left(
\alpha \right) }$, $\mathbf{\pi }^{\left( \alpha \right) }$ recursively in
our Berry phases.

Starting with $\mathbf{r}^{\left( \alpha \right) }$, at our order of
approximation and using $\mathbf{R}^{\left( \alpha \right) }=\mathbf{\mathbf{%
r}^{\left( \alpha \right) }-}\emph{A}_{N}^{\mathbf{R}^{\left( \alpha \right)
}}\left( \mathbf{r}^{\left( \alpha \right) }\mathbf{,\pi }^{\left( \alpha
\right) }\right) $ :
\begin{eqnarray*}
\mathbf{r}^{\left( \alpha \right) } &=&\mathbf{R}^{\left( \alpha \right) }%
\mathbf{+}\emph{A}_{N}^{\mathbf{R}^{\left( \alpha \right) }}\left( \mathbf{r}%
^{\left( \alpha \right) }\mathbf{,\pi }^{\left( \alpha \right) }\right)
-\left( \emph{A}_{N}^{\mathbf{\Pi }^{\left( \alpha \right) }}.\mathbf{\nabla
}_{\mathbf{\pi }^{\left( \alpha \right) }}\right) \emph{A}_{N}^{\mathbf{R}%
}\left( \mathbf{r}^{\left( \alpha \right) }\mathbf{,\pi }^{\left( \alpha
\right) }\right) \\
&\equiv &\mathbf{R}^{\left( \alpha \right) }\mathbf{+}\tilde{\emph{A}}^{%
\mathbf{R}^{\left( \alpha \right) }}\left( \mathbf{r}^{\left( \alpha \right)
}\mathbf{,\pi }^{\left( \alpha \right) }\right)
\end{eqnarray*}
(we have neglected the terms in $\mathbf{\nabla }_{\mathbf{r}^{\left( \alpha
\right) }}$ that do not appear at order $\hbar ^{2}$ and skipped the index $%
N $ for the sake of simplicity). We do the same thing for the momentum by
starting to write it as a function of $\mathbf{k}^{\left( \alpha \right) }$,
$\mathbf{r}^{\left( \alpha \right) }$. Given the definition of the momentum,
we explained before that it was not given by $\mathbf{k}^{\left( \alpha
\right) }-e\mathbf{A}\left( \mathbf{r}^{\left( \alpha \right) }\right) $,
but rather $\mathbf{k}^{\left( \alpha \right) }-e\mathbf{A}\left( \mathbf{r}%
^{\left( \alpha \right) }\right) -\frac{\hbar ^{2}}{8}e\nabla _{R_{k}}%
\mathbf{\nabla }A_{l}\left( \mathbf{R},\mathbf{P}\right) \mathcal{P}%
_{N}\left( \mathcal{A}_{0}^{R_{l}}\mathcal{A}_{0}^{R_{k}}+\mathcal{A}%
_{0}^{R_{k}}\mathcal{A}_{0}^{R_{l}}\right) $. It was explained that this
variable had a physical meaning (the subtracted term was there to compensate
a non gauge invariant term). \ This last term cancels one term in the Berry
phase for $\mathbf{K}^{\left( \alpha \right) }$. As a consequence,
\begin{eqnarray*}
\mathbf{\pi }^{\left( \alpha \right) } &=&\mathbf{k}^{\left( \alpha \right)
}-e\mathbf{A}\left( \mathbf{r}^{\left( \alpha \right) }\right) =\mathbf{K}%
^{\left( \alpha \right) }+\emph{A}_{N}^{\mathbf{K}^{\left( \alpha \right)
}}\left( \mathbf{R}^{\left( \alpha \right) }\mathbf{,\Pi }^{\left( \alpha
\right) }\right) -e\mathbf{A}\left( \mathbf{r}^{\left( \alpha \right)
}\right) \\
&&-\frac{\hbar ^{2}}{8}e\nabla _{R_{k}}\mathbf{\nabla }A_{l}\left( \mathbf{R}%
,\mathbf{P}\right) \mathcal{P}_{N}\left( \mathcal{A}_{0}^{R_{l}}\mathcal{A}%
_{0}^{R_{k}}+\mathcal{A}_{0}^{R_{k}}\mathcal{A}_{0}^{R_{l}}\right) \\
&=&\mathbf{K}^{\left( \alpha \right) }-e\mathbf{A}\left( \mathbf{r}^{\left(
\alpha \right) }\right) +\frac{1}{2}e\left[ \mathbf{\nabla }A_{l}\left(
\mathbf{R}^{\left( \alpha \right) }\right) \emph{A}_{N}^{R_{l}}\left(
\mathbf{R}^{\left( \alpha \right) }\mathbf{,\Pi }^{\left( \alpha \right)
}\right) +\emph{A}_{N}^{R_{l}}\left( \mathbf{R}^{\left( \alpha \right) }%
\mathbf{,\Pi }^{\left( \alpha \right) }\right) \mathbf{\nabla }A_{l}\left(
\mathbf{R}^{\left( \alpha \right) }\right) \right] \\
&&+\frac{\hbar ^{2}}{4}e\mathbf{\nabla }\nabla _{R_{k}}A_{l}\left( \mathbf{R}%
,\mathbf{P}\right) \left( \mathcal{P}_{N}\left[ \mathcal{A}_{0}^{R_{l}}%
\right] \mathcal{P}_{N}\left[ \mathcal{A}_{0}^{R_{k}}\right] +\mathcal{P}_{N}%
\left[ \mathcal{A}_{0}^{R_{k}}\right] \mathcal{P}_{N}\left[ \mathcal{A}%
_{0}^{R_{l}}\right] \right) \\
&=&\mathbf{K}^{\left( \alpha \right) }-e\mathbf{A}\left( \mathbf{r}^{\left(
\alpha \right) }\right) +\frac{1}{2}e\left[ \mathbf{\nabla }A_{l}\left(
\mathbf{R}^{\left( \alpha \right) }\right) \tilde{\emph{A}}^{R_{l}}\left(
\mathbf{r}^{\left( \alpha \right) }\mathbf{,\pi }^{\left( \alpha \right)
}\right) +\tilde{\emph{A}}^{R_{l}}\left( \mathbf{r}^{\left( \alpha \right) }%
\mathbf{,\pi }^{\left( \alpha \right) }\right) \mathbf{\nabla }A_{l}\left(
\mathbf{R}^{\left( \alpha \right) }\right) \right] \\
&&-\frac{\hbar ^{2}}{4}e\mathbf{\nabla }\nabla _{R_{k}}A_{l}\left( \mathbf{R}%
,\mathbf{P}\right) \left( \mathcal{P}_{N}\left[ \mathcal{A}_{0}^{R_{l}}%
\right] \mathcal{P}_{N}\left[ \mathcal{A}_{0}^{R_{k}}\right] +\mathcal{P}_{N}%
\left[ \mathcal{A}_{0}^{R_{k}}\right] \mathcal{P}_{N}\left[ \mathcal{A}%
_{0}^{R_{l}}\right] \right) \\
&\equiv &\mathbf{K}^{\left( \alpha \right) }-e\mathbf{A}\left( \mathbf{r}%
^{\left( \alpha \right) }\right) +\frac{1}{2}e\left[ \mathbf{\nabla }%
A_{l}\left( \mathbf{R}^{\left( \alpha \right) }\right) \tilde{\emph{A}}%
^{R_{l}}\left( \mathbf{r}^{\left( \alpha \right) }\mathbf{,\pi }^{\left(
\alpha \right) }\right) +\tilde{\emph{A}}^{R_{l}}\left( \mathbf{r}^{\left(
\alpha \right) }\mathbf{,\pi }^{\left( \alpha \right) }\right) \mathbf{%
\nabla }A_{l}\left( \mathbf{R}^{\left( \alpha \right) }\right) \right] \\
&&+\hbar ^{2}e\mathbf{\nabla }\nabla _{R_{k}}A_{l}\left( \mathbf{r}^{\left(
\alpha \right) }\right) \tilde{\emph{A}}_{lk}
\end{eqnarray*}
Actually, as seen on their definition, the variables $\mathbf{\pi }^{\left(
\alpha \right) }$ and $\mathbf{r}^{\left( \alpha \right) }$ depend on $%
\mathbf{A}\left( \mathbf{r}^{\left( \alpha \right) }\right) $. Rather than
solving these circular equations where the variables of interest appear on
both side as functions of the canonical variables $\mathbf{R}^{\left( \alpha
\right) }$, $\mathbf{P}^{\left( \alpha \right) }$ we rather unsolved the
dependence of the variables in $\mathbf{A}$ and keep this circularity to
compute the derivatives of interest, since it will allow to express all the
results as functions of the variables $\mathbf{r}^{\left( \alpha \right) }$,
$\mathbf{\pi }^{\left( \alpha \right) }$.

Keeping this in mind, we start by considering a simple example that will
allow to find some general rules for the computation of the functional
derivatives with respect to the gauge potential. Let $G$ be a function
depending on a variable $x$ (in our context $\mathbf{r}^{\left( \alpha
\right) }$, $\mathbf{\pi }^{\left( \alpha \right) }$) depending itself on a
function $a$ of $x$ (here the gauge field) through the relation $x=f\left(
a\left( x\right) \right) $.

We aim at computing first $\frac{\delta }{\delta a\left( y\right) }G\left(
x\right) $ where $y$ is an independent variable (i.e. the space parameter in
our context).

Start first with $\frac{\delta }{\delta a\left( y\right) }x=\frac{\delta }{%
\delta a\left( y\right) }f\left( a\left( x\right) \right) $ :
\begin{equation*}
\frac{\delta }{\delta a\left( y\right) }x=\frac{\delta }{\delta a\left(
y\right) }f\left( a\left( x\right) \right) =\frac{\delta a\left( x\right) }{%
\delta a\left( y\right) }f^{\prime }\left( a\left( x\right) \right)
\end{equation*}
It would be wrong to conclude that $\frac{\delta a\left( x\right) }{\delta
a\left( y\right) }$ reduces to $\delta \left( x-y\right) $ because $y$ is an
independent variables but $x$ is not and depends on $a$. One rather has to
consider that $a\left( x\right) $ is an infinite series of composition $%
a\left( x\right) =a\left( f\left( a\left( x\right) \right) \right) =f\left(
a\left( f\left( a\left( x\right) \right) \right) \right) =f\left( a\left(
f\left( a\left( f\left( a\left( x\right) \right) \right) \right) \right)
\right) $ and so on. A slight functional variation of $a$ propagates along
all the series and has to be taken into account to write symbolically the
infinite series :
\begin{eqnarray*}
\delta a\left( x\right) &=&\delta \left[ a(f(a(f(a(f(a(f(a...\right] =\delta
a(f(a(f(a(f(a(f(a...+a(f(\delta a(f(a(f(a(f(a... \\
&&+a(f(a(f(\delta a(f(a(f(a...+a(f(a(f(a(f(\delta
a(f(a...+a(f(a(f(a(f(a(f(\delta a...+...
\end{eqnarray*}
where the inserted $\delta $ acts solely on the $a$ directly on its right.
The variation $\delta a\left( x\right) $ can be rewritten as :
\begin{equation*}
\delta a\left( x\right) =\delta a\left( x\right) +a(f(\delta a\left(
x\right) )+a(f(a(f(\delta a\left( x\right) )))+a(f(a(f(a(f(\delta a\left(
x\right) ))))+...
\end{equation*}
where now $x$ can be seen as a frozen variable (that is not depending on $a$%
) everywhere in the right hand side. A direct application of the chain rule
yields directly :
\begin{equation*}
\frac{\delta a\left( x\right) }{\delta a\left( y\right) }=\delta \left(
x-y\right) \left[ 1+a^{\prime }\left( x\right) f^{\prime }\left( a\left(
x\right) \right) +\left( a^{\prime }\left( x\right) f^{\prime }\left(
a\left( x\right) \right) \right) ^{2}+...\right]
\end{equation*}
so that :
\begin{equation*}
\frac{\delta }{\delta a\left( y\right) }x=f^{\prime }\left( a\left( x\right)
\right) \frac{\delta a\left( x\right) }{\delta a\left( y\right) }=\delta
\left( x-y\right) f^{\prime }\left( a\left( x\right) \right) \left[
1+a^{\prime }\left( x\right) f^{\prime }\left( a\left( x\right) \right)
+\left( a^{\prime }\left( x\right) f^{\prime }\left( a\left( x\right)
\right) \right) ^{2}+...\right]
\end{equation*}
and ultimately :
\begin{equation*}
\frac{\delta }{\delta a\left( y\right) }G\left( x\right) =\delta \left(
x-y\right) G^{\prime }\left( x\right) f^{\prime }\left( a\left( x\right)
\right) \left[ 1+a^{\prime }\left( x\right) f^{\prime }\left( a\left(
x\right) \right) +\left( a^{\prime }\left( x\right) f^{\prime }\left(
a\left( x\right) \right) \right) ^{2}+...\right]
\end{equation*}
Of course, this expansion is formal, and assumed to converge. Moreover, in
the sequel the sum will always be truncated to a finite order.

We can now translate this results in our context to compute the derivative $%
\mathbf{-}\frac{\delta }{\delta A^{j}\left( \mathbf{x,t}\right) }H_{mat}$
since it involves only a generalization to several variables. The formula
are a bit more involved since the circularity depends on three variables.
Actually $\mathbf{r}^{\left( \alpha \right) }$ depends on $\mathbf{\pi }%
^{\left( \alpha \right) }$ through the Berry phase, $\mathbf{\pi }^{\left(
\alpha \right) }$ depends on $\mathbf{A}$, that depends on $\mathbf{r}%
^{\left( \alpha \right) }$. products like $a^{\prime }\left( x\right)
f^{\prime }\left( a\left( x\right) \right) $ will be now replaced by
products of three types of derivatives, $\frac{d\mathbf{r}^{\left( \alpha
\right) }}{d\mathbf{A}}$, $\frac{d\mathbf{\pi }^{\left( \alpha \right) }}{d%
\mathbf{A}}$, $\frac{\partial \mathbf{A}}{\partial \mathbf{r}^{\left( \alpha
\right) }}=$ $\mathbf{\nabla A}$. However, a simplification arises here.
Actually, since the beginning, we have assumed that the magnetic part of the
interaction between the particles is relatively weak with respect to the
electrostatic potential. We will thus assume that the current is relatively
weak, and the expansion will be performed only at the first order in the
field. As a consequence, at this order of approximation, it is useful for
the sequel to note that the composition series for $\frac{\delta A^{k}\left(
\mathbf{r}^{\left( \alpha \right) }\mathbf{,t}\right) }{\delta A^{i}\left(
\mathbf{x,t}\right) }$\ reduces to $\left[ \delta _{i}^{k}+\nabla
_{l}A^{k}\left( \mathbf{r}^{\left( \alpha \right) }\right) \frac{%
dr_{l}^{\left( \alpha \right) }}{dA^{i}\left( \mathbf{r}^{\left( \alpha
\right) }\right) }\right] $.

Introducing the needed indices, taking into account this approximation and
going to the second order in $\hbar $ we are led to the following results
for the various terms involved in $\mathbf{-}\frac{\delta }{\delta
A^{j}\left( \mathbf{x,t}\right) }H_{mat}$\ :
\begin{eqnarray*}
&&-\sum_{\alpha }\frac{\delta \left( \hat{\varepsilon}_{0N^{\left( \alpha
\right) }}^{\left( \alpha \right) }\left( \mathbf{\pi }^{\left( \alpha
\right) }\right) +D^{\left( \alpha \right) }\right) }{\delta A^{i}\left(
\mathbf{x,t}\right) } \\
&=&-\sum_{\alpha }\frac{\partial }{\partial \pi _{j}^{\left( \alpha \right)
}\left( \mathbf{x,t}\right) }\left( \hat{\varepsilon}_{0N^{\left( \alpha
\right) }}^{\left( \alpha \right) }\left( \mathbf{\pi }^{\left( \alpha
\right) }\right) +D^{\left( \alpha \right) }\right) \left[ \frac{d\pi
_{j}^{\left( \alpha \right) }}{dA^{k}\left( \mathbf{r}^{\left( \alpha
\right) }\right) }\right] \left[ \delta _{i}^{k}+\nabla _{l}A^{k}\left(
\mathbf{r}^{\left( \alpha \right) }\right) \frac{dr_{l}^{\left( \alpha
\right) }}{dA^{i}\left( \mathbf{r}^{\left( \alpha \right) }\right) }\right]
\delta \left( \mathbf{x-r}^{\left( \alpha \right) }\right)
\end{eqnarray*}
and
\begin{equation*}
-\sum_{\alpha }\frac{\delta }{\delta A^{i}\left( \mathbf{x,t}\right) }%
v\left( \mathbf{r}^{\left( \alpha \right) }\right) =-\sum_{\alpha }\frac{%
\partial v\left( \mathbf{r}^{\left( \alpha \right) }\right) }{\partial
r_{j}^{\left( \alpha \right) }}\frac{dr_{j}^{\left( \alpha \right) }}{%
dA^{k}\left( \mathbf{r}^{\left( \alpha \right) }\right) }\left[ \delta
_{i}^{k}+\nabla _{l}A^{k}\left( \mathbf{r}^{\left( \alpha \right) }\right)
\frac{dr_{l}^{\left( \alpha \right) }}{dA^{i}\left( \mathbf{r}^{\left(
\alpha \right) }\right) }\right] \delta \left( \mathbf{x-r}^{\left( \alpha
\right) }\right)
\end{equation*}
and also
\begin{eqnarray*}
&&-\frac{\delta }{\delta A^{i}\left( \mathbf{x,t}\right) }\left[ -\frac{1}{2}%
\sum_{\alpha }\left( \mathbf{\mu }\left( \mathbf{x}^{\left( \alpha \right)
}\right) \mathbf{.B}^{\left( \alpha \right) }\left( \mathbf{x}^{\left(
\alpha \right) }\right) +\mathbf{B}^{\left( \alpha \right) }\left( \mathbf{x}%
^{\left( \alpha \right) }\right) \mathbf{.\mu }\left( \mathbf{x}\right)
\right) -\sum_{\alpha }\mathbf{\hat{\mu}}^{k}\left( \mathbf{x}\right)
\mathbf{.}\nabla _{R_{k}}\mathbf{B}^{\left( \alpha \right) }\left( \mathbf{x}%
^{\left( \alpha \right) }\right) \right. \\
&&\left. +\mathbf{B}^{\left( \alpha \right) }\mathbf{.\tilde{\mu}.B}^{\left(
\alpha \right) }\right] \\
&=&\left[ \frac{1}{2}\sum_{\alpha }\left( \left[ \frac{d\pi _{j}^{\left(
\alpha \right) }}{dA^{i}\left( \mathbf{r}^{\left( \alpha \right) }\right) }%
\frac{\partial }{\partial \pi _{j}^{\left( \alpha \right) }\left( \mathbf{x,t%
}\right) }+\frac{dr_{j}^{\left( \alpha \right) }}{dA^{i}\left( \mathbf{r}%
^{\left( \alpha \right) }\right) }\frac{\partial }{\partial r_{j}^{\left(
\alpha \right) }}\right] \left( \mathbf{\mu }\left( \mathbf{x}^{\left(
\alpha \right) }\right) +\mathbf{\hat{\mu}}^{k}\left( \mathbf{x}\right)
\mathbf{.}\nabla _{R_{k}}\right) \right) .\mathbf{B}^{\left( \alpha \right)
}\left( \mathbf{x}^{\left( \alpha \right) }\right) \right] \\
&&\times \delta \left( \mathbf{x-r}^{\left( \alpha \right) }\right) +H.C. \\
&&+\frac{1}{2}\sum_{\alpha }\left[ \left( \mathbf{\mu }\left( \mathbf{x}%
^{\left( \alpha \right) }\right) \times \mathbf{\nabla }\right) _{p}.\left(
\delta _{i}^{p}\mathbf{+}\nabla _{l}A^{p}\left( \mathbf{r}^{\left( \alpha
\right) }\right) \frac{dr_{l}^{\left( \alpha \right) }}{dA^{i}\left( \mathbf{%
r}^{\left( \alpha \right) }\right) }\right) \delta \left( \mathbf{x-r}%
^{\left( \alpha \right) }\right) \right. \\
&&\left. +\left[ \mathbf{\hat{\mu}}^{k}\left( \mathbf{x}^{\left( \alpha
\right) }\right) \mathbf{\times }\nabla _{R_{k}}\mathbf{\nabla }\delta
\left( \mathbf{x-r}^{\left( \alpha \right) }\right) \right] _{i}\right] +H.C.
\\
&&-\frac{1}{2}\left[ \mathbf{B}\left( \mathbf{x}^{\left( \alpha \right)
}\right) \mathbf{.\tilde{\mu}}\left( \mathbf{x}^{\left( \alpha \right)
}\right) \times \mathbf{\nabla }\delta \left( \mathbf{x-R}^{\left( \alpha
\right) }\right) -\mathbf{\nabla }\delta \left( \mathbf{x-R}^{\left( \alpha
\right) }\right) \times \mathbf{\tilde{\mu}}\left( \mathbf{x}^{\left( \alpha
\right) }\right) \mathbf{.B}\left( \mathbf{x}^{\left( \alpha \right)
}\right) \right] _{i}
\end{eqnarray*}
These expressions involve some derivatives that are computed as follows, as
directly implied by our previous remarks. $\frac{dr_{l}^{\left( \alpha
\right) }}{dA^{i}\left( \mathbf{r}^{\left( \alpha \right) }\right) }$, $%
\frac{d\pi _{j}^{\left( \alpha \right) }}{dA^{k}\left( \mathbf{r}^{\left(
\alpha \right) }\right) }$ cover the dependence of the transformed dynamical
variables in $A^{i}\left( \mathbf{r}^{\left( \alpha \right) }\right) $.
Recall that they have to be computed such that in $A^{i}\left( \mathbf{r}%
^{\left( \alpha \right) }\right) $, the $\mathbf{r}^{\left( \alpha \right) }$
is frozen, so that $A^{i}\left( \mathbf{r}^{\left( \alpha \right) }\right) $
is seen as an ordinary, or independent variable.

Starting with $\frac{dr_{l}^{\left( \alpha \right) }}{dA^{i}\left( \mathbf{r}%
^{\left( \alpha \right) }\right) }$ the dependence in the field comes from
the Berry phase of $r_{l}^{\left( \alpha \right) }$, \textbf{$\tilde{\emph{A}%
}$}$^{R_{l}^{\left( \alpha \right) }}$ which is a function of the two
variables $\mathbf{\pi }^{\left( \alpha \right) }$ and $\mathbf{r}^{\left(
\alpha \right) }$. We can thus write :
\begin{equation*}
\frac{dr_{l}^{\left( \alpha \right) }}{dA^{i}\left( \mathbf{r}^{\left(
\alpha \right) }\right) }=\left( \frac{\partial \mathbf{\tilde{\emph{A}}}%
^{R_{l}^{\left( \alpha \right) }}}{\partial \pi _{n}^{\left( \alpha \right) }%
}\frac{d\pi _{n}^{\left( \alpha \right) }}{dA^{i}\left( \mathbf{r}^{\left(
\alpha \right) }\right) }+\frac{\partial \mathbf{\tilde{\emph{A}}}%
^{R_{l}^{\left( \alpha \right) }}}{\partial r_{n}^{\left( \alpha \right) }}%
\frac{dr_{n}^{\left( \alpha \right) }}{dA^{i}\left( \mathbf{r}^{\left(
\alpha \right) }\right) }+\frac{\partial \mathbf{\tilde{\emph{A}}}%
^{R_{l}^{\left( \alpha \right) }}}{\partial A^{i}\left( \mathbf{r}^{\left(
\alpha \right) }\right) }\right)
\end{equation*}
the second term in the right hand side is of order $\hbar ^{3}$ since $\frac{%
\partial \mathbf{\tilde{\emph{A}}}^{R_{l}^{\left( \alpha \right) }}}{%
\partial r_{n}^{\left( \alpha \right) }}$ is of order $\hbar ^{2}$ and $%
\frac{dr_{n}^{\left( \alpha \right) }}{dA^{i}\left( \mathbf{r}^{\left(
\alpha \right) }\right) }$ is of order $\hbar $. At order $\hbar ^{2}$ one
thus has :
\begin{equation*}
\frac{dr_{l}^{\left( \alpha \right) }}{dA^{i}\left( \mathbf{r}^{\left(
\alpha \right) }\right) }=\frac{\partial \mathbf{\tilde{\emph{A}}}%
^{R_{l}^{\left( \alpha \right) }}}{\partial \pi _{n}^{\left( \alpha \right) }%
}\frac{d\pi _{n}^{\left( \alpha \right) }}{dA^{i}\left( \mathbf{r}^{\left(
\alpha \right) }\right) }+\frac{\partial \mathbf{\tilde{\emph{A}}}%
^{R_{l}^{\left( \alpha \right) }}}{\partial A^{i}\left( \mathbf{r}^{\left(
\alpha \right) }\right) }
\end{equation*}
Practically, the partial derivative $\frac{\partial \mathbf{\tilde{\emph{A}}}%
^{R_{l}^{\left( \alpha \right) }}}{\partial A^{i}\left( \mathbf{r}^{\left(
\alpha \right) }\right) }$ with respect to the field is obtained by
decomposing :
\begin{eqnarray*}
\frac{\partial \mathbf{\tilde{\emph{A}}}^{R_{l}^{\left( \alpha \right) }}}{%
\partial A^{i}\left( \mathbf{r}^{\left( \alpha \right) }\right) } &=&\frac{%
\partial \mathbf{\tilde{\emph{A}}}^{R_{l}^{\left( \alpha \right) }}}{%
\partial A^{i}\left( \mathbf{r}^{\left( \alpha \right) }\right) }-\frac{%
\partial }{\partial A^{i}\left( \mathbf{r}^{\left( \alpha \right) }\right) }%
\left( \emph{A}_{N}^{\mathbf{\Pi }^{\left( \alpha \right) }}.\mathbf{\nabla }%
_{\mathbf{\pi }^{\left( \alpha \right) }}\right) \emph{A}_{N}^{R_{l}^{\left(
\alpha \right) }}\left( \mathbf{r}^{\left( \alpha \right) }\mathbf{,\pi }%
^{\left( \alpha \right) }\right) \\
&=&\frac{\partial \mathbf{\emph{A}}^{R_{l}^{\left( \alpha \right) }}}{%
\partial A^{i}\left( \mathbf{r}^{\left( \alpha \right) }\right) }-\left(
\mathbf{\nabla }_{\mathbf{r}^{\left( \alpha \right) }}\times \mathbf{\nabla }%
_{\mathbf{\pi }^{\left( \alpha \right) }}\right) \emph{A}_{N}^{R_{l}^{\left(
\alpha \right) }}\left( \mathbf{r}^{\left( \alpha \right) }\mathbf{,\pi }%
^{\left( \alpha \right) }\right)
\end{eqnarray*}
and $\frac{\partial \mathbf{\emph{A}}^{R_{l}^{\left( \alpha \right) }}}{%
\partial A^{i}\left( \mathbf{r}^{\left( \alpha \right) }\right) }$ is
carried by the following term computed previously in the one particle case :
\begin{eqnarray*}
\mathcal{A}_{1N}^{\mathbf{R}^{\left( \alpha \right) }} &=&-\frac{\hbar ^{2}}{%
8}\sum_{P}\left\{ e\mathbf{B}.\left( \left( \mathcal{A}_{0}^{\mathbf{R}%
^{\left( \alpha \right) }}\right) _{NP}\times \nabla _{\mathbf{\Pi }^{\left(
\alpha \right) }}\right) \left( \mathcal{A}_{0}^{\mathbf{R}^{\left( \alpha
\right) }}\right) _{PN}\right\} \\
&&-e\frac{\hbar ^{2}}{4}\left( \mathbf{B}.\left( \left( \mathcal{A}_{0}^{%
\mathbf{R}^{\left( \alpha \right) }}\right) _{NN}\times \nabla _{\mathbf{\Pi
}^{\left( \alpha \right) }}\right) \mathcal{A}_{0NN}^{\mathbf{R}^{\left(
\alpha \right) }}+H.C.\right) \\
&&-\frac{\hbar ^{2}}{2}\left( \sum_{P}\left( e\mathbf{B}.\left( \left(
\mathcal{A}_{0}^{\mathbf{R}^{\left( \alpha \right) }}\right) _{NP}\times
\nabla _{\mathbf{\Pi }^{\left( \alpha \right) }}\right) \right) \frac{\left(
\hat{\varepsilon}_{0P}+\hat{\varepsilon}_{0N}\right) }{\hat{\varepsilon}%
_{0P}-\hat{\varepsilon}_{0N}}\hat{\delta}_{PN}\left( \mathcal{A}_{0}^{%
\mathbf{R}^{\left( \alpha \right) }}\right) _{PN}-H.C.\right) + \\
&&\frac{ie\hbar ^{2}}{8}\sum_{P}\left( \left( \mathcal{A}_{0}^{\mathbf{R}%
^{\left( \alpha \right) }}\right) _{NP}\left( \left( \mathcal{A}_{0}^{%
\mathbf{R}^{\left( \alpha \right) }}\right) _{PN}\times \left( \mathcal{A}%
_{0}^{\mathbf{R}^{\left( \alpha \right) }}\right) _{NN}\right) -\left(
\left( \mathcal{A}_{0}^{\mathbf{R}^{\left( \alpha \right) }}\right)
_{NP}\times \left( \mathcal{A}_{0}^{\mathbf{R}^{\left( \alpha \right)
}}\right) _{PP}\left( \mathcal{A}_{0}^{\mathbf{R}^{\left( \alpha \right)
}}\right) _{PN}\right) .\mathbf{B}\right) \hat{\delta}_{PN} \\
&&+H.C.
\end{eqnarray*}
$\frac{\partial }{\partial \mathbf{A}\left( \mathbf{x}\right) }\mathcal{A}%
_{1N}^{\mathbf{R}^{\left( \alpha \right) }}$ is a two tensor whose
components $\left( i,l\right) $ is given by : $\frac{\partial \mathbf{\emph{A%
}}^{R_{l}^{\left( \alpha \right) }}}{\partial A^{i}\left( \mathbf{r}^{\left(
\alpha \right) }\right) }=\varepsilon ^{ijk}X_{i}^{l\left( \alpha \right)
}\nabla _{k}$
\begin{eqnarray*}
X_{i}^{l\left( \alpha \right) } &=&-\frac{\hbar ^{2}}{8}\sum_{P}\left\{
e\left( \left( \mathcal{A}_{0}^{\mathbf{R}^{\left( \alpha \right) }}\right)
_{NP}\times \nabla _{\mathbf{\Pi }}\right) _{j}\left( \mathcal{A}%
_{0}^{R_{l}^{\left( \alpha \right) }}\right) _{PN}\right\} -e\frac{\hbar ^{2}%
}{4}\left( \left( \left( \mathcal{A}_{0}^{\mathbf{R}^{\left( \alpha \right)
}}\right) _{NN}\times \nabla _{\mathbf{\Pi }}\right) _{j}\mathcal{A}%
_{0NN}^{R_{l}^{\left( \alpha \right) }}+H.C.\right) \\
&&-\frac{\hbar ^{2}}{2}\left( \sum_{P}e\left( \left( \mathcal{A}_{0}^{%
\mathbf{R}^{\left( \alpha \right) }}\right) _{NP}\times \nabla _{\mathbf{\Pi
}}\right) _{j}\frac{\left( \hat{\varepsilon}_{0P}+\hat{\varepsilon}%
_{0N}\right) }{\hat{\varepsilon}_{0P}-\hat{\varepsilon}_{0N}}\hat{\delta}%
_{PN}\left( \mathcal{A}_{0}^{R_{l}^{\left( \alpha \right) }}\right)
_{PN}-H.C.\right) \\
&&+i\frac{e\hbar ^{2}}{8}\sum_{P}\left( \left( \mathcal{A}%
_{0}^{R_{l}^{\left( \alpha \right) }}\right) _{NP}\left( \left( \mathcal{A}%
_{0}^{\mathbf{R}^{\left( \alpha \right) }}\right) _{PN}\times \left(
\mathcal{A}_{0}^{\mathbf{R}^{\left( \alpha \right) }}\right) _{NN}\right)
_{j}\right. \\
&&\left. -\left( \left( \mathcal{A}_{0}^{\mathbf{R}^{\left( \alpha \right)
}}\right) _{NP}\times \left( \mathcal{A}_{0}^{\mathbf{R}^{\left( \alpha
\right) }}\right) _{PP}\left( \mathcal{A}_{0}^{R_{l}^{\left( \alpha \right)
}}\right) _{PN}\right) _{j}\right) \hat{\delta}_{PN}+H.C.
\end{eqnarray*}
and the gradient acts on $\delta \left( \mathbf{x-r}^{\left( \alpha \right)
}\right) $.

Having given some expanded formula for $\frac{dr_{l}^{\left( \alpha \right) }%
}{dA^{i}\left( \mathbf{r}^{\left( \alpha \right) }\right) }$ it thus remains
to compute $\frac{d\pi _{n}^{\left( \alpha \right) }}{dA^{i}\left( \mathbf{r}%
^{\left( \alpha \right) }\right) }$. To do so, recall again that the
variable $\mathbf{r}^{\left( \alpha \right) }$ is \textquotedblright
frozen\textquotedblright\ with respect to $\mathbf{A}$, which is thus
thought as an independent variable. We use also our previous result on the
Berry phase :
\begin{eqnarray*}
\pi _{m}^{\left( \alpha \right) } &=&K_{m}^{\left( \alpha \right) }-e\mathbf{%
A}\left( \mathbf{r}^{\left( \alpha \right) }\right) +\frac{1}{2}e\left[
\nabla _{m}A_{l}\left( \mathbf{R}^{\left( \alpha \right) }\right) \tilde{%
\emph{\ A}}^{R_{l}}+\tilde{\emph{A}}^{R_{l}}\nabla _{m}A_{l}\left( \mathbf{R}%
^{\left( \alpha \right) }\right) \right] \\
&&+e\nabla _{m}\nabla _{R_{n}}A_{r}\left( \mathbf{r}^{\left( \alpha \right)
}\right) \tilde{\emph{A}}_{nr} \\
\text{with }\mathbf{\tilde{\emph{A}}}_{n,r} &=&-\frac{\hbar ^{2}}{4}\sum_{P}%
\left[ \left( \mathcal{A}_{0}^{R_{n}^{\left( \alpha \right) }}\right)
_{PN}\left( \mathcal{A}_{0}^{R_{r}^{\left( \alpha \right) }}\right)
_{NP}+\left( \mathcal{A}_{0}^{R_{r}^{\left( \alpha \right) }}\right)
_{PN}\left( \mathcal{A}_{0}^{R_{n}^{\left( \alpha \right) }}\right) _{NP}%
\right]
\end{eqnarray*}
Define also $\mathbf{\emph{A}}_{m,i}=e\mathbf{\frac{d\tilde{\emph{A}}_{n,r}}{%
dA^{i}\left( \mathbf{r}^{\left( \alpha \right) }\right) }}\nabla _{\mathbf{r}%
_{n}^{\left( \alpha \right) }}\nabla _{\mathbf{r}_{m}^{\left( \alpha \right)
}}A_{r}\left( \mathbf{r}^{\left( \alpha \right) }\right) +e$\textbf{$\tilde{%
\emph{\ A}}$}$_{n,i}\nabla _{\mathbf{r}_{n}^{\left( \alpha \right) }}\nabla
_{\mathbf{r}_{m}^{\left( \alpha \right) }}$ so that one has :
\begin{eqnarray*}
-\frac{1}{e}\frac{d\pi _{m}^{\left( \alpha \right) }}{dA^{i}\left( \mathbf{r}%
^{\left( \alpha \right) }\right) } &=&\delta _{i}^{m}-\tilde{\emph{A}}%
\mathbf{^{R_{i}^{\left( \alpha \right) }}}\nabla _{\mathbf{r}_{m}^{\left(
\alpha \right) }}-\nabla _{\mathbf{r}_{m}^{\left( \alpha \right)
}}A^{n}\left( \mathbf{r}^{\left( \alpha \right) }\right) \frac{d\mathbf{%
\tilde{\emph{A}}^{R_{n}^{\left( \alpha \right) }}}}{dA^{i}\left( \mathbf{r}%
^{\left( \alpha \right) }\right) }\mathbf{+\emph{A}}_{m,i} \\
&=&\delta _{i}^{m}-\tilde{\emph{A}}\mathbf{^{R_{i}^{\left( \alpha \right) }}}%
\nabla _{\mathbf{r}_{m}^{\left( \alpha \right) }}-\nabla _{\mathbf{r}%
_{m}^{\left( \alpha \right) }}A^{n}\left( \mathbf{r}^{\left( \alpha \right)
}\right) \frac{dr_{n}^{\left( \alpha \right) }}{dA^{i}\left( \mathbf{r}%
^{\left( \alpha \right) }\right) }+\mathbf{\emph{A}}_{m,i} \\
&=&\delta _{i}^{m}-\tilde{\emph{A}}\mathbf{^{R_{i}^{\left( \alpha \right) }}}%
\nabla _{\mathbf{r}_{m}^{\left( \alpha \right) }}+\mathbf{\emph{A}}_{m,i} \\
&&-\nabla _{\mathbf{r}_{m}^{\left( \alpha \right) }}A^{n}\left( \mathbf{r}%
^{\left( \alpha \right) }\right) \times \left( \frac{\partial \tilde{\emph{A}%
}^{R_{n}^{\left( \alpha \right) }}}{\partial \pi _{r}^{\left( \alpha \right)
}}\frac{d\pi _{r}^{\left( \alpha \right) }}{dA^{i}\left( \mathbf{r}^{\left(
\alpha \right) }\right) }+\frac{\partial \tilde{\emph{A}}^{R_{n}^{\left(
\alpha \right) }}}{\partial r_{r}^{\left( \alpha \right) }}\frac{\partial
\tilde{\emph{A}}^{R_{r}^{\left( \alpha \right) }}}{\partial \pi _{q}^{\left(
\alpha \right) }}\frac{d\pi _{q}^{\left( \alpha \right) }}{dA^{i}\left(
\mathbf{r}^{\left( \alpha \right) }\right) }+\frac{\partial \tilde{\emph{A}}%
^{R_{n}^{\left( \alpha \right) }}}{\partial A^{i}\left( \mathbf{r}^{\left(
\alpha \right) }\right) }\right) \\
&=&\delta _{i}^{m}-\tilde{\emph{A}}\mathbf{^{R_{i}^{\left( \alpha \right) }}}%
\nabla _{\mathbf{r}_{m}^{\left( \alpha \right) }}-\nabla _{\mathbf{r}%
_{m}^{\left( \alpha \right) }}A^{n}\left( \mathbf{r}^{\left( \alpha \right)
}\right) \left( \frac{\partial \tilde{\emph{A}}^{R_{n}^{\left( \alpha
\right) }}}{\partial \pi _{r}^{\left( \alpha \right) }}\frac{d\pi
_{r}^{\left( \alpha \right) }}{dA^{i}\left( \mathbf{r}^{\left( \alpha
\right) }\right) }+\frac{\partial \tilde{\emph{A}}^{R_{n}^{\left( \alpha
\right) }}}{\partial A^{i}\left( \mathbf{r}^{\left( \alpha \right) }\right) }%
\right) +\mathbf{\emph{A}}_{m,i}
\end{eqnarray*}
to our order of approximation.

We can rewrite the last equation as :
\begin{equation*}
\frac{d\pi _{u}^{\left( \alpha \right) }}{dA^{i}\left( \mathbf{r}^{\left(
\alpha \right) }\right) }\left[ \delta _{u}^{m}-e\nabla _{\mathbf{r}%
_{m}^{\left( \alpha \right) }}A^{n}\left( \mathbf{r}^{\left( \alpha \right)
}\right) \frac{\partial \tilde{\emph{A}}^{R_{n}^{\left( \alpha \right) }}}{%
\partial \pi _{u}^{\left( \alpha \right) }}\right] =-e\left( \delta _{i}^{m}-%
\tilde{\emph{A}}\mathbf{^{R_{i}^{\left( \alpha \right) }}}\nabla _{\mathbf{r}%
_{m}^{\left( \alpha \right) }}-\nabla _{\mathbf{r}_{m}^{\left( \alpha
\right) }}A^{n}\left( \mathbf{r}^{\left( \alpha \right) }\right) \frac{%
\partial \tilde{\emph{A}}^{R_{n}^{\left( \alpha \right) }}}{\partial
A^{i}\left( \mathbf{r}^{\left( \alpha \right) }\right) }+\mathbf{\emph{A}}%
_{m,i}\right)
\end{equation*}
which is solved to the second order in $\hbar $ and first order in field by
:
\begin{eqnarray*}
-\frac{1}{e}\frac{d\pi _{m}^{\left( \alpha \right) }}{dA^{i}\left( \mathbf{r}%
^{\left( \alpha \right) }\right) } &=&\delta _{i}^{m}-\tilde{\emph{A}}%
\mathbf{^{R_{i}^{\left( \alpha \right) }}}\nabla _{\mathbf{r}_{m}^{\left(
\alpha \right) }}-\nabla _{\mathbf{r}_{m}^{\left( \alpha \right)
}}A^{n}\left( \mathbf{r}^{\left( \alpha \right) }\right) \frac{\partial
\tilde{\emph{A}}^{R_{n}^{\left( \alpha \right) }}}{\partial A^{i}\left(
\mathbf{r}^{\left( \alpha \right) }\right) } \\
&&+e\nabla _{\mathbf{r}_{m}^{\left( \alpha \right) }}A^{n}\left( \mathbf{r}%
^{\left( \alpha \right) }\right) \frac{\partial \tilde{\emph{A}}%
^{R_{n}^{\left( \alpha \right) }}}{\partial \pi _{u}^{\left( \alpha \right) }%
}\left( \delta _{i}^{u}-\tilde{\emph{\ A}}\mathbf{^{R_{i}^{\left( \alpha
\right) }}}\nabla _{\mathbf{r}_{u}^{\left( \alpha \right) }}\right) +\mathbf{%
\emph{A}}_{m,i} \\
&=&\delta _{i}^{m}-\tilde{\emph{A}}\mathbf{^{R_{i}^{\left( \alpha \right) }}}%
\nabla _{\mathbf{r}_{m}^{\left( \alpha \right) }}-\nabla _{\mathbf{r}%
_{m}^{\left( \alpha \right) }}A^{n}\left( \mathbf{r}^{\left( \alpha \right)
}\right) \frac{\partial \tilde{\emph{A}}^{R_{n}^{\left( \alpha \right) }}}{%
\partial A^{i}\left( \mathbf{r}^{\left( \alpha \right) }\right) } \\
&&+e\nabla _{\mathbf{r}_{m}^{\left( \alpha \right) }}A^{n}\left( \mathbf{r}%
^{\left( \alpha \right) }\right) \left( \frac{\partial \tilde{\emph{\ A}}%
^{R_{n}^{\left( \alpha \right) }}}{\partial \pi _{i}^{\left( \alpha \right) }%
}-\frac{\partial \tilde{\emph{A}}^{R_{n}^{\left( \alpha \right) }}}{\partial
\pi _{u}^{\left( \alpha \right) }}\tilde{\emph{A}}\mathbf{^{R_{i}^{\left(
\alpha \right) }}}\nabla _{\mathbf{r}_{u}^{\left( \alpha \right) }}\right) +%
\mathbf{\emph{A}}_{m,i}
\end{eqnarray*}
the gradient appearing at the end of the right hand side has to be
understood as acting on the delta function $\delta \left( \mathbf{x-r}%
^{\left( \alpha \right) }\right) $ in the expression for $-\sum_{\alpha }%
\frac{\delta }{\delta A^{i}\left( \mathbf{x,t}\right) }\left( \hat{%
\varepsilon}_{0N^{\left( \alpha \right) }}^{\left( \alpha \right) }\left(
\mathbf{\pi }^{\left( \alpha \right) }\right) +D^{\left( \alpha \right)
}\right) $.

\subsubsection{Solution for the Electromagnetic Field}

With the current at hand, it is now possible to find the expression for the
electromagnetic field as a function of the particles dynamical variables.
However, note that in the expression for $\mathbf{j}\left( \mathbf{x}\right)
$ the electromagnetic field appears everywhere through the Berry phases and
explicitly through the magnetic field. This means that in fact reinserting $%
\mathbf{j}\left( \mathbf{x}\right) $ in the expression for $\mathbf{A}\left(
\mathbf{x}\right) $ allows only to compute this last potential
perturbatively. Assuming, as before that the current is relatively weak, the
right hand side can be expanded at the first order in the field. The
expressions depending on the field can be put to the left of the Maxwell
equation. which can be now written :
\begin{equation*}
\left[ \square +\delta O\left( \mathbf{x}\right) \right] \mathbf{A}\left(
\mathbf{x,}t\right) =\int d^{3}y\left[ \delta ^{ij}\delta ^{3}\left( \mathbf{%
x-y}\right) +\frac{\partial ^{2}}{\partial x^{i}\partial x^{j}}\frac{1}{4\pi
\left\vert \mathbf{x-y}\right\vert }\right] \mathbf{\hat{\jmath}}\left(
\mathbf{y,}t\right) _{\mid \mathbf{A}=0}
\end{equation*}
with :
\begin{eqnarray*}
\hat{\jmath}_{i}\left( \mathbf{x}\right) &=&-\left( \frac{\delta }{\delta
\mathbf{A}\left( \mathbf{x}\right) }H_{mat}\right) _{\mid \mathbf{A}=0} \\
&=&e\sum_{\alpha }\frac{\partial \left( \hat{\varepsilon}_{0N^{\left( \alpha
\right) }}^{\left( \alpha \right) }\left( \mathbf{\pi }^{\left( \alpha
\right) }\right) +D^{\left( \alpha \right) }\right) }{\partial \pi
_{j}^{\left( \alpha \right) }\left( \mathbf{x,t}\right) }\left[ \left(
\delta _{i}^{j}-\tilde{\emph{A}}\mathbf{^{R_{i}^{\left( \alpha \right) }}}%
\nabla _{\mathbf{r}_{j}^{\left( \alpha \right) }}+\mathbf{\emph{A}}_{j,i\mid
\mathbf{A}=0}\right) \right] _{\mid \mathbf{A}=0}\delta \left( \mathbf{x-r}%
^{\left( \alpha \right) }\right) \\
&&-\sum_{\alpha }\frac{\partial v\left( \mathbf{r}^{\left( \alpha \right)
}\right) }{\partial r_{j}^{\left( \alpha \right) }}\left[ \frac{\partial
\tilde{\emph{A}}^{R_{j}^{\left( \alpha \right) }}}{\partial \pi _{n}^{\left(
\alpha \right) }}\left( \delta _{i}^{n}-\tilde{\emph{A}}\mathbf{%
^{R_{i}^{\left( \alpha \right) }}}\nabla _{\mathbf{r}_{n}^{\left( \alpha
\right) }}+\mathbf{\emph{A}}_{n,i\mid \mathbf{A}=0}\right) +\frac{\partial
\tilde{\emph{A}}^{R_{j}^{\left( \alpha \right) }}}{\partial A^{i}\left(
\mathbf{r}^{\left( \alpha \right) }\right) }\right] _{\mid \mathbf{A}%
=0}\delta \left( \mathbf{x-r}^{\left( \alpha \right) }\right) \\
&&-\sum_{\alpha }\left[ \left( \mathbf{\mu }\left( \mathbf{x}^{\left( \alpha
\right) }\right) \times \mathbf{\nabla }\right) _{i}\delta \left( \mathbf{x-r%
}^{\left( \alpha \right) }\right) +\left[ \mathbf{\hat{\mu}}^{k}\left(
\mathbf{x}^{\left( \alpha \right) }\right) \mathbf{\times }\nabla _{R_{k}}%
\mathbf{\nabla }\delta \left( \mathbf{x-r}^{\left( \alpha \right) }\right) %
\right] _{i}\right] _{\mid \mathbf{A}=0}
\end{eqnarray*}
The operator $\delta O\left( \mathbf{x}\right) $ will in fact yields
negligible contributions in our applications. We give nevertheless its form
in the appendix for the sake of completeness.\ Note that the correction $%
\delta O\left( \mathbf{x}\right) $ to the Dalembertian are of order $\hbar
^{2}$ due to the definition of the terms involved.

As a consequence, we can thus write for the electromagnetic field :
\begin{equation*}
A^{i}\mathbf{\left( \mathbf{x},t\right) =}\int G_{ij}\left( \mathbf{x-y,t-t}%
^{\prime }\right) \hat{\jmath}^{j}\mathbf{\left( \mathbf{y},t^{\prime
}\right) }dy^{3}dt^{\prime }
\end{equation*}
with :
\begin{eqnarray*}
&&G_{ij}\left( \mathbf{x-y,t-t}^{\prime }\right) \\
&\mathbf{=}&\int \left[ \frac{1}{\frac{\partial ^{2}}{\partial t^{2}}-%
\mathbf{\nabla }_{\mathbf{x}}^{2}-\delta O\left( \mathbf{x}\right) }\left(
\mathbf{x-z,t-t}^{\prime }\right) \left[ \delta ^{ij}\delta ^{3}\left(
\mathbf{z-y}\right) +\frac{\partial ^{2}}{\partial x^{i}\partial x^{j}}\frac{%
1}{4\pi \left\vert \mathbf{z-y}\right\vert }\right] \right] _{ij}dz
\end{eqnarray*}
A more tractable for the effective Green function $G_{ij}\left( \mathbf{%
x-y,t-t}^{\prime }\right) $ can be found if we move to the Fourier
transform. Actually, one can write :
\begin{eqnarray*}
G_{ij}\left( \mathbf{x-y,t-t}^{\prime }\right) &=&\int \exp \left( i\mathbf{%
p.}\left( \mathbf{x-y}\right) -i\omega \left( t-t^{\prime }\right) \right)
\left( \delta _{ir}-\frac{p_{i}p_{r}}{\mathbf{p}^{2}}\right) \left( \frac{1}{%
\omega ^{2}-\left( \mathbf{p}^{2}-\delta O\left( \mathbf{p}\right) \right) }%
\right) _{rj}d\mathbf{p}d\omega \\
&\simeq &\int \exp \left( i\mathbf{p.}\left( \mathbf{x-y}\right) -i\omega
\left( t-t^{\prime }\right) \right) \frac{\delta _{ir}-\frac{p_{i}p_{r}}{%
\mathbf{p}^{2}}}{\omega ^{2}-\mathbf{p}^{2}}\left( \delta _{rj}+\frac{1}{2}%
\frac{\delta O_{rj}\left( \mathbf{p}\right) }{\omega ^{2}-\mathbf{p}^{2}}%
\right) d\mathbf{p}d\omega \\
&\equiv &\int \exp \left( i\mathbf{p.}\left( \mathbf{x-y}\right) -i\omega
\left( t-t^{\prime }\right) \right) G_{ij}\left( \mathbf{p,}\omega \right) d%
\mathbf{p}d\omega
\end{eqnarray*}
and the operators $G^{\left( 0\right) }$ $\delta O,\delta G$ are given in
Fourier components by :
\begin{eqnarray*}
G_{ij}^{\left( 0\right) }\left( \mathbf{p,}\omega \right) &=&\frac{\delta
_{ij}-\frac{p_{i}p_{j}}{\mathbf{p}^{2}}}{\omega ^{2}-\mathbf{p}^{2}} \\
\delta O_{ij}\left( \mathbf{p}\right) &=&\frac{\delta _{ir}-\frac{p_{i}p_{r}%
}{\mathbf{p}^{2}}}{\omega ^{2}-\mathbf{p}^{2}}\delta O_{1rj}\left( \mathbf{p}%
\right) \\
G_{ij}\left( \mathbf{p,}\omega \right) &=&G_{ir}^{\left( 0\right) }\left(
\mathbf{p,}\omega \right) \left( \delta _{rj}+\frac{1}{2}\frac{\delta
O_{rj}\left( \mathbf{p}\right) }{\omega ^{2}-\mathbf{p}^{2}}\right)
\end{eqnarray*}
and the Fourier transform $\delta O_{1}\left( \mathbf{p}\right) $ is given
in the appendix.

Now that we have given the effective photon propagator, note that we can
decompose the vector potential at order $\hbar ^{2}$ at a point $\mathbf{x}$
:
\begin{eqnarray*}
A_{i}\left( \mathbf{x},t\right) &=&\int G_{ij}\left( \mathbf{x-y,t-t}%
^{\prime }\right) \jmath _{1}^{j}\mathbf{\left( \mathbf{y},t^{\prime
}\right) }dy^{3}dt^{\prime } \\
&&-\int dt^{\prime }\sum_{\alpha }\left[ \mu _{l}\left( \mathbf{x}^{\left(
\alpha \right) }\right) +\hat{\mu}_{l}^{k}\left( \mathbf{x}^{\left( \alpha
\right) }\right) \nabla _{R_{k}}\right] \varepsilon ^{lmj}\nabla
_{m}G_{ij}\left( \mathbf{x-x}^{\left( \alpha \right) }\left( t^{\prime
}\right) ,t-t^{\prime }\right) dt^{\prime } \\
&=&\int G_{ij}\left( \mathbf{x-y,t-t}^{\prime }\right) \jmath _{1}^{j}%
\mathbf{\left( \mathbf{y},t^{\prime }\right) }dy^{3}dt^{\prime }-\int
dt^{\prime }\mathcal{M}_{l}^{ef}\varepsilon ^{lmj}\nabla _{m}^{\mathbf{x}%
}G_{ij}\left( \mathbf{x-x}^{\left( \alpha \right) }\left( t^{\prime }\right)
,t-t^{\prime }\right)
\end{eqnarray*}
where $\mathcal{M}_{l}^{ef}$ is the effective magnetization differential
operator :
\begin{equation*}
\mathcal{M}_{m}^{ef}=\sum_{\alpha }\left[ \mu _{m}\left( \mathbf{x}^{\left(
\alpha \right) }\right) +\hat{\mu}_{m}^{k}\left( \mathbf{x}^{\left( \alpha
\right) }\right) \nabla _{R_{k}}\right] _{\mid \mathbf{A}=0}
\end{equation*}
and the current $\mathbf{\jmath }_{1}$ is given by :
\begin{eqnarray*}
\mathbf{\jmath }_{1} &=&e\sum_{\alpha }\frac{\partial \left( \hat{\varepsilon%
}_{0N^{\left( \alpha \right) }}^{\left( \alpha \right) }\left( \mathbf{\pi }%
^{\left( \alpha \right) }\right) +D^{\left( \alpha \right) }\right) }{%
\partial \pi _{j}^{\left( \alpha \right) }\left( \mathbf{x,t}\right) }\left[
\left( \delta _{i}^{j}-\tilde{\emph{A}}\mathbf{^{R_{i}^{\left( \alpha
\right) }}}\nabla _{\mathbf{r}_{j}^{\left( \alpha \right) }}+\mathbf{\emph{A}%
}_{j,i\mid \mathbf{A}=0}\right) \right] \delta \left( \mathbf{x-r}^{\left(
\alpha \right) }\right) \\
&&-\sum_{\alpha }\frac{\partial v\left( \mathbf{r}^{\left( \alpha \right)
}\right) }{\partial r_{j}^{\left( \alpha \right) }}\left[ \frac{\partial
\tilde{\emph{A}}^{R_{j}^{\left( \alpha \right) }}}{\partial \pi _{n}^{\left(
\alpha \right) }}\left( \delta _{i}^{n}-\tilde{\emph{A}}\mathbf{%
^{R_{i}^{\left( \alpha \right) }}}\nabla _{\mathbf{r}_{n}^{\left( \alpha
\right) }}+\mathbf{\emph{A}}_{n,i\mid \mathbf{A}=0}\right) +\frac{\partial
\tilde{\emph{A}}^{R_{j}^{\left( \alpha \right) }}}{\partial A^{i}\left(
\mathbf{r}^{\left( \alpha \right) }\right) }\right] \delta \left( \mathbf{x-r%
}^{\left( \alpha \right) }\right)
\end{eqnarray*}
In $\nabla _{m}^{\mathbf{x}}$ recalls that the gradient is taken with
respect to the variable $\mathbf{x}$.

\subsubsection{Electromagnetic part of the Hamiltonian}

We can now compute the Hamiltonian for the electromagnetic field as :
\begin{eqnarray*}
\frac{1}{2}\int d^{3}x\Pi _{em}^{2}+\frac{1}{2}\int d^{3}x\left( \mathbf{%
\nabla \times A}\right) ^{2} &=&\frac{1}{2}\int dy^{3}dx^{3}A^{i}\mathbf{%
\left( \mathbf{x},t\right) }\left[ \delta ^{ij}\left( \frac{\partial ^{2}}{%
\partial t^{2}}-\mathbf{\nabla }^{2}\right) +\frac{\partial ^{2}}{\partial
x^{i}\partial x^{j}}\right] A^{j}\mathbf{\left( \mathbf{x},t\right) } \\
&=&\frac{1}{2}\int dy^{3}dz^{3}dt^{\prime }dt^{\prime \prime }dx^{3}\hat{%
\jmath}^{l}\mathbf{\left( \mathbf{y},t^{\prime }\right) }G_{il}\left(
\mathbf{x-y,}t-t^{\prime }\right) \\
&&\times \left[ \delta ^{ij}\left( \frac{\partial ^{2}}{\partial t^{2}}-%
\mathbf{\nabla }^{2}\right) +\frac{\partial ^{2}}{\partial x^{i}\partial
x^{j}}\right] G_{jm}\left( \mathbf{x-z,}t-t^{\prime \prime }\right) \hat{%
\jmath}^{m}\mathbf{\left( \mathbf{z},t^{\prime \prime }\right) } \\
&\equiv &\frac{1}{2}\int dy^{3}dx^{3}dt^{\prime }dt\hat{\jmath}^{i}\mathbf{%
\left( \mathbf{x},t\right) }\hat{G}\left( \mathbf{x-y,}t-t^{\prime }\right)
\hat{\jmath}^{j}\mathbf{\left( \mathbf{y},t^{\prime }\right) }
\end{eqnarray*}
where the effective propagator $\hat{G}\left( \mathbf{x-y,}t-t^{\prime
}\right) $ is given through its Fourier transform :
\begin{equation*}
\hat{G}\left( \mathbf{x-y,}t-t^{\prime }\right) =\int \exp \left( i\mathbf{p.%
}\left( \mathbf{x-y}\right) -i\omega \left( t-t^{\prime }\right) \right)
\hat{G}_{ij}\left( \mathbf{p,}\omega \right) d\mathbf{p}d\omega
\end{equation*}
with :
\begin{eqnarray*}
\hat{G}_{ij}\left( \mathbf{p,}\omega \right) &=&G_{il}\left( \mathbf{p,}%
\omega \right) \left[ \left( \delta _{lm}\left( \mathbf{p}^{2}-\omega
^{2}\right) -p_{l}p_{m}\right) \right] G_{mj}\left( \mathbf{p,}\omega \right)
\\
\hat{G}_{ij}\left( \mathbf{p,}\omega \right) &=&\left( \delta _{ir}+\frac{1}{%
2}\frac{\delta O_{ir}\left( \mathbf{p}\right) }{\mathbf{p}^{2}-\omega ^{2}}%
\right) \frac{\delta _{rl}-\frac{p_{r}p_{l}}{\mathbf{p}^{2}}}{\mathbf{p}%
^{2}-\omega ^{2}}\left( \delta _{lm}\left( \mathbf{p}^{2}-\omega ^{2}\right)
-p_{l}p_{m}\right) \frac{\delta _{ms}-\frac{p_{m}p_{s}}{\mathbf{p}^{2}}}{%
\mathbf{p}^{2}-\omega ^{2}}\left( \delta _{sj}+\frac{1}{2}\frac{\delta
O_{sj}\left( \mathbf{p}\right) }{\mathbf{p}^{2}-\omega ^{2}}\right) \\
&&\frac{\delta _{ij}-\frac{p_{i}p_{j}}{\mathbf{p}^{2}}}{\mathbf{p}%
^{2}-\omega ^{2}}+\frac{1}{2}\frac{\delta O_{ir}\left( \mathbf{p}\right) }{%
\mathbf{p}^{2}-\omega ^{2}}\frac{\delta _{rj}-\frac{p_{r}p_{j}}{\mathbf{p}%
^{2}}}{\mathbf{p}^{2}-\omega ^{2}}+\frac{1}{2}\frac{\delta _{is}-\frac{%
p_{i}ps}{\mathbf{p}^{2}}}{\mathbf{p}^{2}-\omega ^{2}}\frac{\delta
O_{sj}\left( \mathbf{p}\right) }{\mathbf{p}^{2}-\omega ^{2}}
\end{eqnarray*}
If we neglect the retardation effect, which is legitimate since we work in
the non relativistic limit, we can discard the dependence in $\omega $, so
that :
\begin{equation*}
\hat{G}_{ij}\left( \mathbf{p}\right) =\frac{\delta _{ij}-\frac{p_{i}p_{j}}{%
\mathbf{p}^{2}}}{\mathbf{p}^{2}}+\frac{1}{2}\frac{\delta O_{ir}\left(
\mathbf{p}\right) }{\mathbf{p}^{2}}\frac{\delta _{rj}-\frac{p_{r}p_{j}}{%
\mathbf{p}^{2}}}{\mathbf{p}^{2}}+\frac{1}{2}\frac{\delta _{is}-\frac{p_{i}ps%
}{\mathbf{p}^{2}}}{\mathbf{p}^{2}}\frac{\delta O_{sj}\left( \mathbf{p}%
\right) }{\mathbf{p}^{2}}
\end{equation*}
and :
\begin{equation*}
\frac{1}{2}\int d^{3}x\Pi _{em}^{2}+\frac{1}{2}\int d^{3}x\left( \mathbf{%
\nabla \times A}\right) ^{2}=\frac{1}{2}\int dy^{3}dx^{3}\hat{\jmath}^{i}%
\mathbf{\left( \mathbf{x},t\right) }\hat{G}\left( \mathbf{x-y}\right) \hat{%
\jmath}^{j}\mathbf{\left( \mathbf{y},t\right) }
\end{equation*}

\subsubsection{Electric potential}

Now, the case of the electrostatic potential is a bit peculiar. As we said
before, we replace it through the potential equation in the Coulomb gauge : $%
-\mathbf{\nabla }^{2}A_{0}=J_{0}$, so that

$\int d^{3}xeA_{0}\left( \mathbf{x}\right) J_{0}\left( \mathbf{x}\right) =%
\frac{1}{2}\int d^{3}x\int d^{3}yJ_{0}\left( \mathbf{x}\right) \frac{1}{4\pi
\left| \mathbf{x-y}\right| }J_{0}\left( \mathbf{y}\right) =\int d^{3}x\left(
\mathbf{\nabla }A_{0}\right) ^{2}$

Considering $v\left( \mathbf{r}^{\left( \alpha \right) }\right) $ alone, one
has first :
\begin{eqnarray*}
&&v\left( \mathbf{r}^{\left( \alpha \right) }\right) -\frac{1}{2}\int
d^{3}x\left( \mathbf{\nabla }A_{0}\right) ^{2} \\
&=&\int d^{3}\mathbf{x}eA_{0}\left( \mathbf{x}\right) e\delta \left( \mathbf{%
x-r}^{\left( \alpha \right) }\right) -\frac{1}{2}\int d^{3}x\left( \mathbf{%
\nabla }A_{0}\right) ^{2} \\
&=&\frac{1}{2}\int d^{3}\mathbf{x}eA_{0}\left( \mathbf{x}\right) e\delta
\left( \mathbf{x-r}^{\left( \alpha \right) }\right) \\
&=&\frac{1}{2}\int d^{3}x\int d^{3}yJ_{0}\left( \mathbf{x}\right) \frac{1}{%
4\pi \left\vert \mathbf{x-y}\right\vert }J_{0}\left( \mathbf{y}\right) \\
&=&\frac{1}{2}\sum_{\alpha \neq \beta }\frac{1}{4\pi \left\vert \mathbf{r}%
^{\left( \alpha \right) }\mathbf{-r}^{\left( \beta \right) }\right\vert } \\
&=&\frac{1}{2}\sum_{\alpha \neq \beta }V\left( \mathbf{r}^{\left( \alpha
\right) }\mathbf{-r}^{\left( \beta \right) }\right)
\end{eqnarray*}
where $V\left( \mathbf{r}^{\left( \alpha \right) }\mathbf{-r}^{\left( \beta
\right) }\right) $ is the usual Coulomb interaction. The same token is
applied to the derivatives of $v\left( \mathbf{r}^{\left( \alpha \right)
}\right) $ appearing in the Hamiltonian, so that the part involving the
electrostatic potential can be written :
\begin{eqnarray*}
H_{\text{elec}} &=&\sum_{\alpha }v\left( \mathbf{r}^{\left( \alpha \right)
}\right) +\frac{\hbar ^{2}}{4}\sum_{\alpha }\left[ \sum_{M^{\left( \alpha
\right) }}\left( \mathcal{A}_{0}^{R_{l}^{\left( \alpha \right) }}\right)
_{N^{\left( \alpha \right) }M^{\left( \alpha \right) }}\left( \mathcal{A}%
_{0}^{R_{m}^{\alpha }}\right) _{M^{\left( \alpha \right) }N^{\left( \alpha
\right) }}\right] \nabla _{R_{m}^{\left( \alpha \right) }}\nabla
_{R_{l}^{\left( \alpha \right) }}v\left( \mathbf{r}^{\left( \alpha \right)
}\right) \\
&&+\frac{\hbar ^{2}}{4}\sum_{\alpha }\frac{\left[ \left[ \varepsilon
_{0}^{\left( \alpha \right) }\left( \mathbf{x}\right) ,\mathcal{A}%
_{0}^{R_{m}^{\left( \alpha \right) }}\right] _{N^{\left( \alpha \right)
}M^{\left( \alpha \right) }}\left( \mathcal{A}_{0}^{R_{l}^{\left( \alpha
\right) }}\right) _{M^{\left( \alpha \right) }N^{\left( \alpha \right)
}}\nabla _{R_{m}^{\left( \alpha \right) }}\nabla _{R_{l}^{\left( \alpha
\right) }}\right] v\left( \mathbf{r}^{\left( \alpha \right) }\right) }{\hat{%
\varepsilon}_{0M^{\left( \alpha \right) }}^{\left( \alpha \right) }-\hat{%
\varepsilon}_{0N^{\left( \alpha \right) }}} \\
&&-\frac{1}{2}\int d^{3}x\left( \mathbf{\nabla }A_{0}\right) ^{2} \\
&=&\frac{1}{2}\sum_{\alpha }V\left( \mathbf{r}^{\left( \alpha \right) }-%
\mathbf{r}^{\left( \beta \right) }\right) \\
&&+\frac{1}{4}\sum_{\alpha }\left[ \sum_{M^{\left( \alpha \right) }}\left(
\mathcal{A}_{0}^{R_{l}^{\left( \alpha \right) }}\right) _{N^{\left( \alpha
\right) }M^{\left( \alpha \right) }}\left( \mathcal{A}_{0}^{R_{m}^{\alpha
}}\right) _{M^{\left( \alpha \right) }N^{\left( \alpha \right) }}\right]
\nabla _{R_{m}^{\left( \alpha \right) }}\nabla _{R_{l}^{\left( \alpha
\right) }}\sum_{\beta \neq \alpha }V\left( \mathbf{r}^{\left( \alpha \right)
}-\mathbf{r}^{\left( \beta \right) }\right) \\
&&+\frac{1}{4}\sum_{\alpha }\frac{\left[ \left[ \varepsilon _{0}^{\left(
\alpha \right) }\left( \mathbf{x}\right) ,\mathcal{A}_{0}^{R_{m}^{\left(
\alpha \right) }}\right] _{N^{\left( \alpha \right) }M^{\left( \alpha
\right) }}\left( \mathcal{A}_{0}^{R_{l}^{\left( \alpha \right) }}\right)
_{M^{\left( \alpha \right) }N^{\left( \alpha \right) }}\nabla
_{R_{m}^{\left( \alpha \right) }}\nabla _{R_{l}^{\left( \alpha \right) }}%
\right] \sum_{\beta \neq \alpha }V\left( \mathbf{r}^{\left( \alpha \right) }-%
\mathbf{r}^{\left( \beta \right) }\right) }{\hat{\varepsilon}_{0M^{\left(
\alpha \right) }}^{\left( \alpha \right) }-\hat{\varepsilon}_{0N^{\left(
\alpha \right) }}}
\end{eqnarray*}
We can now come back to our previous remark about the replacing the
electrostatic potential before or after diagonalization. Had we chosen to
replace this potential before applying the diagonalization procedure would
have in fact led to the same contribution to the Band Hamiltonian. Actually,
it is straightforward to check that this early replacement would have led to
consider the potential $\frac{1}{2}\sum_{\alpha }V\left( \mathbf{R}^{\left(
\alpha \right) }-\mathbf{R}^{\left( \beta \right) }\right) $ in the
Hamiltonian (and no more the potential $v$). This expression mixing the
particles at an early stage would have altered the diagonalization process.
Actually, the diagonalization matrix wouldn't have been the simple tensor
product of individual particles diagonalization matrices, but would have
mixed the various spaces at first order in $\hbar $. The result would be, at
the order $\hbar ^{2}$, in crossed contributions in the Berry phase for $%
\mathbf{K}^{\left( \alpha \right) }$ (not for $\mathbf{R}^{\left( \alpha
\right) }$ as can directly been checked). Namely, the Berry phase for $%
\mathbf{K}^{\left( \alpha \right) }$ would acquire a supplementary
contribution proportional to $\nabla _{R_{m}^{\left( \alpha \right) }}\nabla
_{R_{l}^{\left( \beta \right) }}\sum_{\beta \neq \alpha }V\left( \mathbf{R}%
^{\left( \alpha \right) }-\mathbf{R}^{\left( \beta \right) }\right) \left(
\mathcal{A}_{0}^{R_{m}^{\left( \beta \right) }}\right) _{N^{\left( \beta
\right) }M^{\left( \beta \right) }}\hat{\delta}_{N^{\left( \beta \right)
}M^{\left( \beta \right) }}$. Given our generalized results for the
Diagonalized Hamiltonian, this term will not contribute to the Band
Hamiltonian nor to the dynamical operators at order $\hbar ^{2}$. This is
the reason why we choose to treat the electrostatic potential in a less
rigorous, but symmetric with respect to the vector potential, way.

\subsubsection{$P$ particles Hamiltonian}

We can now gather all our results to obtain the final form for the
Hamiltonian

\begin{eqnarray*}
H_{N} &=&\sum_{\alpha }\hat{\varepsilon}_{0N^{\left( \alpha \right)
}}^{\left( \alpha \right) }\left( \mathbf{\pi }^{\left( \alpha \right)
}\right) +\frac{1}{2}\sum_{\alpha }V\left( \mathbf{r}^{\left( \alpha \right)
}-\mathbf{r}^{\left( \beta \right) }\right) -\frac{1}{2}\sum_{\alpha }\left(
\mathbf{\mu }\left( \mathbf{x}^{\left( \alpha \right) }\right) \mathbf{.B}%
^{\left( \alpha \right) }\left( \mathbf{x}^{\left( \alpha \right) }\right) +%
\mathbf{B}^{\left( \alpha \right) }\left( \mathbf{x}^{\left( \alpha \right)
}\right) \mathbf{.\mu }\left( \mathbf{x}\right) \right) \\
&&-\mathbf{\hat{\mu}}^{k}\left( \mathbf{x}\right) \mathbf{.}\nabla _{R_{k}}%
\mathbf{B}^{\left( \alpha \right) }\left( \mathbf{x}^{\left( \alpha \right)
}\right) +\mathbf{B}^{\left( \alpha \right) }\mathbf{.\tilde{\mu}.B}^{\left(
\alpha \right) } \\
&&+\frac{\hbar ^{2}}{4}\sum_{\alpha }\left[ \sum_{M^{\left( \alpha \right)
}}\left( \mathcal{A}_{0}^{R_{l}^{\left( \alpha \right) }}\right) _{N^{\left(
\alpha \right) }M^{\left( \alpha \right) }}\left( \mathcal{A}%
_{0}^{R_{m}^{\alpha }}\right) _{M^{\left( \alpha \right) }N^{\left( \alpha
\right) }}\right] \nabla _{R_{m}^{\left( \alpha \right) }}\nabla
_{R_{l}^{\left( \alpha \right) }}\sum_{\beta \neq \alpha }V\left( \mathbf{r}%
^{\left( \alpha \right) }-\mathbf{r}^{\left( \beta \right) }\right) \\
&&+\frac{\hbar ^{2}}{4}\sum_{\alpha }\frac{\left[ \left[ \varepsilon
_{0}^{\left( \alpha \right) }\left( \mathbf{x}\right) ,\mathcal{A}%
_{0}^{R_{m}^{\left( \alpha \right) }}\right] _{N^{\left( \alpha \right)
}M^{\left( \alpha \right) }}\left( \mathcal{A}_{0}^{R_{l}^{\left( \alpha
\right) }}\right) _{M^{\left( \alpha \right) }N^{\left( \alpha \right)
}}\nabla _{R_{m}^{\left( \alpha \right) }}\nabla _{R_{l}^{\left( \alpha
\right) }}\right] \sum_{\beta \neq \alpha }V\left( \mathbf{r}^{\left( \alpha
\right) }-\mathbf{r}^{\left( \beta \right) }\right) }{\hat{\varepsilon}%
_{0M^{\left( \alpha \right) }}^{\left( \alpha \right) }-\hat{\varepsilon}%
_{0N^{\left( \alpha \right) }}} \\
+ &&\frac{1}{2}\int dy^{3}dx^{3}dt\hat{\jmath}^{i}\mathbf{\left( \mathbf{x}%
,t\right) }\hat{G}\left( \mathbf{x-y}\right) \hat{\jmath}^{j}\mathbf{\left(
\mathbf{y},t\right) }
\end{eqnarray*}

\subsection{Application : Hamiltonian at the lowest order}

We have seen in the previous section that, due to the complexity to the
Maxwell equation, the Full Hamiltonian to the order $\hbar ^{2}$ cannot be
computed exactly but has in fact to be computed as a power series in the
current. The Formula we gave included first order corrections for the Green
function. Here, to give an application of our method we will neglect these
first order corrections and consider the Hamiltonian at the lowest order in
current, that is quadratic in current. We will see that it will result in a
kind of generalized Breit Hamiltonian, including magnetization-magnetization
interaction. To do so, we divide our work in two steps. First computing the
Hamiltonian without electromagnetic interaction and second including the
electromagnetic corrections.

\subsubsection{ Coulombian interaction}

If we neglect, the magnetic interactions, three simplifications arise.
First, we can cancel all contributions to the magnetic field. Second, in the
first order Berry phases, $\mathcal{A}_{0}^{R_{l}^{\left( \alpha \right)
\beta }}$ is null, except if $\alpha =\beta $. As before, we will note $%
\mathcal{A}_{0}^{R_{l}^{\left( \alpha \right) }}$ the zeroth order Berry
phase, keeping in mind there are no crossed terms. Third, the crystal
momentum dynamical variables are unchanged through the transformation. Thus $%
\mathbf{k}^{\left( \beta \right) }=\mathbf{K}^{\left( \beta \right) }$. The
Hamiltonian reduces thus to :
\begin{eqnarray*}
H_{N} &=&\sum_{\alpha }\hat{\varepsilon}_{0N^{\left( \alpha \right)
}}^{\left( \alpha \right) }\left( \mathbf{k}^{\left( \alpha \right) }\right)
+\frac{1}{2}\sum_{\alpha }V\left( \mathbf{r}^{\left( \alpha \right) }-%
\mathbf{r}^{\left( \beta \right) }\right) \\
&&+\frac{\hbar ^{2}}{4}\sum_{\alpha }\left[ \sum_{M^{\left( \alpha \right)
}}\left( \mathcal{A}_{0}^{R_{l}^{\left( \alpha \right) }}\right) _{N^{\left(
\alpha \right) }M^{\left( \alpha \right) }}\left( \mathcal{A}%
_{0}^{R_{m}^{\alpha }}\right) _{M^{\left( \alpha \right) }N^{\left( \alpha
\right) }}\right] \nabla _{R_{m}^{\left( \alpha \right) }}\nabla
_{R_{l}^{\left( \alpha \right) }}\sum_{\beta \neq \alpha }V\left( \mathbf{r}%
^{\left( \alpha \right) }-\mathbf{r}^{\left( \beta \right) }\right) \\
&&+\frac{\hbar ^{2}}{4}\sum_{\alpha }\frac{\left[ \left[ \varepsilon
_{0}^{\left( \alpha \right) }\left( \mathbf{x}\right) ,\mathcal{A}%
_{0}^{R_{m}^{\left( \alpha \right) }}\right] _{N^{\left( \alpha \right)
}M^{\left( \alpha \right) }}\left( \mathcal{A}_{0}^{R_{l}^{\left( \alpha
\right) }}\right) _{M^{\left( \alpha \right) }N^{\left( \alpha \right)
}}\nabla _{R_{m}^{\left( \alpha \right) }}\nabla _{R_{l}^{\left( \alpha
\right) }}\right] \sum_{\beta \neq \alpha }V\left( \mathbf{r}^{\left( \alpha
\right) }-\mathbf{r}^{\left( \beta \right) }\right) }{\hat{\varepsilon}%
_{0M^{\left( \alpha \right) }}^{\left( \alpha \right) }-\hat{\varepsilon}%
_{0N^{\left( \alpha \right) }}}
\end{eqnarray*}
where the dynamical variables in that set up are given by : \
\begin{eqnarray*}
\mathbf{r}_{N}^{\left( \alpha \right) } &=&\mathbf{R}^{\left( \alpha \right)
}+\emph{A}_{N^{\left( \alpha \right) }}^{\mathbf{R}\left( ^{\alpha }\right) }
\\
&\equiv &\mathbf{R}^{\left( \alpha \right) }+\mathcal{A}_{NN}^{\mathbf{R}%
^{\left( \alpha \right) }} \\
\mathbf{k}_{N}^{\left( \alpha \right) } &=&\mathbf{K}^{\left( \alpha \right)
}
\end{eqnarray*}
with
\begin{eqnarray*}
\emph{A}_{N^{\left( \alpha \right) }}^{\mathbf{R}\left( ^{\alpha }\right) }
&=&\left( \mathcal{A}^{\mathbf{R}}\right) _{N^{\left( \alpha \right)
}N^{\left( \alpha \right) }}=\left( \hbar \mathcal{A}_{0}^{\mathbf{R}%
}\right) _{N^{\left( \alpha \right) }N^{\left( \alpha \right) }}+\frac{\hbar
^{2}}{8}\left( \mathcal{A}_{0}^{R_{l}}\right) _{N^{\left( \alpha \right)
}P}\nabla _{R_{l}}\left( \mathcal{A}_{0}^{\mathbf{R}}\right) _{PN^{\left(
\alpha \right) }} \\
&&+\frac{\hbar ^{2}}{2}\left( \left( \mathcal{A}_{0}^{\mathbf{R}}\right)
_{N^{\left( \alpha \right) }P^{\left( \alpha \right) }}\left( \mathcal{A}%
_{0}^{R_{l}}\right) _{P^{\left( \alpha \right) }N^{\left( \alpha \right)
}}+\left( \mathcal{A}_{0}^{R_{l}}\right) _{N^{\left( \alpha \right)
}P^{\left( \alpha \right) }}\left( \mathcal{A}_{0}^{\mathbf{R}}\right)
_{P^{\left( \alpha \right) }N^{\left( \alpha \right) }}\right) \hat{\delta}%
_{P^{\left( \alpha \right) }N^{\left( \alpha \right) }}\frac{\nabla
_{R_{l}}V\left( \mathbf{R}\right) }{\hat{\varepsilon}_{0P^{\left( \alpha
\right) }}-\hat{\varepsilon}_{0N^{\left( \alpha \right) }}} \\
&&+H.C.
\end{eqnarray*}

\subsubsection{ Electromagnetic interaction at the lowest order and
magnetization-magnetization interaction}

We now turn to the electromagnetic part of the interaction in our
Hamiltonian for $P$ particles. It is found by isolating, at the second order
in the current the interaction terms particles-electromagnetic field \ plus
the internal Field Hamiltonian.

In the non relativistic limit which is of interest for us, it takes the form
:
\begin{eqnarray*}
&&H_{\text{magnetic}}=-\sum_{\alpha }e\left( \frac{\partial \left( \hat{%
\varepsilon}_{0N^{\left( \alpha \right) }}^{\left( \alpha \right) }\left(
\mathbf{k}^{\left( \alpha \right) }\right) +D^{\left( \alpha \right)
}\right) +\sum_{\beta \neq \alpha }V\left( \mathbf{r}^{\left( \alpha \right)
}-\mathbf{r}^{\left( \beta \right) }\right) }{\partial k_{j}^{\left( \alpha
\right) }}\right) _{\mid \mathbf{A}=0}A^{j}\left( \mathbf{r}^{\left( \alpha
\right) }\right) \\
&&+\frac{1}{2}e^{2}\left( \frac{\partial ^{2}\left( \hat{\varepsilon}%
_{0N^{\left( \alpha \right) }}^{\left( \alpha \right) }\left( \mathbf{k}%
^{\left( \alpha \right) }\right) +D^{\left( \alpha \right) }\right)
+\sum_{\beta \neq \alpha }V\left( \mathbf{r}^{\left( \alpha \right) }-%
\mathbf{r}^{\left( \beta \right) }\right) }{\partial k_{j}^{\left( \alpha
\right) }\partial k_{k}^{\left( \alpha \right) }}\right) _{\mid \mathbf{A}%
=0}A^{j}\left( \mathbf{r}^{\left( \alpha \right) }\right) A^{k}\left(
\mathbf{r}^{\left( \alpha \right) }\right) \\
- &&\frac{1}{2}\sum_{\alpha }\left( \mathbf{\mu }\left( \mathbf{x}^{\left(
\alpha \right) }\right) \mathbf{.B}^{\left( \alpha \right) }\left( \mathbf{x}%
^{\left( \alpha \right) }\right) +\mathbf{B}^{\left( \alpha \right) }\left(
\mathbf{x}^{\left( \alpha \right) }\right) \mathbf{.\mu }\left( \mathbf{x}%
\right) \right) -\mathbf{\hat{\mu}}^{k}\left( \mathbf{x}\right) \mathbf{.}%
\nabla _{R_{k}}\mathbf{B}^{\left( \alpha \right) }\left( \mathbf{x}^{\left(
\alpha \right) }\right) +\mathbf{B}^{\left( \alpha \right) }\mathbf{.\tilde{%
\mu}.B}^{\left( \alpha \right) } \\
&&+\frac{1}{2}\int dy^{3}dx^{3}dt\hat{\jmath}^{i}\mathbf{\left( \mathbf{x}%
,t\right) }\hat{G}\left( \mathbf{x-y}\right) \hat{\jmath}^{j}\mathbf{\left(
\mathbf{y},t\right) }
\end{eqnarray*}
Moreover, in first approximation, $\hat{G}\left( \mathbf{x-y}\right) $ can
be approximated by $G_{0}\left( \mathbf{x-y}\right) $ so that
\begin{eqnarray*}
\frac{1}{2}\int dy^{3}dx^{3}dt\hat{\jmath}^{i}\mathbf{\left( \mathbf{x}%
,t\right) }\hat{G}\left( \mathbf{x-y}\right) \hat{\jmath}^{j}\mathbf{\left(
\mathbf{y},t\right) } &\simeq &\frac{1}{2}\int dy^{3}dx^{3}\hat{\jmath}_{i}%
\mathbf{\left( \mathbf{x},t\right) }G_{0}^{ij}\left( \mathbf{x-y}\right)
\hat{\jmath}_{j}\mathbf{\left( \mathbf{y},t\right) } \\
&=&\int d^{3}\mathbf{x}d^{3}\mathbf{y}\hat{\jmath}_{i}\mathbf{\left( \mathbf{%
x},t\right) }\frac{\delta ^{ij}+\frac{\left( \mathbf{x-y}\right) _{i}\left(
\mathbf{x-y}\right) _{j}}{\left\vert \mathbf{x-y}\right\vert ^{2}}}{%
2\left\vert \mathbf{x-y}\right\vert }\hat{\jmath}_{j}\mathbf{\left( \mathbf{y%
},t\right) }
\end{eqnarray*}
As seen before, the current decomposes as :
\begin{equation*}
\hat{\jmath}_{i}\mathbf{\left( \mathbf{y},t\right) =\jmath }_{1i}\mathbf{%
\left( \mathbf{y},t\right) }+\mathcal{M}_{l}^{ef}\mathbf{\left( \mathbf{y}%
,t\right) }\varepsilon ^{lmi}\nabla _{m}^{\left( \mathbf{y}\right) }
\end{equation*}
(we have used the fact that in the definition of the magnetization, the
gradient of a delta acts as minus a derivative) where the magnetization $%
\mathcal{M}_{i}^{ef}$\ and the current $\mathbf{\jmath }_{1}$ is given by :
\begin{eqnarray*}
\mathcal{M}_{i}^{ef}\left( \mathbf{y}\right) &=&\sum_{\alpha }\mathcal{M}%
_{i}^{ef\left( \alpha \right) }\left( \mathbf{x}^{\left( \alpha \right)
}\right) \delta \left( \mathbf{y-r}^{\left( \alpha \right) }\right) \\
&=&\sum_{\alpha }\left[ \mu _{i}\left( \mathbf{x}^{\left( \alpha \right)
}\right) +\hat{\mu}_{i}^{k}\left( \mathbf{x}^{\left( \alpha \right) }\right)
\nabla _{R_{k}}\right] \delta \left( \mathbf{y-r}^{\left( \alpha \right)
}\right)
\end{eqnarray*}
with $\mathcal{M}_{i}^{ef\left( \alpha \right) }$ is the particle $\alpha $
individual particle magnetization, and
\begin{eqnarray*}
\mathbf{\jmath }_{1i}\mathbf{\left( \mathbf{y},t\right) } &=&e\sum_{\alpha }%
\frac{\partial \left( \hat{\varepsilon}_{0N^{\left( \alpha \right)
}}^{\left( \alpha \right) }\left( \mathbf{\pi }^{\left( \alpha \right)
}\right) +D^{\left( \alpha \right) }\right) }{\partial k_{j}^{\left( \alpha
\right) }}\left[ \left( \delta _{i}^{j}-\tilde{\emph{A}}\mathbf{%
^{R_{i}^{\left( \alpha \right) }}}\nabla _{\mathbf{r}_{j}^{\left( \alpha
\right) }}+\mathbf{\emph{A}}_{j,i}\right) \right] _{\mid \mathbf{A}=0}\delta
\left( \mathbf{y-r}^{\left( \alpha \right) }\right) \\
&&-\sum_{\alpha }\frac{\partial v\left( \mathbf{r}^{\left( \alpha \right)
}\right) }{\partial r_{j}^{\left( \alpha \right) }}\left[ \frac{\partial
\tilde{\emph{A}}^{R_{j}^{\left( \alpha \right) }}}{\partial \pi _{n}^{\left(
\alpha \right) }}\left( \delta _{i}^{n}-\tilde{\emph{A}}\mathbf{%
^{R_{i}^{\left( \alpha \right) }}}\nabla _{\mathbf{r}_{n}^{\left( \alpha
\right) }}+\mathbf{\emph{A}}_{n,i\mid \mathbf{A}=0}\right) +\frac{\partial
\tilde{\emph{A}}^{R_{j}^{\left( \alpha \right) }}}{\partial A^{i}\left(
\mathbf{r}^{\left( \alpha \right) }\right) }\right] _{\mid \mathbf{A}%
=0}\delta \left( \mathbf{y-r}^{\left( \alpha \right) }\right)
\end{eqnarray*}
ultimately, as seen in the previous section $v\left( \mathbf{r}^{\left(
\alpha \right) }\right) $ can be replaced by $\sum_{\beta }V\left( \mathbf{r}%
^{\left( \alpha \right) }-\mathbf{r}^{\left( \beta \right) }\right) $.

These formula allow to derive directly the potential and the magnetic field
in our approximation :
\begin{eqnarray*}
\mathbf{A}\left( \mathbf{x}\right) &=&\sum_{\alpha }\frac{\jmath _{1}\mathbf{%
\left( \mathbf{r}^{\left( \alpha \right) }\right) +}\frac{\left[ \jmath _{1}%
\mathbf{\left( \mathbf{r}^{\left( \alpha \right) }\right) .}\left( \mathbf{%
x-r}^{\left( \alpha \right) }\right) \right] \left( \mathbf{x-r}^{\left(
\alpha \right) }\right) }{\left| \mathbf{x-r}^{\left( \alpha \right)
}\right| ^{2}}}{2\left| \mathbf{x-r}^{\left( \alpha \right) }\right| }+\frac{%
\mathcal{M}_{m}^{ef}\left( \mathbf{r}^{\left( \alpha \right) }\right) \times
\left( \mathbf{x-r}^{\left( \alpha \right) }\right) }{\left| \mathbf{x-r}%
^{\left( \alpha \right) }\right| ^{2}} \\
\mathbf{B}\left( \mathbf{x}\right) &=&\sum_{\alpha }\frac{\mathbf{\jmath }%
_{1}\times \left( \mathbf{x-r}^{\left( \alpha \right) }\right) }{\left|
\mathbf{x-r}^{\left( \alpha \right) }\right| ^{3}}+\mathbf{\nabla \times }%
\frac{\mathcal{M}_{m}^{ef}\left( \mathbf{r}^{\left( \alpha \right) }\right)
\times \left( \mathbf{x-r}^{\left( \alpha \right) }\right) }{\left| \mathbf{%
x-r}^{\left( \alpha \right) }\right| ^{3}}
\end{eqnarray*}
We can now compute the various terms appearing in $H_{\text{magnetic}}$. The
first term, involving the magnetization is given by :

\begin{eqnarray*}
&&-\frac{1}{2}\sum_{\alpha }\left( \mathbf{\mu }\left( \mathbf{x}^{\left(
\alpha \right) }\right) \mathbf{.B}\left( \mathbf{x}^{\left( \alpha \right)
}\right) +\mathbf{B}\left( \mathbf{x}^{\left( \alpha \right) }\right)
\mathbf{.\mu }\left( \mathbf{x}^{\left( \alpha \right) }\right) \right) -%
\mathbf{\hat{\mu}}^{k}\left( \mathbf{x}^{\left( \alpha \right) }\right)
\mathbf{.}\nabla _{R_{k}}\mathbf{B}\left( \mathbf{x}^{\left( \alpha \right)
}\right) +\mathbf{B.\tilde{\mu}\left( \mathbf{x}^{\left( \alpha \right)
}\right) .B}^{\left( \alpha \right) } \\
&=&-\frac{1}{2}\sum_{\alpha ,\beta ,\alpha \neq \beta }\left( \mu _{l}\left(
\mathbf{r}^{\left( \alpha \right) }\right) \varepsilon ^{lmi}\nabla
_{m}G_{0}^{ij}\left( \mathbf{r}^{\left( \alpha \right) }\mathbf{-r}^{\left(
\beta \right) }\right) \jmath _{1j}\mathbf{\left( \mathbf{r}^{\left( \beta
\right) }\right) +H.C.}\right. \\
&&\left. \mathbf{-}\hat{\mu}_{l}^{k}\left( \mathbf{r}^{\left( \alpha \right)
}\right) \varepsilon ^{lmi}\nabla _{k}\nabla _{m}G_{0}^{ij}\left( \mathbf{r}%
^{\left( \alpha \right) }\mathbf{-r}^{\left( \beta \right) }\right) \jmath
_{1j}\mathbf{\left( \mathbf{r}^{\left( \beta \right) }\right) }\right) \\
&&+\sum_{\alpha ,\beta ,\alpha \neq \beta }\tilde{\mu}_{lp}\mathbf{\left(
\mathbf{r}^{\left( \alpha \right) }\right) }\varepsilon ^{lmi}\nabla
_{m}G_{0}^{ij}\left( \mathbf{r}^{\left( \alpha \right) }\mathbf{-r}^{\left(
\beta \right) }\right) \jmath _{1j}\mathbf{\left( \mathbf{r}^{\left( \beta
\right) }\right) }\varepsilon ^{pmi}\nabla _{m}G_{0}^{ij}\left( \mathbf{r}%
^{\left( \alpha \right) }\mathbf{-r}^{\left( \gamma \right) }\right) \jmath
_{1j}\mathbf{\left( \mathbf{r}^{\left( \gamma \right) }\right) } \\
&&+\sum_{\alpha ,\beta ,\alpha \neq \beta }\mu _{l}\left( \mathbf{r}^{\left(
\alpha \right) }\right) \varepsilon ^{lri}\varepsilon ^{jmp}\nabla
_{r}\nabla _{p}G_{0}^{ij}\left( \mathbf{r}^{\left( \alpha \right) }\mathbf{-r%
}^{\left( \beta \right) }\right) \mu _{m}\left( \mathbf{r}^{\left( \beta
\right) }\right)
\end{eqnarray*}
The gradient, in all these expressions and in the sequel, is understood as
acting on the first variable (here $\mathbf{r}^{\left( \alpha \right) }$).

The second relevant term is the energy of the internal magnetic field :
\begin{eqnarray*}
&&\frac{1}{2}\int dy^{3}dx^{3}\hat{\jmath}_{i}\mathbf{\left( \mathbf{x}%
,t\right) }G_{0}^{ij}\left( \mathbf{x-y}\right) \hat{\jmath}_{j}\mathbf{%
\left( \mathbf{y},t\right) } \\
&=&\frac{1}{2}\sum_{\alpha ,\beta ,\alpha \neq \beta }\jmath _{1i}\mathbf{%
\left( \mathbf{r}^{\left( \alpha \right) }\right) }G_{0}^{ij}\left( \mathbf{r%
}^{\left( \alpha \right) }\mathbf{-r}^{\left( \beta \right) }\right) \jmath
_{1j}\mathbf{\left( \mathbf{r}^{\left( \beta \right) }\right) } \\
&&\mathbf{+}\frac{1}{2}\sum_{\alpha ,\beta ,\alpha \neq \beta }\left( \mu
_{l}\left( \mathbf{r}^{\left( \alpha \right) }\right) +\hat{\mu}%
_{l}^{k}\left( \mathbf{r}^{\left( \alpha \right) }\right) \nabla _{k}\right)
\varepsilon ^{lmi}\nabla _{m}G_{0}^{ij}\left( \mathbf{r}^{\left( \alpha
\right) }\mathbf{-r}^{\left( \beta \right) }\right) \jmath _{1j}\mathbf{%
\left( \mathbf{r}^{\left( \beta \right) }\right) +H.C.} \\
&&-\frac{1}{2}\sum_{\alpha ,\beta ,\alpha \neq \beta }\mu _{l}\left( \mathbf{%
r}^{\left( \alpha \right) }\right) \varepsilon ^{lri}\varepsilon
^{jmp}\nabla _{r}\nabla _{p}G_{0}^{ij}\left( \mathbf{x-y}\right) \mu
_{m}\left( \mathbf{r}^{\left( \beta \right) }\right)
\end{eqnarray*}
We gather these two contributions to get :
\begin{eqnarray*}
&&-\frac{1}{2}\sum_{\alpha }\left( \mathbf{\mu }\left( \mathbf{x}^{\left(
\alpha \right) }\right) \mathbf{.B}^{\left( \alpha \right) }\left( \mathbf{x}%
^{\left( \alpha \right) }\right) +\mathbf{B}^{\left( \alpha \right) }\left(
\mathbf{x}^{\left( \alpha \right) }\right) \mathbf{.\mu }\left( \mathbf{x}%
\right) \right) -\mathbf{\hat{\mu}}^{k}\left( \mathbf{x}\right) \mathbf{.}%
\nabla _{R_{k}}\mathbf{B}^{\left( \alpha \right) }\left( \mathbf{x}^{\left(
\alpha \right) }\right) +\mathbf{B}^{\left( \alpha \right) }\mathbf{.\tilde{%
\mu}.B}^{\left( \alpha \right) } \\
&&+\frac{1}{2}\int dy^{3}dx^{3}dt\hat{\jmath}^{i}\mathbf{\left( \mathbf{x}%
,t\right) }\hat{G}\left( \mathbf{x-y}\right) \hat{\jmath}^{j}\mathbf{\left(
\mathbf{y},t\right) } \\
&=&\frac{1}{2}\sum_{\alpha ,\beta ,\alpha \neq \beta }\jmath _{1i}\mathbf{%
\left( \mathbf{r}^{\left( \alpha \right) }\right) }G_{0}^{ij}\left( \mathbf{r%
}^{\left( \alpha \right) }\mathbf{-r}^{\left( \beta \right) }\right) \jmath
_{1j}\mathbf{\left( \mathbf{r}^{\left( \beta \right) }\right) } \\
&&+\frac{1}{2}\sum_{\alpha ,\beta ,\alpha \neq \beta }\mu _{l}\left( \mathbf{%
r}^{\left( \alpha \right) }\right) \varepsilon ^{lri}\varepsilon
^{jmp}\nabla _{r}\nabla _{p}G_{0}^{ij}\left( \mathbf{x-y}\right) \mu
_{m}\left( \mathbf{r}^{\left( \beta \right) }\right) \\
&&+\sum_{\alpha ,\beta ,\alpha \neq \beta }\tilde{\mu}_{lp}\mathbf{\left(
\mathbf{r}^{\left( \alpha \right) }\right) }\varepsilon ^{lmi}\nabla
_{m}G_{0}^{ij}\left( \mathbf{r}^{\left( \alpha \right) }\mathbf{-r}^{\left(
\beta \right) }\right) \jmath _{1j}\mathbf{\left( \mathbf{r}^{\left( \beta
\right) }\right) }\varepsilon ^{pmi}\nabla _{m}G_{0}^{ij}\left( \mathbf{r}%
^{\left( \alpha \right) }\mathbf{-r}^{\left( \gamma \right) }\right) \jmath
_{1j}\mathbf{\left( \mathbf{r}^{\left( \gamma \right) }\right) }
\end{eqnarray*}
The last term of interest for us, can be developer in the following way :
\begin{eqnarray*}
&&-\sum_{\alpha }e\left( \frac{\partial \left( \hat{\varepsilon}_{0N^{\left(
\alpha \right) }}^{\left( \alpha \right) }\left( \mathbf{k}^{\left( \alpha
\right) }\right) +D^{\left( \alpha \right) }\right) +\sum_{\beta \neq \alpha
}V\left( \mathbf{r}^{\left( \alpha \right) }-\mathbf{r}^{\left( \beta
\right) }\right) }{\partial k_{j}^{\left( \alpha \right) }}\right) _{\mid
\mathbf{A}=0}A^{j}\left( \mathbf{r}^{\left( \alpha \right) }\right) \\
&&+\frac{1}{2}e^{2}\left( \frac{\partial ^{2}\left( \hat{\varepsilon}%
_{0N^{\left( \alpha \right) }}^{\left( \alpha \right) }\left( \mathbf{k}%
^{\left( \alpha \right) }\right) +D^{\left( \alpha \right) }\right)
+\sum_{\beta \neq \alpha }V\left( \mathbf{r}^{\left( \alpha \right) }-%
\mathbf{r}^{\left( \beta \right) }\right) }{\partial k_{j}^{\left( \alpha
\right) }\partial k_{k}^{\left( \alpha \right) }}\right) _{\mid \mathbf{A}%
=0}A^{j}\left( \mathbf{r}^{\left( \alpha \right) }\right) A^{k}\left(
\mathbf{r}^{\left( \alpha \right) }\right) \\
&=&-\sum_{\alpha }\mathbf{\jmath }_{1}\mathbf{\left( \mathbf{r}^{\left(
\alpha \right) }\right) A}\left( \mathbf{r}^{\left( \alpha \right) }\right) +%
\frac{1}{2}e^{2}\left( \frac{\partial ^{2}\left( \hat{\varepsilon}%
_{0N^{\left( \alpha \right) }}^{\left( \alpha \right) }\left( \mathbf{k}%
^{\left( \alpha \right) }\right) +D^{\left( \alpha \right) }\right) }{%
\partial k_{j}^{\left( \alpha \right) }\partial k_{k}^{\left( \alpha \right)
}}\right) _{\mid \mathbf{A}=0}A^{j}\left( \mathbf{r}^{\left( \alpha \right)
}\right) A^{k}\left( \mathbf{r}^{\left( \alpha \right) }\right) \\
&=&-\sum_{\alpha ,\beta ,\alpha \neq \beta }\jmath _{1i}\mathbf{\left(
\mathbf{r}^{\left( \alpha \right) }\right) }G_{0}^{ij}\left( \mathbf{r}%
^{\left( \alpha \right) }\mathbf{-r}^{\left( \beta \right) }\right) \jmath
_{1j}\mathbf{\left( \mathbf{r}^{\left( \beta \right) }\right) } \\
&&\mathbf{-}\frac{1}{2}\left[ \sum_{\alpha ,\beta ,\alpha \neq \beta }\left(
\mu _{l}\left( \mathbf{r}^{\left( \alpha \right) }\right) +\hat{\mu}%
_{l}^{k}\left( \mathbf{r}^{\left( \alpha \right) }\right) \nabla _{k}\right)
\varepsilon ^{lmi}\nabla _{m}G_{0}^{ij}\left( \mathbf{r}^{\left( \alpha
\right) }\mathbf{-r}^{\left( \beta \right) }\right) \jmath _{1j}\mathbf{%
\left( \mathbf{r}^{\left( \beta \right) }\right) +H.C.}\right] \\
&&+\frac{1}{2}e^{2}\left( \frac{\partial ^{2}\left( \hat{\varepsilon}%
_{0N^{\left( \alpha \right) }}^{\left( \alpha \right) }\left( \mathbf{k}%
^{\left( \alpha \right) }\right) +D^{\left( \alpha \right) }\right) }{%
\partial k_{j}^{\left( \alpha \right) }\partial k_{k}^{\left( \alpha \right)
}}\right) _{\mid \mathbf{A}=0}A^{j}\left( \mathbf{r}^{\left( \alpha \right)
}\right) A^{k}\left( \mathbf{r}^{\left( \alpha \right) }\right)
\end{eqnarray*}
The first equality is implied by construction of $\mathbf{\jmath }_{1}%
\mathbf{\left( \mathbf{r}^{\left( \alpha \right) }\right) }$ since this last
quantity was precisely defined as $\sum_{\alpha }e\left( \frac{\partial
\left( \hat{\varepsilon}_{0N^{\left( \alpha \right) }}^{\left( \alpha
\right) }\left( \mathbf{k}^{\left( \alpha \right) }\right) +D^{\left( \alpha
\right) }\right) +\sum_{\beta \neq \alpha }V\left( \mathbf{r}^{\left( \alpha
\right) }-\mathbf{r}^{\left( \beta \right) }\right) }{\partial k_{j}^{\left(
\alpha \right) }}\right) _{\mid \mathbf{A}=0}$ .

We can now use the following facts :
\begin{eqnarray*}
G_{0}^{ij}\left( \mathbf{x-y}\right) &=&\frac{\delta ^{ij}+\frac{\left(
\mathbf{x-y}\right) _{i}\left( \mathbf{x-y}\right) _{j}}{\left| \mathbf{x-y}%
\right| ^{2}}}{2\left| \mathbf{x-y}\right| } \\
\varepsilon ^{lmi}\nabla _{m}G_{0}^{ij}\left( \mathbf{x-y}\right)
&=&\varepsilon ^{lji}\frac{\left( \mathbf{x-y}\right) _{i}}{\left| \mathbf{%
x-y}\right| ^{3}} \\
\varepsilon ^{lri}\varepsilon ^{jmp}\nabla _{r}\nabla _{p}G_{0}^{ij}\left(
\mathbf{x-y}\right) &=&\frac{\delta ^{ij}-3\frac{\left( \mathbf{x-y}\right)
_{i}\left( \mathbf{x-y}\right) _{j}}{\left| \mathbf{x-y}\right| ^{2}}-\frac{%
8\pi }{3}\delta ^{ij}\delta \left( \mathbf{x-y}\right) }{\left| \mathbf{x-y}%
\right| ^{3}}
\end{eqnarray*}
Reintroducing the effective individual magnetization operator as :
\begin{equation*}
\mathcal{M}^{ef\left( \alpha \right) }\left( \mathbf{r}^{\left( \alpha
\right) }\right) =\mathbf{\mu }\left( \mathbf{r}^{\left( \alpha \right)
}\right) +\mathbf{\mu }^{k}\left( \mathbf{r}^{\left( \alpha \right) }\right)
\nabla _{k}
\end{equation*}
one can rewrite ultimately at the order $\hbar ^{2}$ :

\begin{eqnarray*}
H_{\text{magnetic}} &=&-\sum_{\alpha \neq \beta }\frac{1}{2}\frac{\mathbf{%
\jmath }_{1}\mathbf{\left( \mathbf{r}^{\left( \alpha \right) }\right)
.\jmath }_{1}\mathbf{\left( \mathbf{r}^{\left( \beta \right) }\right) }+%
\frac{\jmath _{1}\mathbf{\left( \mathbf{r}^{\left( \alpha \right) }\right) .}%
\left( \mathbf{r}^{\left( \alpha \right) }\mathbf{-r}^{\left( \beta \right)
}\right) \jmath _{1}\mathbf{\left( \mathbf{r}^{\left( \beta \right) }\right)
}.\left( \mathbf{r}^{\left( \alpha \right) }\mathbf{-r}^{\left( \beta
\right) }\right) }{\left| \mathbf{r}^{\left( \alpha \right) }\mathbf{-r}%
^{\left( \beta \right) }\right| ^{2}}}{2\left| \mathbf{r}^{\left( \alpha
\right) }\mathbf{-r}^{\left( \beta \right) }\right| } \\
&&\mathbf{+}\frac{1}{2}\sum_{\alpha \neq \beta }\left[ \mathcal{M}^{ef\left(
\alpha \right) }.\left( \frac{\left( \mathbf{r}^{\left( \alpha \right) }%
\mathbf{-r}^{\left( \beta \right) }\right) }{\left| \mathbf{r}^{\left(
\alpha \right) }\mathbf{-r}^{\left( \beta \right) }\right| ^{3}}\times
\mathbf{\jmath }_{1}\mathbf{\left( \mathbf{r}^{\left( \beta \right) }\right)
}\right) \mathbf{+H.C.}\right] \\
&&+\frac{1}{2}\sum_{\alpha \neq \beta }\frac{\mathcal{M}^{ef\left( \alpha
\right) }.\mathcal{M}^{ef\left( \beta \right) }-3\frac{\mathcal{M}%
_{i}^{ef\left( \alpha \right) }.\left( \mathbf{r}^{\left( \alpha \right) }%
\mathbf{-r}^{\left( \beta \right) }\right) \mathcal{M}^{ef\left( \beta
\right) }.\left( \mathbf{r}^{\left( \alpha \right) }\mathbf{-r}^{\left(
\beta \right) }\right) }{\left| \mathbf{r}^{\left( \alpha \right) }\mathbf{-r%
}^{\left( \beta \right) }\right| ^{2}}-\frac{8\pi }{3}\mathcal{M}^{ef\left(
\alpha \right) }.\mathcal{M}^{ef\left( \beta \right) }\delta \left( \mathbf{r%
}^{\left( \alpha \right) }\mathbf{-r}^{\left( \beta \right) }\right) }{%
\left| \mathbf{r}^{\left( \alpha \right) }\mathbf{-r}^{\left( \beta \right)
}\right| ^{3}} \\
&&\mathbf{+}\sum_{\alpha \neq \beta ,\alpha \neq \gamma ,\beta \neq \gamma
}\left( \jmath _{1i}\mathbf{\left( \mathbf{r}^{\left( \gamma \right)
}\right) \times \frac{\left( \mathbf{r}^{\left( \alpha \right) }\mathbf{-r}%
^{\left( \beta \right) }\right) }{\left| \mathbf{r}^{\left( \alpha \right) }%
\mathbf{-r}^{\left( \beta \right) }\right| ^{3}}}\right) \mathbf{.}\tilde{\mu%
}\mathbf{\left( \mathbf{r}^{\left( \alpha \right) }\right) .}\left( \frac{%
\left( \mathbf{r}^{\left( \alpha \right) }\mathbf{-r}^{\left( \beta \right)
}\right) }{\left| \mathbf{r}^{\left( \alpha \right) }\mathbf{-r}^{\left(
\beta \right) }\right| ^{3}}\mathbf{\times }\jmath _{1j}\mathbf{\left(
\mathbf{r}^{\left( \beta \right) }\right) }\right) \\
&&+\sum_{\alpha \neq \beta }\frac{1}{2}\left( \frac{\jmath _{1}\mathbf{%
\left( \mathbf{r}^{\left( \alpha \right) }\right) -}\frac{\jmath _{1}\mathbf{%
\left( \mathbf{r}^{\left( \alpha \right) }\right) .}\left( \mathbf{r}%
^{\left( \alpha \right) }\mathbf{-r}^{\left( \beta \right) }\right) \left(
\mathbf{r}^{\left( \alpha \right) }\mathbf{-r}^{\left( \beta \right)
}\right) }{\left| \mathbf{r}^{\left( \alpha \right) }\mathbf{-r}^{\left(
\beta \right) }\right| ^{2}}}{2\left| \mathbf{r}^{\left( \alpha \right) }%
\mathbf{-r}^{\left( \beta \right) }\right| }+\mathcal{M}^{ef\left( \alpha
\right) }\times \frac{\left( \mathbf{r}^{\left( \alpha \right) }\mathbf{-r}%
^{\left( \beta \right) }\right) }{\left| \mathbf{r}^{\left( \alpha \right) }%
\mathbf{-r}^{\left( \beta \right) }\right| ^{3}}\right) _{j} \\
&&\times \frac{\partial ^{2}\left( \hat{\varepsilon}_{0N^{\left( \alpha
\right) }}^{\left( \alpha \right) }\left( \mathbf{k}^{\left( \alpha \right)
}\right) +D^{\left( \alpha \right) }\right) }{\partial k_{j}^{\left( \alpha
\right) }\partial k_{k}^{\left( \alpha \right) }} \\
&&\times \left( \frac{\jmath _{1}\mathbf{\left( \mathbf{r}^{\left( \alpha
\right) }\right) -}\frac{\jmath _{1}\mathbf{\left( \mathbf{r}^{\left( \alpha
\right) }\right) .}\left( \mathbf{r}^{\left( \alpha \right) }\mathbf{-r}%
^{\left( \beta \right) }\right) \left( \mathbf{r}^{\left( \alpha \right) }%
\mathbf{-r}^{\left( \beta \right) }\right) }{\left| \mathbf{r}^{\left(
\alpha \right) }\mathbf{-r}^{\left( \beta \right) }\right| ^{2}}}{2\left|
\mathbf{r}^{\left( \alpha \right) }\mathbf{-r}^{\left( \beta \right)
}\right| }+\mathcal{M}^{ef\left( \alpha \right) }\times \frac{\left( \mathbf{%
r}^{\left( \alpha \right) }\mathbf{-r}^{\left( \beta \right) }\right) }{%
\left| \mathbf{r}^{\left( \alpha \right) }\mathbf{-r}^{\left( \beta \right)
}\right| ^{3}}\right) _{k}
\end{eqnarray*}
Note here that the double scalar product involving the 2 tensor $\tilde{\mu}%
\mathbf{\left( \mathbf{r}^{\left( \alpha \right) }\right) }$ is performed on
each of its indices separately.

The interpretation of the various terms can done by comparison to the very
similar \ Breit Hamiltonian for the Dirac electron. Given the current is
decomposed in two parts due to the velocity of the particles and their
magnetization, the magnetic part of the Hamiltonian is mainly a
current-current interaction. The first term is the usual current-current
interaction involving $\mathbf{\jmath }_{1}\mathbf{\left( \mathbf{r}^{\left(
\alpha \right) }\right) }$ only, whereas the second term mixing $\mathbf{%
\jmath }_{1}\mathbf{\left( \mathbf{r}^{\left( \alpha \right) }\right) }$ and
the magnetization of other particles is a magnetization-orbit coupling
between different particles. This interaction is formally similar to the
spin-orbit interaction term, except that here we are here in a non
relativistic context, and that the magnetization arise as a band phenomenon.
The third term is for the magnetization-magnetization interaction term which
is formally similar to the spin-spin interaction term. The two last terms
are new with respect to the usual Breit Hamiltonian. The fourth one is again
a magnetization orbit coupling but of higher order, since it involves
triples of particles. The last term is a correction to the energy due to the
development to the second order of the free energy. It is a shift of the
individual particles energy operator due to the field created by other
particles. It should be present in the Breit Hamiltonian but is in fact
neglected while considering the quantum field derivation of this last one
since it is of order higher than $\frac{1}{m^{2}}$ for the Dirac case ( \cite%
{Landau1}).

\subsubsection{Full Hamiltonian}

Since our development was in the second order in the currents we can simply
gather the magnetic part and the electric part of the Hamiltonian to obtain
ultimately the full $P$ particles Hamiltonian :
\begin{eqnarray*}
H_{N} &=&\sum_{\alpha }\hat{\varepsilon}_{0N^{\left( \alpha \right)
}}^{\left( \alpha \right) }\left( \mathbf{k}^{\left( \alpha \right) }\right)
+\frac{1}{2}\sum_{\alpha \neq \beta }V\left( \mathbf{r}^{\left( \alpha
\right) }-\mathbf{r}^{\left( \beta \right) }\right) \\
&&+\frac{\hbar ^{2}}{4}\sum_{\alpha }\left[ \sum_{M^{\left( \alpha \right)
}}\left( \mathcal{A}_{0}^{R_{l}^{\left( \alpha \right) }}\right) _{N^{\left(
\alpha \right) }M^{\left( \alpha \right) }}\left( \mathcal{A}%
_{0}^{R_{m}^{\alpha }}\right) _{M^{\left( \alpha \right) }N^{\left( \alpha
\right) }}\right] \nabla _{R_{m}^{\left( \alpha \right) }}\nabla
_{R_{l}^{\left( \alpha \right) }}\sum_{\beta \neq \alpha }V\left( \mathbf{r}%
^{\left( \alpha \right) }-\mathbf{r}^{\left( \beta \right) }\right) \\
&&+\frac{\hbar ^{2}}{4}\sum_{\alpha }\frac{\left[ \left[ \hat{\varepsilon}%
_{0N^{\left( \alpha \right) }}^{\left( \alpha \right) }\left( \mathbf{k}%
^{\left( \alpha \right) }\right) ,\mathcal{A}_{0}^{R_{m}^{\left( \alpha
\right) }}\right] _{N^{\left( \alpha \right) }M^{\left( \alpha \right)
}}\left( \mathcal{A}_{0}^{R_{l}^{\left( \alpha \right) }}\right) _{M^{\left(
\alpha \right) }N^{\left( \alpha \right) }}\nabla _{R_{m}^{\left( \alpha
\right) }}\nabla _{R_{l}^{\left( \alpha \right) }}\right] \sum_{\beta \neq
\alpha }V\left( \mathbf{r}^{\left( \alpha \right) }-\mathbf{r}^{\left( \beta
\right) }\right) }{\hat{\varepsilon}_{0M^{\left( \alpha \right) }}^{\left(
\alpha \right) }-\hat{\varepsilon}_{0N^{\left( \alpha \right) }}} \\
&&-\sum_{\alpha \neq \beta }\frac{1}{2}\frac{\mathbf{\jmath }_{1}\mathbf{%
\left( \mathbf{r}^{\left( \alpha \right) }\right) .\jmath }_{1}\mathbf{%
\left( \mathbf{r}^{\left( \beta \right) }\right) }+\frac{\jmath _{1}\mathbf{%
\left( \mathbf{r}^{\left( \alpha \right) }\right) .}\left( \mathbf{r}%
^{\left( \alpha \right) }\mathbf{-r}^{\left( \beta \right) }\right) \jmath
_{1}\mathbf{\left( \mathbf{r}^{\left( \beta \right) }\right) }.\left(
\mathbf{r}^{\left( \alpha \right) }\mathbf{-r}^{\left( \beta \right)
}\right) }{\left| \mathbf{r}^{\left( \alpha \right) }\mathbf{-r}^{\left(
\beta \right) }\right| ^{2}}}{2\left| \mathbf{r}^{\left( \alpha \right) }%
\mathbf{-r}^{\left( \beta \right) }\right| } \\
&&\mathbf{+}\frac{1}{2}\sum_{\alpha \neq \beta }\left[ \left( \mathcal{M}%
^{ef\left( \alpha \right) }\times \frac{\left( \mathbf{r}^{\left( \alpha
\right) }\mathbf{-r}^{\left( \beta \right) }\right) }{\left| \mathbf{r}%
^{\left( \alpha \right) }\mathbf{-r}^{\left( \beta \right) }\right| ^{3}}%
\right) .\mathbf{\jmath }_{1}\mathbf{\left( \mathbf{r}^{\left( \beta \right)
}\right) +H.C.}\right] \\
&&+\frac{1}{2}\sum_{\alpha \neq \beta }\frac{\mathcal{M}^{ef\left( \alpha
\right) }.\mathcal{M}^{ef\left( \beta \right) }-3\frac{\mathcal{M}%
_{i}^{ef\left( \alpha \right) }.\left( \mathbf{r}^{\left( \alpha \right) }%
\mathbf{-r}^{\left( \beta \right) }\right) \mathcal{M}^{ef\left( \beta
\right) }.\left( \mathbf{r}^{\left( \alpha \right) }\mathbf{-r}^{\left(
\beta \right) }\right) }{\left| \mathbf{r}^{\left( \alpha \right) }\mathbf{-r%
}^{\left( \beta \right) }\right| ^{2}}-\frac{8\pi }{3}\mathcal{M}^{ef\left(
\alpha \right) }.\mathcal{M}^{ef\left( \beta \right) }\delta \left( \mathbf{r%
}^{\left( \alpha \right) }\mathbf{-r}^{\left( \beta \right) }\right) }{%
\left| \mathbf{r}^{\left( \alpha \right) }\mathbf{-r}^{\left( \beta \right)
}\right| ^{3}} \\
&&\mathbf{+}\sum_{\alpha \neq \beta }\jmath _{1i}\mathbf{\left( \mathbf{r}%
^{\left( \alpha \right) }\right) .}\left( \mathbf{\frac{\left( \mathbf{r}%
^{\left( \alpha \right) }\mathbf{-r}^{\left( \beta \right) }\right) }{\left|
\mathbf{r}^{\left( \alpha \right) }\mathbf{-r}^{\left( \beta \right)
}\right| ^{3}}\times }\tilde{\mu}\mathbf{\left( \mathbf{r}^{\left( \alpha
\right) }\right) \times }\frac{\left( \mathbf{r}^{\left( \alpha \right) }%
\mathbf{-r}^{\left( \beta \right) }\right) }{\left| \mathbf{r}^{\left(
\alpha \right) }\mathbf{-r}^{\left( \beta \right) }\right| ^{3}}\right)
.\jmath _{1j}\mathbf{\left( \mathbf{r}^{\left( \beta \right) }\right) } \\
&&+\sum_{\alpha \neq \beta }\frac{1}{2}\left( \frac{\jmath _{1}\mathbf{%
\left( \mathbf{r}^{\left( \alpha \right) }\right) -}\frac{\jmath _{1}\mathbf{%
\left( \mathbf{r}^{\left( \alpha \right) }\right) .}\left( \mathbf{r}%
^{\left( \alpha \right) }\mathbf{-r}^{\left( \beta \right) }\right) \left(
\mathbf{r}^{\left( \alpha \right) }\mathbf{-r}^{\left( \beta \right)
}\right) }{\left| \mathbf{r}^{\left( \alpha \right) }\mathbf{-r}^{\left(
\beta \right) }\right| ^{2}}}{2\left| \mathbf{r}^{\left( \alpha \right) }%
\mathbf{-r}^{\left( \beta \right) }\right| }+\mathcal{M}^{ef\left( \alpha
\right) }\times \nabla _{k}\frac{\left( \mathbf{r}^{\left( \alpha \right) }%
\mathbf{-r}^{\left( \beta \right) }\right) }{\left| \mathbf{r}^{\left(
\alpha \right) }\mathbf{-r}^{\left( \beta \right) }\right| ^{3}}\right) _{j}
\\
&&\times \frac{\partial ^{2}\left( \hat{\varepsilon}_{0N^{\left( \alpha
\right) }}^{\left( \alpha \right) }\left( \mathbf{k}^{\left( \alpha \right)
}\right) +D^{\left( \alpha \right) }\right) }{\partial k_{j}^{\left( \alpha
\right) }\partial k_{k}^{\left( \alpha \right) }} \\
&&\times \left( \frac{\jmath _{1}\mathbf{\left( \mathbf{r}^{\left( \alpha
\right) }\right) -}\frac{\jmath _{1}\mathbf{\left( \mathbf{r}^{\left( \alpha
\right) }\right) .}\left( \mathbf{r}^{\left( \alpha \right) }\mathbf{-r}%
^{\left( \beta \right) }\right) \left( \mathbf{r}^{\left( \alpha \right) }%
\mathbf{-r}^{\left( \beta \right) }\right) }{\left| \mathbf{r}^{\left(
\alpha \right) }\mathbf{-r}^{\left( \beta \right) }\right| ^{2}}}{2\left|
\mathbf{r}^{\left( \alpha \right) }\mathbf{-r}^{\left( \beta \right)
}\right| }+\mathcal{M}^{ef\left( \alpha \right) }\times \nabla _{k}\frac{%
\left( \mathbf{r}^{\left( \alpha \right) }\mathbf{-r}^{\left( \beta \right)
}\right) }{\left| \mathbf{r}^{\left( \alpha \right) }\mathbf{-r}^{\left(
\beta \right) }\right| ^{3}}\right) _{i}
\end{eqnarray*}

with :
\begin{eqnarray*}
\mathbf{\jmath }_{1i}\mathbf{\left( \mathbf{r}^{\left( \alpha \right)
}\right) } &=&e\sum_{\alpha }\frac{\partial \left( \hat{\varepsilon}%
_{0N^{\left( \alpha \right) }}^{\left( \alpha \right) }\left( \mathbf{k}%
^{\left( \alpha \right) }\right) +D^{\left( \alpha \right) }\right) }{%
\partial k_{j}^{\left( \alpha \right) }}\left[ \left( \delta _{i}^{j}-%
\mathbf{\emph{A}^{R_{i}^{\left( \alpha \right) }}}\nabla _{\mathbf{r}%
_{j}^{\left( \alpha \right) }}+\mathbf{\emph{A}}_{j,i}\right) \right] _{\mid
\mathbf{A}=0} \\
&&-\sum_{\alpha }\frac{\partial \sum_{\beta }V\left( \mathbf{r}^{\left(
\alpha \right) }-\mathbf{r}^{\left( \beta \right) }\right) }{\partial
r_{j}^{\left( \alpha \right) }}\left[ \frac{\partial \tilde{\emph{A}}
^{R_{j}^{\left( \alpha \right) }}}{\partial \pi _{n}^{\left( \alpha \right) }%
}\left( \delta _{i}^{n}-\mathbf{\emph{A}^{R_{i}^{\left( \alpha \right) }}}%
\nabla _{\mathbf{r}_{n}^{\left( \alpha \right) }}+\mathbf{\emph{A}}_{n,i\mid
\mathbf{A}=0}\right) +\frac{\partial \tilde{\emph{A}}^{R_{j}^{\left( \alpha
\right) }}}{\partial A^{i}\left( \mathbf{r}^{\left( \alpha \right) }\right) }%
\right] _{\mid \mathbf{A}=0} \\
\mathcal{M}^{ef\left( \alpha \right) }\left( \mathbf{r}^{\left( \alpha
\right) }\right) &=&\mathbf{\mu }\left( \mathbf{r}^{\left( \alpha \right)
}\right) +\mathbf{\mu }^{k}\left( \mathbf{r}^{\left( \alpha \right) }\right)
\nabla _{k}
\end{eqnarray*}
Recall at that point that we have performed a development in currents (or
field) to the lowest order. As such in all our expression the
electromagnetic field is set to $0$. As a consequence, the variables $%
\mathbf{k}^{\left( \alpha \right) }$, $\mathbf{r}^{\left( \alpha \right) }$
are computed with $\mathbf{A}=0$.

The dynamical variables are thus given by : \
\begin{eqnarray*}
\mathbf{r}_{N}^{\left( \alpha \right) } &=&\mathbf{R}^{\left( \alpha \right)
}+\emph{A}_{N^{\left( \alpha \right) }}^{\mathbf{R}\left( ^{\alpha }\right) }
\\
\mathbf{k}_{N}^{\left( \alpha \right) } &=&\mathbf{K}^{\left( \alpha \right)
}
\end{eqnarray*}
Apart from the magnetic part of the Hamiltonian, already discussed, we have
included the usual electrostatic part, as well as the Darwin term, which is
analogous to the eponymous term in the Breit Hamiltonian. Note however that
this term in our context is not only a contact term (i.e. of Dirac delta
type) since its form depends ultimately on the form of the Berry phases that
depend on the structure of the system at stake.

Let us also remark ultimately that the expansion performed, is different
from the one chosen in \cite{Landau1} to derive the Breit Hamiltonian.
Actually, the derivation of this last one through the context of quantum
field involves a power expansion in $\frac{1}{m}$, which is in fact
performed at the second order.

\section{Conclusion}

The diagonalization of the Hamiltonian for a Bloch electron in a magnetic
field is an old problem is solid state physics initiated principally by
Blount \cite{Blount}, who developed a general procedure for the removal of
the interband matrix elements based on a asymptotic series expansion in the
fields strength. In this way, the actual effective one-band Hamiltonian was
obtained to the second order in the magnetic field. In this paper,we came
back, in a way, to this old problem, but by trying instead to derive an
in-band Hamiltonian as a series expansion in the Planck constant. A first
attempt in this direction was done when we provided a procedure at order $%
\hbar $ for an arbitrary matrix valued Hamiltonian \cite{PIERRESEMIDIAG}.
This method resulted in an effective diagonal Hamiltonian in terms of
gauge-covariant but noncanonical, actually noncommutative, coordinates. It
has also revealed \ that a generalized Peierls substitution taking into
account a Berry phase term must be considered for the semiclassical
treatment of electrons in a magnetic field \cite{PIERREEBS}. In particular,
the Bohr-Sommerfeld quantization condition when reformulated in terms of the
generalized Peierls substitution leads to a modification of the
semiclassical quantization rules as well as to a generalization of the
cross-sectional area derived by Roth \cite{ROTH}. Soon later, the
semiclassical diagonalization was extended to any order, when we developed
another method of diagonalization where $\hbar $ was considered as a running
parameter. This method allowed us to solve formally the problem in terms of
a differential equation for the diagonal Hamiltonian with respect to Planck
constant $\hbar $, which could, in principle, be solved recursively by a
series expansion in $\hbar $ \cite{SERIESPIERRE}. As an example, the energy
spectrum of a Bloch electron in an electric field was derived to all order
in $\hbar .$ However, in the presence of a magnetic field, the differential
approach beyond the semiclassical turned out to be very complicated.

In this paper, we solved the Hamiltonian diagonalization for a Bloch
electron in an electromagnetic field to the second order in $\hbar $ by
applying a radically different procedure, which was recently developed and
provides a general diagonal expression for any kind of matrix valued
Hamiltonian \cite{DIAGOEXACT} (this diagonal expression is also solution of
the differential equation of \cite{SERIESPIERRE}). The main advantage of our
method with respect to Blount's procedure is obviously that our result is
valid even in the presence of strong external electromagnetic fields. But a
second important advantage with an expansion in $\hbar $ is that it allows
us to consider particles in interaction. Indeed, although the removal of the
interband matrix elements to the second order in $\hbar $ for a Bloch
electron is important in itself, the principal objective of this paper was
the determination of the effective Hamiltonian of interacting Bloch
electrons living in different energy bands. As we have shown, even if the
electrostatic interaction dominates the magnetic one, effects like magnetic
moment-moment interaction mediated by the magnetic part of the full
electromagnetic interaction requires a computation to second order in $\hbar
$. Comparison with other methods\ is difficult because to our knowledge
other results for interacting Bloch electrons do not exist. Nevertheless
because of the strong analogy with the Dirac equation it is possible to try
a comparison with the Breit Hamiltonian for the Dirac electrons \cite{BREIT}%
. As for Breit, we found that the electronic current is made of two
contributions: one comes from the velocity and the other is a magnetic
moment current similar to the spin current for Dirac \cite{PIERREDIRAC}. It
is this last one which is responsible for the magnetic moment-moment
interaction similar to the spin-spin interaction for Dirac. Another
important interaction revealed by our approach is the moment-orbit coupling
by analogy with the spin-orbit one. From the results of this paper, our goal
in the future is to consider the physical relevance of interactions bringing
into play magnetic moments.

\section{Appendix}

Since the right hand side of the Maxwell equation involves $-\int d^{3}y%
\left[ \delta ^{ij}\delta ^{3}\left( \mathbf{x-y}\right) +\frac{\partial ^{2}%
}{\partial x^{i}\partial x^{j}}\frac{1}{4\pi \left\vert \mathbf{x-y}%
\right\vert }\right] \frac{\delta }{\delta A^{j}\left( \mathbf{y,t}\right) }%
H_{mat}$ the operator $\delta O\left( \mathbf{x}\right) $ acts on $\mathbf{A}%
\left( \mathbf{x,}t\right) $ as :
\begin{equation*}
\delta O\left( \mathbf{x}\right) \mathbf{A}\left( \mathbf{x,}t\right)
=-\sum_{\alpha }\int d^{3}y\left[ \delta ^{ir}\delta ^{3}\left( \mathbf{x-y}%
\right) +\frac{\partial ^{2}}{\partial x^{i}\partial x^{r}}\frac{1}{4\pi
\left\vert \mathbf{x-y}\right\vert }\right] \delta O_{1}\left( \mathbf{y}%
\right) \mathbf{A}\left( \mathbf{y,}t\right)
\end{equation*}
where the operator $\delta O_{1}\left( \mathbf{x}\right) $ is obtained by
isolating in $-\frac{\delta }{\delta A^{j}\left( \mathbf{y,t}\right) }%
H_{mat} $ the linear term in the electromagnetic potential. Collecting the
relevant terms yields directly :
\begin{eqnarray*}
&&\left( \delta O_{1}\left( \mathbf{x}\right) \mathbf{A}\left( \mathbf{x,}%
t\right) \right) _{i} \\
&=&-\sum_{\alpha }\frac{\partial }{\partial \pi _{j}^{\left( \alpha \right)
}\left( \mathbf{x,t}\right) }\left( \hat{\varepsilon}_{0N^{\left( \alpha
\right) }}^{\left( \alpha \right) }\left( \mathbf{\pi }^{\left( \alpha
\right) }\right) +D^{\left( \alpha \right) }\right) \\
&&\times \left[ \left( \delta _{k}^{j}-\tilde{\emph{A}}\mathbf{%
^{R_{k}^{\left( \alpha \right) }}}\nabla _{\mathbf{r}_{j}^{\left( \alpha
\right) }}\right) \right] \left[ \frac{\partial \tilde{\emph{A}}%
^{R_{l}^{\left( \alpha \right) }}}{\partial \pi _{n}^{\left( \alpha \right) }%
}\left( \delta _{i}^{n}-\mathbf{\emph{A}^{R_{i}^{\left( \alpha \right) }}}%
\nabla _{\mathbf{r}_{n}^{\left( \alpha \right) }}\right) \right] \nabla
_{l}A^{k}\left( \mathbf{r}^{\left( \alpha \right) }\right) \\
&&+e\sum_{\alpha }\frac{\partial }{\partial \pi _{j}^{\left( \alpha \right)
}\left( \mathbf{x,t}\right) }\left( \hat{\varepsilon}_{0N^{\left( \alpha
\right) }}^{\left( \alpha \right) }\left( \mathbf{\pi }^{\left( \alpha
\right) }\right) +D^{\left( \alpha \right) }\right) \times \\
&&\left( -\nabla _{\mathbf{r}_{j}^{\left( \alpha \right) }}\left( \frac{%
\partial \tilde{\emph{\ A}}^{R_{n}^{\left( \alpha \right) }}}{\partial
A^{i}\left( \mathbf{r}^{\left( \alpha \right) }\right) }\right) +e\nabla _{%
\mathbf{r}_{m}^{\left( \alpha \right) }}\left( \frac{\partial \tilde{\emph{A}%
}^{R_{n}^{\left( \alpha \right) }}}{\partial \pi _{i}^{\left( \alpha \right)
}}-\frac{\partial \tilde{\emph{A}}^{R_{n}^{\left( \alpha \right) }}}{%
\partial \pi _{u}^{\left( \alpha \right) }}\tilde{\emph{A}}\mathbf{%
^{R_{i}^{\left( \alpha \right) }}}\nabla _{\mathbf{r}_{u}^{\left( \alpha
\right) }}\right) +\mathbf{\frac{d\mathbf{\emph{A}}_{u,r}}{dA^{i}\left(
\mathbf{r}^{\left( \alpha \right) }\right) }}\nabla _{\mathbf{r}_{u}^{\left(
\alpha \right) }}\nabla _{\mathbf{r}_{n}^{\left( \alpha \right) }}\right)
A^{n}\left( \mathbf{r}^{\left( \alpha \right) }\right) \\
&&-e\sum_{\alpha }\frac{\partial v\left( \mathbf{r}^{\left( \alpha \right)
}\right) }{\partial r_{j}^{\left( \alpha \right) }}\frac{\partial \tilde{%
\emph{\ A}}^{R_{j}^{\left( \alpha \right) }}}{\partial \pi _{n}^{\left(
\alpha \right) }}\times \\
&&\left( \nabla _{\mathbf{r}_{m}^{\left( \alpha \right) }}\left( \frac{%
\partial \tilde{\emph{A}}^{R_{n}^{\left( \alpha \right) }}}{\partial
A^{i}\left( \mathbf{r}^{\left( \alpha \right) }\right) }\right) -e\nabla _{%
\mathbf{r}_{m}^{\left( \alpha \right) }}\left( \frac{\partial \tilde{\emph{A}%
}^{R_{n}^{\left( \alpha \right) }}}{\partial \pi _{i}^{\left( \alpha \right)
}}-\frac{\partial \tilde{\emph{A}}^{R_{n}^{\left( \alpha \right) }}}{%
\partial \pi _{u}^{\left( \alpha \right) }}\mathbf{\emph{A}^{R_{i}^{\left(
\alpha \right) }}}\nabla _{\mathbf{r}_{u}^{\left( \alpha \right) }}\right) +%
\mathbf{\frac{d\mathbf{\emph{A}}_{u,r}}{dA^{i}\left( \mathbf{r}^{\left(
\alpha \right) }\right) }}\nabla _{\mathbf{r}_{u}^{\left( \alpha \right)
}}\nabla _{\mathbf{r}_{n}^{\left( \alpha \right) }}\right) A^{n}\left(
\mathbf{r}^{\left( \alpha \right) }\right) \\
&&+e\sum_{\alpha }\frac{\partial v\left( \mathbf{r}^{\left( \alpha \right)
}\right) }{\partial r_{j}^{\left( \alpha \right) }}\left( \delta _{k}^{j}-%
\tilde{\emph{A}}\mathbf{^{R_{k}^{\left( \alpha \right) }}}\nabla _{\mathbf{r}%
_{j}^{\left( \alpha \right) }}\right) \left[ \nabla _{l}\left( -e\frac{%
\partial \tilde{\emph{A}}^{R_{l}^{\left( \alpha \right) }}}{\partial \pi
_{n}^{\left( \alpha \right) }}\left( \delta _{i}^{n}-\tilde{\emph{A}}\mathbf{%
^{R_{i}^{\left( \alpha \right) }}}\nabla _{\mathbf{r}_{n}^{\left( \alpha
\right) }}\right) \right) \right] A^{k}\left( \mathbf{r}^{\left( \alpha
\right) }\right) \\
&&-\frac{1}{2}\sum_{\alpha }\mathbf{B}^{\left( \alpha \right) }\left(
\mathbf{x}^{\left( \alpha \right) }\right) .\left[ \frac{d\pi _{j}^{\left(
\alpha \right) }}{dA^{i}\left( \mathbf{r}^{\left( \alpha \right) }\right) }%
\frac{\partial }{\partial \pi _{j}^{\left( \alpha \right) }\left( \mathbf{x,t%
}\right) }+\frac{dr_{j}^{\left( \alpha \right) }}{dA^{i}\left( \mathbf{r}%
^{\left( \alpha \right) }\right) }\frac{\partial }{\partial r_{j}^{\left(
\alpha \right) }}\right] \left( \mathbf{\mu }\left( \mathbf{x}^{\left(
\alpha \right) }\right) +\mathbf{\hat{\mu}}^{k}\left( \mathbf{x}\right)
\mathbf{.}\nabla _{R_{k}}\mathbf{.B}^{\left( \alpha \right) }\left( \mathbf{x%
}^{\left( \alpha \right) }\right) \right) \\
&&+\frac{1}{2}e\sum_{\alpha }\left[ \left( \mathbf{\mu }\left( \mathbf{x}%
^{\left( \alpha \right) }\right) \times \mathbf{\nabla }\right) _{p}.\left(
\nabla _{l}A^{p}\left( \mathbf{r}^{\left( \alpha \right) }\right) \left(
\frac{\partial \tilde{\emph{A}}^{R_{l}^{\left( \alpha \right) }}}{\partial
\pi _{n}^{\left( \alpha \right) }}\left( \delta _{i}^{n}-\tilde{\emph{A}}%
\mathbf{^{R_{i}^{\left( \alpha \right) }}}\nabla _{\mathbf{r}_{n}^{\left(
\alpha \right) }}\right) \right) \right) \delta \left( \mathbf{x-r}^{\left(
\alpha \right) }\right) \right] \\
&&-\frac{1}{2}\left[ \mathbf{B}\left( \mathbf{x}^{\left( \alpha \right)
}\right) \mathbf{.\tilde{\mu}}\left( \mathbf{x}^{\left( \alpha \right)
}\right) \times \mathbf{\nabla }\delta \left( \mathbf{x-R}^{\left( \alpha
\right) }\right) -\mathbf{\nabla }\delta \left( \mathbf{x-R}^{\left( \alpha
\right) }\right) \times \mathbf{\tilde{\mu}}\left( \mathbf{x}^{\left( \alpha
\right) }\right) \mathbf{.B}\left( \mathbf{x}^{\left( \alpha \right)
}\right) \right] _{i} \\
&&-\left( \frac{\delta }{\delta \mathbf{A}\left( \mathbf{x}\right) }\frac{%
\delta }{\delta \mathbf{A}_{j}\left( \mathbf{x}\right) }H_{mat}\right)
_{\mid \mathbf{A}=0}\mathbf{A}_{j}\left( \mathbf{x}\right) +\mathbf{B}%
^{\left( \alpha \right) }\mathbf{.\tilde{\mu}.B}^{\left( \alpha \right) }
\end{eqnarray*}

The Fourier transform of this operator is straightforwardly obtained as
being equal to :
\begin{eqnarray*}
&&\delta O_{1}\left( \mathbf{p}\right)  \\
&=&-\sum_{\alpha }\frac{\partial }{\partial \pi _{j}^{\left( \alpha \right)
}\left( \mathbf{x,t}\right) }\left( \hat{\varepsilon}_{0N^{\left( \alpha
\right) }}^{\left( \alpha \right) }\left( \mathbf{\pi }^{\left( \alpha
\right) }\right) +D^{\left( \alpha \right) }\right)  \\
&&\times \left[ \left( \delta _{k}^{j}-\tilde{\emph{A}}\mathbf{%
^{R_{k}^{\left( \alpha \right) }}}\nabla _{\mathbf{r}_{j}^{\left( \alpha
\right) }}\right) \right] \left[ \frac{\partial \tilde{\emph{A}}%
^{R_{l}^{\left( \alpha \right) }}}{\partial \pi _{n}^{\left( \alpha \right) }%
}\left( \delta _{i}^{n}-\tilde{\emph{A}}\mathbf{^{R_{i}^{\left( \alpha
\right) }}}\nabla _{\mathbf{r}_{n}^{\left( \alpha \right) }}\right) \right]
p_{l}A^{k}\left( \mathbf{r}^{\left( \alpha \right) }\right)  \\
&&+e\sum_{\alpha }\frac{\partial }{\partial \pi _{j}^{\left( \alpha \right)
}\left( \mathbf{x,t}\right) }\left( \hat{\varepsilon}_{0N^{\left( \alpha
\right) }}^{\left( \alpha \right) }\left( \mathbf{\pi }^{\left( \alpha
\right) }\right) +D^{\left( \alpha \right) }\right) \times  \\
&&\left( -\nabla _{\mathbf{r}_{j}^{\left( \alpha \right) }}\left( \frac{%
\partial \tilde{\emph{\ A}}^{R_{n}^{\left( \alpha \right) }}}{\partial
A^{i}\left( \mathbf{r}^{\left( \alpha \right) }\right) }\right) +e\nabla _{%
\mathbf{r}_{m}^{\left( \alpha \right) }}\left( \frac{\partial \tilde{\emph{A}%
}^{R_{n}^{\left( \alpha \right) }}}{\partial \pi _{i}^{\left( \alpha \right)
}}-\frac{\partial \tilde{\emph{A}}^{R_{n}^{\left( \alpha \right) }}}{%
\partial \pi _{u}^{\left( \alpha \right) }}\tilde{\emph{A}}\mathbf{%
^{R_{i}^{\left( \alpha \right) }}}\nabla _{\mathbf{r}_{u}^{\left( \alpha
\right) }}\right) +\mathbf{\frac{d\mathbf{\emph{A}}_{u,r}}{dA^{i}\left(
\mathbf{r}^{\left( \alpha \right) }\right) }}\nabla _{\mathbf{r}_{u}^{\left(
\alpha \right) }}\nabla _{\mathbf{r}_{n}^{\left( \alpha \right) }}\right)
A^{n}\left( \mathbf{r}^{\left( \alpha \right) }\right)  \\
&&-e\sum_{\alpha }\frac{\partial v\left( \mathbf{r}^{\left( \alpha \right)
}\right) }{\partial r_{j}^{\left( \alpha \right) }}\frac{\partial \tilde{%
\emph{\ A}}^{R_{j}^{\left( \alpha \right) }}}{\partial \pi _{n}^{\left(
\alpha \right) }}\times  \\
&&\left( \nabla _{\mathbf{r}_{m}^{\left( \alpha \right) }}\left( \frac{%
\partial \tilde{\emph{A}}^{R_{n}^{\left( \alpha \right) }}}{\partial
A^{i}\left( \mathbf{r}^{\left( \alpha \right) }\right) }\right) -e\nabla _{%
\mathbf{r}_{m}^{\left( \alpha \right) }}\left( \frac{\partial \tilde{\emph{A}%
}^{R_{n}^{\left( \alpha \right) }}}{\partial \pi _{i}^{\left( \alpha \right)
}}-\frac{\partial \tilde{\emph{A}}^{R_{n}^{\left( \alpha \right) }}}{%
\partial \pi _{u}^{\left( \alpha \right) }}\tilde{\emph{A}}\mathbf{%
^{R_{i}^{\left( \alpha \right) }}}\nabla _{\mathbf{r}_{u}^{\left( \alpha
\right) }}\right) +\mathbf{\frac{d\mathbf{\emph{A}}_{u,r}}{dA^{i}\left(
\mathbf{r}^{\left( \alpha \right) }\right) }}\nabla _{\mathbf{r}_{u}^{\left(
\alpha \right) }}\nabla _{\mathbf{r}_{n}^{\left( \alpha \right) }}\right)
A^{n}\left( \mathbf{r}^{\left( \alpha \right) }\right)  \\
&&+e\sum_{\alpha }\frac{\partial v\left( \mathbf{r}^{\left( \alpha \right)
}\right) }{\partial r_{j}^{\left( \alpha \right) }}\left( \delta _{k}^{j}-%
\tilde{\emph{A}}\mathbf{^{R_{k}^{\left( \alpha \right) }}}\nabla _{\mathbf{r}%
_{j}^{\left( \alpha \right) }}\right) \left[ \nabla _{l}\left( -e\frac{%
\partial \tilde{\emph{A}}^{R_{l}^{\left( \alpha \right) }}}{\partial \pi
_{n}^{\left( \alpha \right) }}\left( \delta _{i}^{n}-\tilde{\emph{A}}\mathbf{%
^{R_{i}^{\left( \alpha \right) }}}\nabla _{\mathbf{r}_{n}^{\left( \alpha
\right) }}\right) \right) \right] A^{k}\left( \mathbf{r}^{\left( \alpha
\right) }\right)  \\
&&-\frac{1}{2}\sum_{\alpha }\mathbf{B}^{\left( \alpha \right) }\left(
\mathbf{x}^{\left( \alpha \right) }\right) \left[ \frac{d\pi _{j}^{\left(
\alpha \right) }}{dA^{i}\left( \mathbf{r}^{\left( \alpha \right) }\right) }%
\frac{\partial }{\partial \pi _{j}^{\left( \alpha \right) }\left( \mathbf{x,t%
}\right) }+\frac{dr_{j}^{\left( \alpha \right) }}{dA^{i}\left( \mathbf{r}%
^{\left( \alpha \right) }\right) }\frac{\partial }{\partial r_{j}^{\left(
\alpha \right) }}\right] \left( \mathbf{\mu }\left( \mathbf{x}^{\left(
\alpha \right) }\right) +\mathbf{\hat{\mu}}^{k}\left( \mathbf{x}\right)
\mathbf{.}\nabla _{R_{k}}\mathbf{.B}^{\left( \alpha \right) }\left( \mathbf{x%
}^{\left( \alpha \right) }\right) \right)  \\
&&+\mathbf{B}^{\left( \alpha \right) }\mathbf{.\tilde{\mu}.B}^{\left( \alpha
\right) }
\end{eqnarray*}

\end{document}